\documentclass[
	english, 
	letterpaper,twoside
]{article}

\def\documenttitle {Inequities in Breast Cancer Outcomes in Chile: An Analysis of Case Fatality and Survival Rates (2007--2018)}
\def\documentsubtitle {}
\def\documentdate {\today}
\def\journalname {Revista, conferencia del artículo}

\def\iitembcirc {\raisebox{.55\height}{\scriptsize$\bullet$}}
\def\iitembsquare {\raisebox{.3\height}{\tiny$\blacksquare$}}
\def\iitemdash {\raisebox{.35\height}{\textendash}}
\def\iitemcirc {\raisebox{.25\height}{\small$\circ$}}

\def\compilertype {pdf2latex}      
\def\documentfontsize {9.5}        
\def\documentinterline {1}         
\def\documentparindent {15}        
\def\documentparskip {0}           
\def\fontdocument {libertine}      
\def\fonttypewriter {tmodern}      
\def\fonturl {same}                
\def\graphicxdraft {false}         
\def\pointdecimal {true}           
\def\showlayoutlines {false}       
\def\showlinenumbers {false}       

\def\hfstyle {style1}              
\def\titleauthorspacing {0.35}     
\def\titleauthormaxwidth {0.85}    
\def\titlebold {true}              
\def\titlestyle {style2}           

\def\captionalignment {justified}  
\def\captionfontsize {small}       
\def\captionlabelformat {simple}   
\def\captionlabelsep {colon}       
\def\captionlessmarginimage {0.1}  
\def\captionlrmargin {2}           
\def\captionlrmarginmc {0.5}       
\def\captionmarginimage {0}        
\def\captionmarginimages {0}       
\def\captionmarginimagesmc {0}     
\def\captionmarginmultimg {0}      
\def\captionnumcode {arabic}       
\def\captionnumequation {arabic}   
\def\captionnumfigure {arabic}     
\def\captionnumsubfigure {alph}    
\def\captionnumsubtable {alph}     
\def\captionnumtable {arabic}      
\def\captionsubchar {.}            
\def\captiontbmarginfigure {9.35}  
\def\captiontbmargintable {7}      
\def\captiontextbold {true}        
\def\captiontextsubnumbold {false} 
\def\codecaptiontop {true}         
\def\equationcaptioncenter {true}  
\def\figurecaptiontop {false}      
\def\marginaligncaptbottom {0.1}   
\def\marginaligncapttop {-0.75}    
\def\marginalignedcaptbottom {0.1} 
\def\marginalignedcapttop {-0.75}  
\def\margineqncaptionbottom {0}    
\def\margineqncaptiontop {-0.7}    
\def\margingathercaptbottom {0.1}  
\def\margingathercapttop {-0.9}    
\def\margingatheredcaptbottom{0.1} 
\def\margingatheredcapttop {-0.7}  
\def\sectioncaptiondelimiter {.}   
\def\showsectioncaptioncode {none} 
\def\showsectioncaptioneqn {none}  
\def\showsectioncaptionfig {none}  
\def\showsectioncaptionmat {none}  
\def\showsectioncaptiontab {none}  
\def\subcaptionfsize{footnotesize} 
\def\subcaptionlabelformat{parens} 
\def\subcaptionlabelsep {space}    
\def\tablecaptiontop {false}        

\def\apacitebothers {et al.}       
\def\apaciterefcitecharclose {]}   
\def\apaciterefcitecharopen {[}    
\def\apaciterefnumber {false}      
\def\apaciterefsep {2}             
\def\apaciteshowurl {false}        
\def\apacitestyle {apacite}        
\def\appendixindepobjnum {true}    
\def\backrefpagecite {false}       
\def\bibtexenvrefsecnum {false}    
\def\bibtexrefsep {2}              
\def\bibtexstyle {apa}          
\def\bibtextextalign {justify}     
\def\fontsizerefbibl {\small}      
\def\natbibrefcitecharclose {]}    
\def\natbibrefcitecharopen {[}     
\def\natbibrefcitecompress {true}  
\def\natbibrefcitesepcomma {true}  
\def\natbibrefcitetype {numbers}   
\def\natbibrefsep {2}              
\def\natbibrefstyle {ieeetr}    
\def\stylecitereferences {natbib}  
\def\twocolumnreferences {false}   

\def\columnsepwidth {2.1}          
\def\defaultimagefolder {img/}     
\def\equationleftalign {false}     
\def\equationrestart {none}        
\def\footnotelmargin {10}          
\def\footnoterestart {none}        
\def\footnoterulefigure {false}    
\def\footnoterulepage {false}      
\def\footnoteruletable {false}     
\def\footnotetwocolumn {false}     
\def\fpremovetopbottomcenter{true} 
\def\imagedefaultplacement {H}     
\def\marginalignbottom {-0.4}      
\def\marginalignedbottom {-0.2}    
\def\marginalignedtop {-0.4}       
\def\marginaligntop {-0.4}         
\def\marginequationbottom {-0.2}   
\def\marginequationtop {0}         
\def\marginfloatimages {-13}       
\def\margingatherbottom {-0.2}     
\def\margingatheredbottom {-0.1}   
\def\margingatheredtop {-0.4}      
\def\margingathertop {-0.4}        
\def\marginimagebottom {-0.15}     
\def\marginimagemultright {0.5}    
\def\marginimagemulttop {-0.3}     
\def\marginimagetop {0}            
\def\numberedequation {true}       
\def\senumerti {\arabic{enumi}.}   
\def\senumertii {\alph{enumii})}   
\def\senumertiii{\roman{enumiii}.} 
\def\senumertiv {\Alph{enumiv})}   
\def\sitemizei {\iitembcirc}       
\def\sitemizeii {\iitemdash}       
\def\sitemizeiii {\iitemcirc}      
\def\sitemizeiv {\iitembsquare}    
\def\sourcecodebtmargin {0}        
\def\sourcecodefontf {\ttfamily}   
\def\sourcecodefonts {\small}      
\def\sourcecodeilfontf {\ttfamily} 
\def\sourcecodeilfonts {\small}    
\def\sourcecodenumbersep {6}       
\def\sourcecodenumbersize {\tiny}  
\def\sourcecodeskipabove {0.75}    
\def\sourcecodeskipbelow {1}       
\def\sourcecodetabsize {3}         
\def\tabledefaultplacement {H}     
\def\tablenotesameline {true}      
\def\tablenotesfontsize{\footnotesize} 
\def\tablepaddingh {0.75}          
\def\tablepaddingv {1.15}          
\def\tikzdefaultplacement {H}      

\def\anumsecaddtocounter {false}   
\def\charaftersectionnum {}        
\def\charappendixsection {}        
\def\paragfontsize {\normalsize}   
\def\paragfontstyle {\bfseries}    
\def\paragspacingbottom {4}        
\def\paragspacingleft {0}          
\def\paragspacingtop {8}           
\def\paragsubfontsize{\normalsize} 
\def\paragsubfontstyle {\bfseries} 
\def\paragsubspacingbottom {4}     
\def\paragsubspacingleft {0}       
\def\paragsubspacingtop {8}        
\def\sectionfontsize {\Large}      
\def\sectionfontstyle {\bfseries}  
\def\sectionspacingbottom {10}     
\def\sectionspacingleft {0}        
\def\sectionspacingtop {15}        
\def\spacingaftersection {\quad}   
\def\ssectionfontsize {\large}     
\def\ssectionfontstyle {\bfseries} 
\def\ssectionspacingbottom {8}     
\def\ssectionspacingleft {0}       
\def\ssectionspacingtop {12}       
\def\sssectionfontsize{\normalsize}
\def\sssectionfontstyle{\bfseries} 
\def\sssectionspacingbottom {6}    
\def\sssectionspacingleft {0}      
\def\sssectionspacingtop {10}      
\def\ssssectionfontstyle{\bfseries}
\def\ssssectionfontsz{\normalsize} 
\def\ssssectionspacingbottom {4}   
\def\ssssectionspacingleft {0}     
\def\ssssectionspacingtop {8}      

\def\captioncolor {black}          
\def\captiontextcolor {black}      
\def\highlightcolor {yellow}       
\def\linenumbercolor {gray}        
\def\linkcolor {black}             
\def\maintextcolor {black}         
\def\numcitecolor {black}          
\def\pagescolor {white}            
\def\paragcolor {black}            
\def\paragsubcolor {black}         
\def\sectioncolor {black}          
\def\showborderonlinks {false}     
\def\sourcecodebgcolor {lgray}     
\def\ssectioncolor {black}         
\def\sssectioncolor {black}        
\def\ssssectioncolor {black}       
\def\tablelinecolor {black}        
\def\tablerowfirstcolor {none}     
\def\tablerowsecondcolor {gray!20} 
\def\urlcolor {blue}            

\def\pagemarginbottom {1.91}       
\def\pagemarginleft {1.27}         
\def\pagemarginright {1.27}        
\def\pagemargintop {1.91}          

\def\cfgbookmarksopenlevel {1}     
\def\cfgpdfbookmarkopen {true}     
\def\cfgpdfcenterwindow {true}     
\def\cfgpdfcopyright {}            
\def\cfgpdfdisplaydoctitle {true}  
\def\cfgpdffitwindow {false}       
\def\cfgpdfkeywords {}             
\def\cfgpdflayout {OneColumn}      
\def\cfgpdfmenubar {true}          
\def\cfgpdfpageview {FitH}         
\def\cfgpdfsecnumbookmarks {true}  
\def\cfgpdftoolbar {true}          
\def\cfgshowbookmarkmenu {false}   
\def\indexdepth {4}                
\def\pdfcompilecompression {9}     
\def\pdfcompileobjcompression {2}  
\def\pdfcompileversion {7}         
\def\usepdfmetadata {true}         

\def\nameabstract {Abstract}           
\def\nameappendixsection {Appendix}     
\def\namechapter {Chapter}           
\def\nameltappendixsection {Appendix}    
\def\nameltcont{Índice de Contenidos} 
\def\namelteqn {Índice de Ecuaciones} 
\def\nameltfigure {Índice de Figuras} 
\def\nameltsrc {Índice de Códigos}    
\def\namelttable {Índice de Tablas}   
\def\nameltwfigure {Figure}           
\def\nameltwsrc {Código}              
\def\nameltwtable {Table}             
\def\namemathcol {Corolario}          
\def\namemathdefn {Definición}        
\def\namemathej {Ejemplo}             
\def\namemathlem {Lema}               
\def\namemathobs {Observación}        
\def\namemathprp {Proposición}        
\def\namemaththeorem {Teorema}        
\def\namepageof { de }                
\def\namereferences {References}     

\let\counterwithin\relax
\let\underbar\relax
\let\underline\relax

\def\unaccentedoperators {}
\def\decimalpoint {}
\def\bibname {}

\makeatletter
\def\underline#1{\relax\ifmmode\@@underline{#1}\else $\@@underline{\hbox{#1}}\m@th$\relax\fi}
\def\underbar#1{\underline{\sbox\tw@{#1}\dp\tw@\z@\box\tw@}}
\makeatother

\usepackage{ifthen}

\usepackage{tracklang}
\IfTrackedLanguage{spanish}{
	\usepackage[es-nosectiondot,es-lcroman,es-noquoting]{babel}
}{ 
	\usepackage{babel}
}

\usepackage{sectsty}

\ifthenelse{\equal{\compilertype}{pdf2latex}}{
	\usepackage[utf8]{inputenc}}{
}

\newcommand{\throwbadconfig}[4][]{
	\ifthenelse{\equal{#1}{noheader}}{
		\errmessage{LaTeX Warning: #4}
	}{
		\ifthenelse{\equal{#1}{noheader-nostop}}{
			\errmessage{LaTeX Warning: #4}
		}{
			\errmessage{LaTeX Warning: #2 (\noexpand #3= #3). Valores esperados: #4}
		}
	}
	\ifthenelse{\equal{#1}{nostop}}{}{
		\ifthenelse{\equal{#1}{noheader-nostop}}{}{
			\stop
		}
	}
}

\ifthenelse{\equal{\equationleftalign}{true}}{
	\usepackage[fleqn]{amsmath}
}{
	\usepackage{amsmath}
}

\usepackage{scrextend}
\usepackage{anyfontsize}
\changefontsizes{\documentfontsize pt}

\usepackage{amssymb}       
\usepackage{amsthm}        
\usepackage{array}         
\usepackage{bigstrut}      
\usepackage{bm}            
\usepackage{booktabs}      
\usepackage{caption}       
\usepackage{changepage}    
\usepackage{chngcntr}      
\usepackage{color}         
\usepackage{datetime}      
\usepackage{floatpag}      
\usepackage{floatrow}      
\usepackage{framed}        
\usepackage{gensymb}       
\usepackage{graphicx}      
\usepackage{lipsum}        
\usepackage{listings}      
\usepackage{longtable}     
\usepackage{mathtools}     
\usepackage{multicol}      
\usepackage{needspace}     
\usepackage{pdflscape}     
\usepackage{pdfpages}      
\usepackage{physics}       
\usepackage{realboxes}     
\usepackage{rotating}      
\usepackage{selinput}      
\usepackage{setspace}      
\usepackage{soul}          
\usepackage{stfloats}      
\usepackage{subfig}        
\usepackage{textcomp}      
\usepackage{wrapfig}       
\usepackage{xspace}        
\usepackage{xurl}          

\usepackage[makeroom]{cancel} 
\usepackage[inline]{enumitem} 
\usepackage[subfigure,titles]{tocloft} 
\usepackage[figure,table,lstlisting]{totalcount} 
\usepackage[normalem]{ulem} 
\usepackage[nointegrals]{wasysym} 
\usepackage[dvipsnames,table,usenames]{xcolor} 

\ifthenelse{\equal{\graphicxdraft}{true}}{
	\usepackage[
		allfiguresdraft,
		filename,
		size={scriptsize},
		style={tt}
	]{draftfigure}}{
}

\ifthenelse{\equal{\compilertype}{pdf2latex}}{
	\usepackage{listingsutf8}}{
}

\ifthenelse{\equal{\footnotetwocolumn}{true}}{
	\usepackage{dblfnote}}{
}

\ifthenelse{\equal{\footnoterulepage}{true}}{
	\usepackage[bottom,hang]{footmisc} 
}{
	\usepackage[bottom,norule,hang]{footmisc}
}

\ifthenelse{\equal{\backrefpagecite}{true}}{
	\usepackage[pdfencoding=auto,psdextra,backref=page]{hyperref} 
}{
	\usepackage[pdfencoding=auto,psdextra]{hyperref} 
}
\ifthenelse{\equal{\stylecitereferences}{natbib}}{ 
	\usepackage[nottoc,notlof,notlot]{tocbibind}
	\ifthenelse{\equal{\natbibrefcitecompress}{true}}{
		\usepackage[sort&compress]{natbib}
	}{
		\usepackage{natbib}
	}
}{
\ifthenelse{\equal{\stylecitereferences}{apacite}}{ 
	\usepackage[nottoc,notlof,notlot]{tocbibind}
	\usepackage{apacite}
}{
\ifthenelse{\equal{\stylecitereferences}{bibtex}}{ 
}{
\ifthenelse{\equal{\stylecitereferences}{custom}}{ 
}{}}}
}

\def\showappendixsecindex {false}
\ifthenelse{\equal{\showappendixsecindex}{true}}{
}{
	\usepackage{appendix}
}

\ifthenelse{\equal{\compilertype}{lualatex}}{ 
	\usepackage[top=\pagemargintop cm,bottom=\pagemarginbottom cm,margin=\pagemarginleft cm]{geometry}
}{ 
	\usepackage{geometry}
}

\ifthenelse{\equal{\tablenotesameline}{true}}{
	\usepackage[para]{threeparttable}
}{
	\usepackage{threeparttable}
}

\usepackage{bookmark}      
\usepackage{fancyhdr}      
\usepackage{float}         
\usepackage{hyperxmp}      
\usepackage{multirow}      
\usepackage{notoccite}     
\usepackage{titlesec}      

\ifthenelse{\equal{\fontdocument}{lmodern}}{
	\usepackage{lmodern}
}{
\ifthenelse{\equal{\fontdocument}{arial}}{
	\usepackage{helvet}
	
}{
\ifthenelse{\equal{\fontdocument}{arial2}}{
	\usepackage{arial}
}{
\ifthenelse{\equal{\fontdocument}{times}}{
	\usepackage{mathptmx}
}{
\ifthenelse{\equal{\fontdocument}{mathptmx}}{
	\usepackage{mathptmx}
}{
\ifthenelse{\equal{\fontdocument}{helvet}}{
	
	\usepackage[scaled=0.95]{helvet}
	\usepackage[helvet]{sfmath}
}{
\ifthenelse{\equal{\fontdocument}{opensans}}{
	\usepackage[default,scale=0.95]{opensans}
}{
\ifthenelse{\equal{\fontdocument}{mathpazo}}{
	\usepackage{mathpazo}
}{
\ifthenelse{\equal{\fontdocument}{cambria}}{
	\usepackage{caladea}
}{
\ifthenelse{\equal{\fontdocument}{libertine}}{
	\usepackage[libertine]{newtxmath}
	\usepackage[tt=false]{libertine}
}{
\ifthenelse{\equal{\fontdocument}{custom}}{
}{

\ifthenelse{\equal{\fontdocument}{accanthis}}{
	\usepackage{accanthis}
}{
\ifthenelse{\equal{\fontdocument}{alegreya}}{
	\usepackage{Alegreya}
	
}{
\ifthenelse{\equal{\fontdocument}{alegreyasans}}{
	\usepackage[sfdefault]{AlegreyaSans}
	
}{
\ifthenelse{\equal{\fontdocument}{algolrevived}}{
	\usepackage{algolrevived}
}{
\ifthenelse{\equal{\fontdocument}{almendra}}{
	\usepackage{almendra}
}{
\ifthenelse{\equal{\fontdocument}{antpolt}}{
	\usepackage{antpolt}
}{
\ifthenelse{\equal{\fontdocument}{antpoltlight}}{
	\usepackage[light]{antpolt}
}{
\ifthenelse{\equal{\fontdocument}{anttor}}{
	\usepackage[math]{anttor}
}{
\ifthenelse{\equal{\fontdocument}{anttorcondensed}}{
	\usepackage[condensed,math]{anttor}
}{
\ifthenelse{\equal{\fontdocument}{anttorlight}}{
	\usepackage[light,math]{anttor}
}{
\ifthenelse{\equal{\fontdocument}{anttorlightcondensed}}{
	\usepackage[light,condensed,math]{anttor}
}{
\ifthenelse{\equal{\fontdocument}{arev}}{

	\usepackage{arev}
}{
\ifthenelse{\equal{\fontdocument}{arimo}}{
	\usepackage[sfdefault]{arimo}
	
}{
\ifthenelse{\equal{\fontdocument}{arvo}}{
	\usepackage{Arvo}
}{
\ifthenelse{\equal{\fontdocument}{baskervald}}{
	\usepackage{baskervald}
}{
\ifthenelse{\equal{\fontdocument}{baskervaldx}}{
	\usepackage[lf]{Baskervaldx}
	\usepackage[bigdelims,vvarbb]{newtxmath}
	\usepackage[cal=boondoxo]{mathalfa}
	
}{
\ifthenelse{\equal{\fontdocument}{berasans}}{
	\usepackage[scaled]{berasans}
	
}{
\ifthenelse{\equal{\fontdocument}{beraserif}}{
	\usepackage{bera}
}{
\ifthenelse{\equal{\fontdocument}{biolinum}}{
	\usepackage{libertine}
	
}{
\ifthenelse{\equal{\fontdocument}{bitter}}{
	\usepackage{bitter}
}{
\ifthenelse{\equal{\fontdocument}{boisik}}{
	
	\usepackage{boisik}
}{
\ifthenelse{\equal{\fontdocument}{bookman}}{
	\usepackage{bookman}
}{
\ifthenelse{\equal{\fontdocument}{cabin}}{
	\usepackage[sfdefault]{cabin}
	
}{
\ifthenelse{\equal{\fontdocument}{cabincondensed}}{
	\usepackage[sfdefault,condensed]{cabin}
	
}{
\ifthenelse{\equal{\fontdocument}{caladea}}{
	\usepackage{caladea}
}{
\ifthenelse{\equal{\fontdocument}{cantarell}}{
	\usepackage[default]{cantarell}
}{
\ifthenelse{\equal{\fontdocument}{carlito}}{
	\usepackage[sfdefault]{carlito}
	
}{
\ifthenelse{\equal{\fontdocument}{charterbt}}{
	\usepackage[bitstream-charter]{mathdesign}
}{
\ifthenelse{\equal{\fontdocument}{chivolight}}{
	\usepackage[familydefault,light]{Chivo}
}{
\ifthenelse{\equal{\fontdocument}{chivoregular}}{
	\usepackage[familydefault,regular]{Chivo}
}{
\ifthenelse{\equal{\fontdocument}{clara}}{
	\usepackage{clara}
}{
\ifthenelse{\equal{\fontdocument}{clearsans}}{
	\usepackage[sfdefault]{ClearSans}
	
}{
\ifthenelse{\equal{\fontdocument}{cochineal}}{
	\usepackage{cochineal}
}{
\ifthenelse{\equal{\fontdocument}{coelacanth}}{
	\usepackage[nf]{coelacanth}
	\let\oldnormalfont\normalfont
	\def\normalfont{\oldnormalfont\mdseries}
}{
\ifthenelse{\equal{\fontdocument}{coelacanthextralight}}{
	\usepackage[el,nf]{coelacanth}
	\let\oldnormalfont\normalfont
	\def\normalfont{\oldnormalfont\mdseries}
}{
\ifthenelse{\equal{\fontdocument}{coelacanthlight}}{
	\usepackage[l,nf]{coelacanth}
	\let\oldnormalfont\normalfont
	\def\normalfont{\oldnormalfont\mdseries}
}{
\ifthenelse{\equal{\fontdocument}{comfortaa}}{
	\usepackage[default]{comfortaa}
}{
\ifthenelse{\equal{\fontdocument}{comicneue}}{
	\usepackage[default]{comicneue}
}{
\ifthenelse{\equal{\fontdocument}{comicneueangular}}{
	\usepackage[default,angular]{comicneue}
}{
\ifthenelse{\equal{\fontdocument}{computerconcrete}}{
	\usepackage{concmath}
}{
\ifthenelse{\equal{\fontdocument}{computerconcreteeuler}}{

	\usepackage{beton}
	\usepackage{euler}
}{
\ifthenelse{\equal{\fontdocument}{computermodern}}{
}{
\ifthenelse{\equal{\fontdocument}{computermodernbright}}{
	\usepackage{cmbright}
}{
\ifthenelse{\equal{\fontdocument}{crimson}}{
	\usepackage{crimson}
}{
\ifthenelse{\equal{\fontdocument}{crimsonpro}}{
	\usepackage{CrimsonPro}
	\let\oldnormalfont\normalfont
	\def\normalfont{\oldnormalfont\mdseries}
}{
\ifthenelse{\equal{\fontdocument}{crimsonproextralight}}{
	\usepackage[extralight]{CrimsonPro}
	\let\oldnormalfont\normalfont
	\def\normalfont{\oldnormalfont\mdseries}
}{
\ifthenelse{\equal{\fontdocument}{crimsonprolight}}{
	\usepackage[light]{CrimsonPro}
	\let\oldnormalfont\normalfont
	\def\normalfont{\oldnormalfont\mdseries}
}{
\ifthenelse{\equal{\fontdocument}{crimsonpromedium}}{
	\usepackage[medium]{CrimsonPro}
	\let\oldnormalfont\normalfont
	\def\normalfont{\oldnormalfont\mdseries}
}{
\ifthenelse{\equal{\fontdocument}{cyklop}}{
	\usepackage{cyklop}
}{
\ifthenelse{\equal{\fontdocument}{dejavusans}}{
	\usepackage{DejaVuSans}
	
}{
\ifthenelse{\equal{\fontdocument}{dejavusanscondensed}}{
	\usepackage{DejaVuSansCondensed}
	
}{
\ifthenelse{\equal{\fontdocument}{domitian}}{
	\usepackage{mathpazo}
	\usepackage{domitian}
	
}{
\ifthenelse{\equal{\fontdocument}{droidsans}}{
	\usepackage[defaultsans]{droidsans}
	
}{
\ifthenelse{\equal{\fontdocument}{electrum}}{
	\usepackage[lf]{electrum}
}{	
\ifthenelse{\equal{\fontdocument}{erewhon}}{
	\usepackage[proportional,scaled=1.064]{erewhon}
	\usepackage[erewhon,vvarbb,bigdelims]{newtxmath}
	
}{
\ifthenelse{\equal{\fontdocument}{fbb}}{
	\usepackage{fbb}
}{
\ifthenelse{\equal{\fontdocument}{fetamont}}{
	\usepackage{fetamont}
	
}{
\ifthenelse{\equal{\fontdocument}{firasans}}{
	\usepackage[sfdefault]{FiraSans}
	
}{
\ifthenelse{\equal{\fontdocument}{firasansnewtxsf}}{
	\usepackage[sfdefault]{FiraSans}
	\usepackage{newtxsf}
}{
\ifthenelse{\equal{\fontdocument}{fourier}}{
	\usepackage{fourier}
}{
\ifthenelse{\equal{\fontdocument}{fouriernc}}{
	\usepackage{fouriernc}
}{
\ifthenelse{\equal{\fontdocument}{gfsartemisia}}{
	
	\usepackage{gfsartemisia}
}{
\ifthenelse{\equal{\fontdocument}{gfsartemisiaeuler}}{

	\usepackage{gfsartemisia-euler}
}{
\ifthenelse{\equal{\fontdocument}{heuristica}}{
	\usepackage{heuristica}
	\usepackage[heuristica,vvarbb,bigdelims]{newtxmath}
	
}{
\ifthenelse{\equal{\fontdocument}{iwona}}{
	\usepackage[math]{iwona}
}{
\ifthenelse{\equal{\fontdocument}{iwonacondensed}}{
	\usepackage[condensed,math]{iwona}
}{
\ifthenelse{\equal{\fontdocument}{iwonalight}}{
	\usepackage[light,math]{iwona}
}{
\ifthenelse{\equal{\fontdocument}{iwonalightcondensed}}{
	\usepackage[light,condensed,math]{iwona}
}{
\ifthenelse{\equal{\fontdocument}{kerkis}}{
	\usepackage{kmath,kerkis}
}{
\ifthenelse{\equal{\fontdocument}{kurier}}{
	\usepackage[math]{kurier}
}{
\ifthenelse{\equal{\fontdocument}{kuriercondensed}}{
	\usepackage[condensed,math]{kurier}
}{
\ifthenelse{\equal{\fontdocument}{kurierlight}}{
	\usepackage[light,math]{kurier}
}{
\ifthenelse{\equal{\fontdocument}{kurierlightcondensed}}{
	\usepackage[light,condensed,math]{kurier}
}{
\ifthenelse{\equal{\fontdocument}{lato}}{
	\usepackage[default]{lato}
}{
\ifthenelse{\equal{\fontdocument}{libertinus}}{
	\usepackage{libertinus}
}{
\ifthenelse{\equal{\fontdocument}{librebaskerville}}{
	\usepackage{librebaskerville}
}{
\ifthenelse{\equal{\fontdocument}{librebodoni}}{
	\usepackage{LibreBodoni}
}{
\ifthenelse{\equal{\fontdocument}{librecaslon}}{
	\usepackage{librecaslon}
}{
\ifthenelse{\equal{\fontdocument}{libris}}{
	\usepackage{libris}
	
}{
\ifthenelse{\equal{\fontdocument}{lxfonts}}{
	\usepackage{lxfonts}
}{
\ifthenelse{\equal{\fontdocument}{merriweather}}{
	\usepackage[sfdefault]{merriweather}
}{
\ifthenelse{\equal{\fontdocument}{merriweatherlight}}{
	\usepackage[sfdefault,light]{merriweather}
}{
\ifthenelse{\equal{\fontdocument}{mintspirit}}{
	\usepackage[default]{mintspirit}
}{
\ifthenelse{\equal{\fontdocument}{mlmodern}}{
	\usepackage{mlmodern}
}{
\ifthenelse{\equal{\fontdocument}{montserratalternatesextralight}}{
	\usepackage[defaultfam,extralight,tabular,lining,alternates]{montserrat}
	
}{
\ifthenelse{\equal{\fontdocument}{montserratalternatesregular}}{
	\usepackage[defaultfam,tabular,lining,alternates]{montserrat}
	
}{
\ifthenelse{\equal{\fontdocument}{montserratalternatesthin}}{
	\usepackage[defaultfam,thin,tabular,lining,alternates]{montserrat}
	
}{
\ifthenelse{\equal{\fontdocument}{montserratextralight}}{
	\usepackage[defaultfam,extralight,tabular,lining]{montserrat}
	
}{
\ifthenelse{\equal{\fontdocument}{montserratlight}}{
	\usepackage[defaultfam,light,tabular,lining]{montserrat}
	
}{
\ifthenelse{\equal{\fontdocument}{montserratregular}}{
	\usepackage[defaultfam,tabular,lining]{montserrat}
	
}{
\ifthenelse{\equal{\fontdocument}{montserratthin}}{
	\usepackage[defaultfam,thin,tabular,lining]{montserrat}
	
}{
\ifthenelse{\equal{\fontdocument}{newpx}}{
	\usepackage{newpxtext,newpxmath}
}{
\ifthenelse{\equal{\fontdocument}{nimbussans}}{
	\usepackage{nimbussans}
	
}{
\ifthenelse{\equal{\fontdocument}{noto}}{
	\usepackage[sfdefault]{noto}
	
}{
\ifthenelse{\equal{\fontdocument}{notoserif}}{
	\usepackage{notomath}
}{
\ifthenelse{\equal{\fontdocument}{opensansserif}}{
	\usepackage[default,oldstyle,scale=0.95]{opensans}
}{
\ifthenelse{\equal{\fontdocument}{overlock}}{
	\usepackage[sfdefault]{overlock}
	
}{
\ifthenelse{\equal{\fontdocument}{paratype}}{
	\usepackage{paratype}
	
}{
\ifthenelse{\equal{\fontdocument}{paratypesanscaption}}{
	\usepackage{PTSansCaption}
	
}{
\ifthenelse{\equal{\fontdocument}{paratypesansnarrow}}{
	\usepackage{PTSansNarrow}
	
}{
\ifthenelse{\equal{\fontdocument}{pxfonts}}{
	\usepackage{pxfonts}
}{
\ifthenelse{\equal{\fontdocument}{quattrocento}}{
	\usepackage[sfdefault]{quattrocento}
}{
\ifthenelse{\equal{\fontdocument}{raleway}}{
	\usepackage[default]{raleway}
}{
\ifthenelse{\equal{\fontdocument}{ralewayblack}}{
	\usepackage[black]{raleway}
}{
\ifthenelse{\equal{\fontdocument}{ralewayextralight}}{
	\usepackage[extralight]{raleway}
}{
\ifthenelse{\equal{\fontdocument}{ralewaymedium}}{
	\usepackage[medium]{raleway}
}{
\ifthenelse{\equal{\fontdocument}{ralewaylight}}{
	\usepackage[light]{raleway}
}{
\ifthenelse{\equal{\fontdocument}{ralewaythin}}{
	\usepackage[thin]{raleway}
}{
\ifthenelse{\equal{\fontdocument}{roboto}}{
	\usepackage[sfdefault]{roboto}
}{
\ifthenelse{\equal{\fontdocument}{robotocondensed}}{
	\usepackage[sfdefault,condensed]{roboto}
}{
\ifthenelse{\equal{\fontdocument}{robotolight}}{
	\usepackage[sfdefault,light]{roboto}
}{
\ifthenelse{\equal{\fontdocument}{robotolightcondensed}}{
	\usepackage[sfdefault,light,condensed]{roboto}
}{
\ifthenelse{\equal{\fontdocument}{robotothin}}{
	\usepackage[sfdefault,thin]{roboto}
}{
\ifthenelse{\equal{\fontdocument}{rosario}}{
	\usepackage[familydefault]{Rosario}
}{
\ifthenelse{\equal{\fontdocument}{sourcesanspro}}{
	\usepackage[default]{sourcesanspro}
}{
\ifthenelse{\equal{\fontdocument}{step}}{
	\usepackage[notext]{stix}
	\usepackage{step}
}{
\ifthenelse{\equal{\fontdocument}{stickstoo}}{
	\usepackage{stickstootext}
	\usepackage[stickstoo,vvarbb]{newtxmath}
}{
\ifthenelse{\equal{\fontdocument}{texgyrebonum}}{
	\usepackage{tgbonum}
}{
\ifthenelse{\equal{\fontdocument}{txfonts}}{
	\usepackage{txfonts}
}{
\ifthenelse{\equal{\fontdocument}{uarial}}{
	\usepackage{uarial}
	
}{
\ifthenelse{\equal{\fontdocument}{ugq}}{

}{
\ifthenelse{\equal{\fontdocument}{universalis}}{
	\usepackage[sfdefault]{universalis}
}{
\ifthenelse{\equal{\fontdocument}{universaliscondensed}}{
	\usepackage[condensed,sfdefault]{universalis}
}{
\ifthenelse{\equal{\fontdocument}{venturis}}{
	\usepackage[lf]{venturis}
	
}{
	\throwbadconfig[nostop]{Fuente desconocida}{\fontdocument}{(Fuentes recomendadas) lmodern,carial,arial2,times,mathptmx,helvet,opensans,mathpazo,cambria,libertine,custom}
	\throwbadconfig[noheader-nostop]{Fuente desconocida}{\fontdocument}{(Fuentes adicionales) accanthis,alegreya,alegreyasans,algolrevived,almendra,antpolt,antpoltlight,anttor,anttorcondensed,anttorlight,anttorlightcondensed,arev,arimo,arvo,baskervald,baskervaldx,berasans,beraserif,biolinum,bitter,boisik,bookman,cabin,cabincondensed,cantarell,caladea,carlito,charterbt,chivolight,chivoregular,clara,clearsans,cochineal,coelacanth,coelacanthextralight,coelacanthlight,comfortaa,comicneue,comicneueangular,computerconcrete,computerconcreteeuler,computermodern,computermodernbright,crimson,crimsonpro,crimsonproextralight,crimsonprolight,crimsonpromedium,cyklop}
	\throwbadconfig[noheader-nostop]{Fuente desconocida}{\fontdocument}{dejavusans,dejavusanscondensed,domitian,droidsans,electrum,erewhon,fbb,fetamont,firasans,firasansnewtxsf,fourier,fouriernc,gfsartemisia,gfsartemisiaeuler,heuristica,iwona,iwonacondensed,iwonalight,iwonalightcondensed,kerkis,kurier,kuriercondensed,kurierlight,kurierlightcondensed,lato,libertinus,librebaskerville,librebodoni,librecaslon,libris,lxfonts}
	\throwbadconfig[noheader]{Fuente desconocida}{\fontdocument}{merriweather,merriweatherlight,mintspirit,mlmodern,montserratalternatesextralight,montserratalternatesregular,montserratalternatesthin,montserratextralight,montserratlight,montserratregular,montserratthin,newpx,nimbussans,noto,notoserif,opensansserif,overlock,paratype,paratypesanscaption,paratypesansnarrow,pxfonts,quattrocento,raleway,ralewayblack,ralewayextralight,ralewaymedium,ralewaylight,ralewaythin,roboto,robotolight,robotolightcondensed,robotothin,rosario,sourcesanspro,step,stickstoo,uarial,texgyrebonum,txfonts,ugq,universalis,universaliscondensed,venturis}
	}}}}}}}}}}}}}}}}}}}}}}}}}}}}}}}}}}}}}}}}}}}}}}}}}}}}}}}}}}}}}}}}}}}}}}}}}}}}}}}}}}}}}}}}}}}}}}}}}}}}}}}}}}}}}}}}}}}}}}}}}}}}}}}}}}}}}}
}

\ifthenelse{\equal{\fonttypewriter}{custom}}{
}{
\ifthenelse{\equal{\fonttypewriter}{tmodern}}{
	
}{
\ifthenelse{\equal{\fonttypewriter}{anonymouspro}}{
	\usepackage[ttdefault=true]{AnonymousPro}
}{
\ifthenelse{\equal{\fonttypewriter}{ascii}}{
	\usepackage{ascii}
	
}{
\ifthenelse{\equal{\fonttypewriter}{beramono}}{
	\usepackage[scaled]{beramono}
}{
\ifthenelse{\equal{\fonttypewriter}{cascadiacode}}{
	\usepackage{cascadia-code}
}{
\ifthenelse{\equal{\fonttypewriter}{cmpica}}{
	\usepackage{addfont}
	\addfont{OT1}{cmpica}{\pica}
	\addfont{OT1}{cmpicab}{\picab}
	\addfont{OT1}{cmpicati}{\picati}
	
}{
\ifthenelse{\equal{\fonttypewriter}{cmodern}}{
}{
\ifthenelse{\equal{\fonttypewriter}{courier}}{
	\usepackage{courier}
}{
\ifthenelse{\equal{\fonttypewriter}{courier10}}{
	\usepackage{courierten}
}{
\ifthenelse{\equal{\fonttypewriter}{cmvtt}}{
	
}{
\ifthenelse{\equal{\fonttypewriter}{dejavusansmono}}{
	\usepackage[scaled]{DejaVuSansMono}
}{
\ifthenelse{\equal{\fonttypewriter}{droidsansmono}}{
	\usepackage[defaultmono]{droidsansmono}
}{
\ifthenelse{\equal{\fonttypewriter}{firamono}}{
	\usepackage[scale=0.85]{FiraMono}
}{
\ifthenelse{\equal{\fonttypewriter}{gomono}}{
	\usepackage[scale=0.85]{GoMono}
}{
\ifthenelse{\equal{\fonttypewriter}{inconsolata}}{
	\usepackage{inconsolata}
}{
\ifthenelse{\equal{\fonttypewriter}{nimbusmono}}{
	\usepackage{nimbusmono}
}{
\ifthenelse{\equal{\fonttypewriter}{newtxtt}}{
	\usepackage[zerostyle=d]{newtxtt}
}{
\ifthenelse{\equal{\fonttypewriter}{nimbusmono}}{
	\usepackage{nimbusmono}
}{
\ifthenelse{\equal{\fonttypewriter}{nimbusmononarrow}}{
	\usepackage{nimbusmononarrow}
}{
\ifthenelse{\equal{\fonttypewriter}{lcmtt}}{
	
}{
\ifthenelse{\equal{\fonttypewriter}{sourcecodepro}}{
	\usepackage[ttdefault=true,scale=0.85]{sourcecodepro}
}{
\ifthenelse{\equal{\fonttypewriter}{texgyrecursor}}{
	\usepackage{tgcursor}
}{
\ifthenelse{\equal{\fonttypewriter}{txtt}}{
	
}{
	\throwbadconfig{Fuente desconocida}{\fonttypewriter}{custom,anonymouspro,ascii,beramono,cascadiacode,cmpica,cmodern,courier,courier10,cvmtt,dejavusansmono,droidsansmono,firamono,gomono,inconsolata,kpmonospaced,lcmtt,newtxtt,nimbusmono,nimbusmononarrow,texgyrecursor,tmodern,txtt}
	}}}}}}}}}}}}}}}}}}}}}}}
}

\usepackage[T1]{fontenc} 
\ifthenelse{\equal{\showlayoutlines}{true}}{ 
	\usepackage{showframe}}{
}
\ifthenelse{\equal{\showlinenumbers}{true}}{ 
	\usepackage[switch,columnwise,running]{lineno}}{
}
\usepackage{csquotes} 
\ifthenelse{\equal{\compilertype}{pdf2latex}}{
	\inputencoding{utf8}}{
}

\global\def\GLOBALcaptiondefn {EMPTY-VAR}       
\global\def\GLOBALchapternumenabled {false}     
\global\def\GLOBALenvappendix {false}           
\global\def\GLOBALenvimageadded {false}         
\global\def\GLOBALenvimageinitialized {false}   
\global\def\GLOBALenvmulticol {false}           
\global\def\GLOBALsectionalph {false}           
\global\def\GLOBALsectionanumenabled {false}    
\global\def\GLOBALsubsectionanumenabled {false} 
\global\def\GLOBALtablerowcolorindex {2}        
\global\def\GLOBALtablerowcolorswitch {false}   

\def\LOCALpercentchar#1{}
\edef\LOCALpercentchar{\expandafter\LOCALpercentchar\string\%}

\newcounter{templateEquations}      
\newcounter{templateFigures}        
\newcounter{templateIndexEquations} 
\newcounter{templateListings}       
\newcounter{templatePageCounter}    
\newcounter{templateTables}         

\newcounter{templateBookmarksLevelPrev}
\addtocounter{templateBookmarksLevelPrev}{-1}

\newcounter{chapter}

\newcommand{\throwerror}[2]{%
	\errmessage{LaTeX Error: \noexpand#1 #2 (linea \the\inputlineno)}%
	\stop
}

\newcommand{\throwwarning}[1]{%
	\errmessage{LaTeX Warning: #1 (linea \the\inputlineno)}%
}

\newcommand{\throwbadconfigondoc}[3]{%
	\errmessage{#1 \noexpand #2=#2. Valores esperados: #3}%
	\stop
}

\newcommand{\checkvardefined}[1]{%
	\ifthenelse{\isundefined{#1}}{%
		\errmessage{LaTeX Warning: Variable \noexpand#1 no definida}%
		\stop}{%
	}%
}

\newcommand{\coretemplatemessage}[1]{%
	\message{Template: #1}%
}

\newcommand{\checkextravarexist}[2]{%
	\ifthenelse{\isundefined{#1}}{%
		\errmessage{LaTeX Warning: Variable \noexpand#1 no definida}%
		\ifx\hfuzz#2\hfuzz%
			\errmessage{LaTeX Warning: Defina la variable en el bloque de INFORMACION DEL DOCUMENTO al comienzo del archivo principal del template}%
		\else%
			\errmessage{LaTeX Warning: #2}%
		\fi}{%
	}
}

\newcommand{\emptyvarerr}[3]{%
	\ifx\hfuzz#2\hfuzz%
		\errmessage{LaTeX Warning: \noexpand#1 #3 (linea \the\inputlineno)}%
	\fi
}

\newcommand{\setcaptionmargincm}[1]{
	\captionsetup{margin=#1cm}
}

\newcommand{\setpagemargincm}[4]{
	\ifthenelse{\equal{\compilertype}{lualatex}}{
	}{
		\newgeometry{left=#1cm, top=#2cm, right=#3cm, bottom=#4cm}
	}
}

\newcommand{\resetindexcaption}{%
	\global\def\GLOBALcaptiondefn {EMPTY-VAR}%
	\hbadness=10000%
}

\newcommand{\changemargin}[2]{%
	\emptyvarerr{\changemargin}{#1}{Margen izquierdo no definido}%
	\emptyvarerr{\changemargin}{#2}{Margen derecho no definido}%
	\list{}{\rightmargin#2\leftmargin#1}\item[]%
}

\newcommand{\checkonlyonenvimage}{%
	\ifthenelse{\equal{\GLOBALenvimageinitialized}{true}}{}{%
		\throwwarning{Funciones \noexpand\addimage o \noexpand\addimageboxed no pueden usarse fuera del entorno \noexpand\images}\stop%
	}%
}

\newcommand{\checkoutsideenvimage}{%
	\ifthenelse{\equal{\GLOBALenvimageinitialized}{true}}{%
		\throwwarning{Esta funcion solo puede usarse fuera del entorno \noexpand\images}%
		\stop}{%
	}%
}

\newcommand{\checkinsidemulticol}{%
	\ifthenelse{\equal{\GLOBALenvmulticol}{false}}{%
		\throwwarning{Esta funcion solo puede usarse dentro de multicols}%
		\stop}{%
	}
}

\newcommand{\checkoutsideappendix}{%
	\ifthenelse{\equal{\GLOBALenvappendix}{true}}{%
		\throwwarning{Esta funcion solo puede usarse fuera de anexo}%
		\stop}{%
	}
}

\newcommand{\corecheckbooleanvar}[1]{%
	\emptyvarerr{\corecheckbooleanvar}{#1}{Variable no definida}%
	\ifthenelse{\equal{#1}{true}}{}{%
	\ifthenelse{\equal{#1}{false}}{}{%
		\throwwarning{Variable debe ser true o false}\stop%
	}}%
}

\newcommand{\verticallycentertext}[1]{%
	\emptyvarerr{\verticallycentertext}{#1}{Texto no definido}%
	\topskip0pt%
	\vspace*{\fill}%
	#1%
	\vspace*{\fill}%
}

\newlength{\coregluevarcm}
\newcommand{\corevspacevarcm}[1]{%
	\vspace{\dimexpr#1 cm plus #1\coregluevarcm minus #1\coregluevarcm}%
}

\makeatletter
\newcommand\addpathimage[1]{%
	\gappto\Ginput@path{{#1}}%
}
\makeatother

\newcommand{\corecheckfontsize}[1]{%
	\ifthenelse{\equal{#1}{normalsize}}{}{%
	\ifthenelse{\equal{#1}{small}}{}{%
	\ifthenelse{\equal{#1}{large}}{}{%
	\ifthenelse{\equal{#1}{Large}}{}{%
	\ifthenelse{\equal{#1}{LARGE}}{}{%
	\ifthenelse{\equal{#1}{huge}}{}{%
	\ifthenelse{\equal{#1}{Huge}}{}{%
	\ifthenelse{\equal{#1}{HUGE}}{}{%
	\ifthenelse{\equal{#1}{footnotesize}}{}{%
	\ifthenelse{\equal{#1}{scriptsize}}{}{%
	\ifthenelse{\equal{#1}{tiny}}{}{%
		\errmessage{LaTeX Warning: Tamano de fuente incorrecto (\noexpand #1= #1). Valores esperados: tiny,scriptsize,footnotesize,small,normalisize,large,Large,LARGE,huge,Huge,HUGE}%
		\stop%
		}}}}}}}}}}%
	}%
}

\newcommand{\aasin}[1][]{%
	\ifx\hfuzz#1\hfuzz
		\ensuremath{\sin^{-1}#1}
	\else
		\ensuremath{{\sin}^{-1}}
	\fi
}

\newcommand{\aacos}[1][]{%
	\ifx\hfuzz#1\hfuzz
		\ensuremath{\cos^{-1}#1}
	\else
		\ensuremath{\cos^{-1}}
	\fi
}

\newcommand{\aatan}[1][]{%
	\ifx\hfuzz#1\hfuzz
		\ensuremath{\tan^{-1}#1}
	\else
		\ensuremath{\tan^{-1}}
	\fi
}

\newcommand{\aacsc}[1][]{%
	\ifx\hfuzz#1\hfuzz
		\ensuremath{\csc^{-1}#1}
	\else
		\ensuremath{\csc^{-1}}
	\fi
}

\newcommand{\aasec}[1][]{%
	\ifx\hfuzz#1\hfuzz
		\ensuremath{\sec^{-1}#1}
	\else
		\ensuremath{\sec^{-1}}
	\fi
}

\newcommand{\aacot}[1][]{%
	\ifx\hfuzz#1\hfuzz
		\ensuremath{\cot^{-1}#1}
	\else
		\ensuremath{\cot^{-1}}
	\fi
}

\makeatletter
\def\longtilde#1{%
	\mathop{\vbox{\m@th\ialign{##\crcr\noalign{\kern3\p@}%
	\sortoftildefill\crcr\noalign{\kern3\p@\nointerlineskip}%
	$\hfil\displaystyle{#1}\hfil$\crcr}}}\limits%
}
\def\sortoftildefill {%
	$\m@th \setbox\z@\hbox{$\braceld$}%
	\braceld\leaders\vrule \@height\ht\z@ \@depth\z@\hfill\braceru$%
}
\makeatother

\ifthenelse{\isundefined{\C}}{\newcommand{\C}{C}}{}
\renewcommand{\C}{\ensuremath{\mathbb{C}}}

\ifthenelse{\isundefined{\G}}{\newcommand{\G}{G}}{}
\renewcommand{\G}{\ensuremath{\mathcal{G}}}

\ifthenelse{\isundefined{\H}}{\newcommand{\H}{H}}{}
\renewcommand{\H}{\ensuremath{\mathcal{H}}}

\ifthenelse{\isundefined{\U}}{\newcommand{\U}{U}}{}
\renewcommand{\U}{\ensuremath{\mathcal{U}}}

\ifthenelse{\equal{\fontdocument}{step}}{}{ 

}

	

\newcommand{\coreafterequationfn}{%
	\hbadness=10000%
}

\newcommand{\equationresize}[2]{%
	\emptyvarerr{\equationresize}{#1}{Dimension no definida}%
	\emptyvarerr{\equationresize}{#2}{Ecuacion a redimensionar no definida}%
	\resizebox{#1\linewidth}{!}{$#2$}%
}

\newcommand{\coreinsertequationcaption}[1]{%
	\begin{changemargin}{\captionlrmargin cm}{\captionlrmargin cm}%
		\ifthenelse{\equal{\equationcaptioncenter}{true}}{%
			\centering%
		}{
			\justifying%
		}%
		\textcolor{\captiontextcolor}{%
			\linespread{0.5}\selectfont{%
				\begin{\captionfontsize}#1\end{\captionfontsize}%
			}
		}%
	\end{changemargin}%
}

\newcommand{\insertequation}[2][]{%
	\emptyvarerr{\insertequation}{#2}{Ecuacion no definida}%
	\ifthenelse{\equal{\numberedequation}{true}}{%
		\corevspacevarcm{\marginequationtop}%
		\begin{samepage}%
		\begin{equation}%
			\text{#1} #2
		\end{equation}
		\corevspacevarcm{\marginequationbottom}%
		\end{samepage}
		\coreafterequationfn%
	}{%
		\ifx\hfuzz#1\hfuzz%
		\else%
			\throwwarning{Label invalido en ecuacion sin numero}
		\fi
		\insertequationanum{#2}%
	}
}

\newcommand{\insertequationanum}[1]{%
	\emptyvarerr{\insertequationanum}{#1}{Ecuacion no definida}%
	\corevspacevarcm{\marginequationtop}%
	\begin{samepage}%
	\begin{equation*}%
		\ensuremath{#1}
	\end{equation*}
	\corevspacevarcm{\marginequationbottom}%
	\end{samepage}
	\coreafterequationfn%
}

\newcommand{\insertindexequation}[3][]{%
	\emptyvarerr{\insertindexequation}{#2}{Ecuacion no definida}%
	\emptyvarerr{\insertindexequation}{#3}{Leyenda no definida}%
	\corevspacevarcm{\margineqnindextop}%
	\begin{samepage}%
	\begin{align}%
		\text{#1} \ensuremath{#2}
	\end{align}
	\myindexequations{#3}%
	\corevspacevarcm{\margineqnindexbottom}%
	\end{samepage}
	\coreinsertequationcaption{\textit{#3}}%
	\addtocounter{templateIndexEquations}{1}%
	\coreafterequationfn%
}

\newcommand{\insertequationleft}[2][]{%
	\emptyvarerr{\insertequationleft}{#2}{Ecuacion no definida}%
	\ifthenelse{\equal{\numberedequation}{true}}{%
		\vspace{\dimexpr\marginequationtop cm - \baselineskip}%
		\begin{samepage}%
		\begin{equation}
			\hfilneg \text{#1} #2 \hspace{10000pt minus 1fil}
		\end{equation}
		\vspace{\dimexpr-0.2\baselineskip + \marginequationbottom cm}%
		\end{samepage}
		\coreafterequationfn%
	}{%
		\ifx\hfuzz#1\hfuzz%
		\else
			\throwwarning{Label invalido en ecuacion sin numero}
		\fi
		\insertequationleftanum{#2}%
	}
}

\newcommand{\insertequationleftanum}[1]{%
	\emptyvarerr{\insertequationleftanum}{#1}{Ecuacion no definida}%
	\vspace{\dimexpr\marginequationtop cm - \baselineskip}%
	\begin{samepage}%
	\begin{equation*}
		\hfilneg \ensuremath{#1} \hspace{10000pt minus 1fil}
	\end{equation*}
	\vspace{\dimexpr-0.2\baselineskip + \marginequationbottom cm}%
	\end{samepage}
	\coreafterequationfn%
}

\newcommand{\insertequationright}[2][]{%
	\emptyvarerr{\insertequationright}{#2}{Ecuacion no definida}%
	\ifthenelse{\equal{\numberedequation}{true}}{%
		\vspace{\dimexpr\marginequationtop cm - \baselineskip}%
		\begin{samepage}%
		\begin{equation}
			\hspace{10000pt minus 1fil} \text{#1} #2 \hfilneg
		\end{equation}
		\vspace{\dimexpr-0.2\baselineskip + \marginequationbottom cm}%
		\end{samepage}
		\coreafterequationfn%
	}{%
		\ifx\hfuzz#1\hfuzz%
		\else
			\throwwarning{Label invalido en ecuacion sin numero}
		\fi
		\insertequationrightanum{#2}%
	}
}

\newcommand{\insertequationrightanum}[1]{%
	\emptyvarerr{\insertequationrightanum}{#1}{Ecuacion no definida}%
	\vspace{\dimexpr\marginequationtop cm - \baselineskip}%
	\begin{samepage}%
	\begin{equation*}
		\hspace{10000pt minus 1fil} \ensuremath{#1} \hfilneg
	\end{equation*}
	\vspace{\dimexpr-0.2\baselineskip + \marginequationbottom cm}%
	\end{samepage}
	\coreafterequationfn%
}

\newcommand{\insertequationcaptioned}[3][]{%
	\emptyvarerr{\insertequationcaptioned}{#2}{Ecuacion no definida}%
	\ifx\hfuzz#3\hfuzz%
		\insertequation[#1]{#2}%
	\else
		\ifthenelse{\equal{\numberedequation}{true}}{%
			\corevspacevarcm{\marginequationtop}%
			\begin{samepage}%
			\begin{equation}
				\text{#1} #2
			\end{equation}
			\corevspacevarcm{\margineqncaptiontop}%
			\coreinsertequationcaption{#3}%
			\corevspacevarcm{\margineqncaptionbottom}%
			\end{samepage}
			\coreafterequationfn%
		}{%
			\ifx\hfuzz#1\hfuzz
			\else
				\throwwarning{Label invalido en ecuacion sin numero}
			\fi
			\insertequationcaptionedanum{#2}{#3}
		}
	\fi
}

\newcommand{\insertequationcaptionedanum}[2]{%
	\emptyvarerr{\insertequationcaptionedanum}{#1}{Ecuacion no definida}%
	\ifx\hfuzz#2\hfuzz%
		\insertequationanum{#1}%
	\else
		\corevspacevarcm{\marginequationtop}%
		\begin{samepage}%
		\begin{equation*}
			\ensuremath{#1}%
		\end{equation*}%
		\corevspacevarcm{\margineqncaptiontop}%
		\coreinsertequationcaption{#2}%
		\corevspacevarcm{\margineqncaptionbottom}%
		\end{samepage}
		\coreafterequationfn%
	\fi
}

\newcommand{\insertgather}[1]{%
	\emptyvarerr{\insertgather}{#1}{Ecuacion no definida}%
	\ifthenelse{\equal{\numberedequation}{true}}{%
		\corevspacevarcm{\margingathertop}%
		\begin{samepage}%
		\begin{gather}%
			\ensuremath{#1}
		\end{gather}
		\corevspacevarcm{\margingatherbottom}%
		\end{samepage}
		\coreafterequationfn%
	}{%
		\insertgatheranum{#1}%
	}
}

\newcommand{\insertgatheranum}[1]{%
	\emptyvarerr{\insertgatheranum}{#1}{Ecuacion no definida}%
	\corevspacevarcm{\margingathertop}%
	\begin{samepage}%
	\begin{gather*}%
		\ensuremath{#1}
	\end{gather*}
	\corevspacevarcm{\margingatherbottom}%
	\end{samepage}
	\coreafterequationfn%
}

\newcommand{\insertgathercaptioned}[2]{%
	\emptyvarerr{\insertgathercaptioned}{#1}{Ecuacion no definida}%
	\ifx\hfuzz#2\hfuzz%
		\insertgather{#1}%
	\else
		\ifthenelse{\equal{\numberedequation}{true}}{%
			\corevspacevarcm{\margingathertop}%
			\begin{samepage}%
			\begin{gather}%
				\ensuremath{#1}
			\end{gather}
			\corevspacevarcm{\margingathercapttop}%
			\coreinsertequationcaption{#2}%
			\corevspacevarcm{\margingathercaptbottom}%
			\end{samepage}
			\coreafterequationfn%
		}{%
			\insertgathercaptionedanum{#1}{#2}%
		}
	\fi
}

\newcommand{\insertgathercaptionedanum}[2]{%
	\emptyvarerr{\insertgathercaptionedanum}{#1}{Ecuacion no definida}%
	\ifx\hfuzz#2\hfuzz%
		\insertgatheranum{#1}%
	\else
		\corevspacevarcm{\margingathertop}%
		\begin{samepage}%
		\begin{gather*}%
			\ensuremath{#1}
		\end{gather*}
		\corevspacevarcm{\margingathercapttop}%
		\coreinsertequationcaption{#2}%
		\corevspacevarcm{\margingathercaptbottom}%
		\end{samepage}
		\coreafterequationfn%
	\fi
}

\newcommand{\insertgathered}[2][]{%
	\emptyvarerr{\insertgathered}{#2}{Ecuacion no definida}%
	\ifthenelse{\equal{\numberedequation}{true}}{%
		\corevspacevarcm{\marginequationtop}%
		\begin{samepage}%
		\begin{equation}
			\begin{gathered}
				\text{#1} \ensuremath{#2}
			\end{gathered}
		\end{equation}
		\corevspacevarcm{\margingatheredbottom}%
		\end{samepage}
	}{%
		\ifx\hfuzz#1\hfuzz
		\else
			\throwwarning{Label invalido en ecuacion (gathered) sin numero}
		\fi
		\corevspacevarcm{\margingatheredtop}%
		\begin{samepage}%
		\begin{gather*}%
			\ensuremath{#2}
		\end{gather*}
		\corevspacevarcm{\margingatheredbottom}%
		\end{samepage}
	}
	\coreafterequationfn%
}

\newcommand{\insertgatheredanum}[1]{%
	\emptyvarerr{\insertgatheredanum}{#1}{Ecuacion no definida}%
	\corevspacevarcm{\margingatheredtop}%
	\begin{samepage}%
	\begin{gather*}
		\ensuremath{#1}
	\end{gather*}
	\vspace{\dimexpr-0.15cm + \margingatheredbottom cm}%
	\end{samepage}
	\coreafterequationfn%
}

\newcommand{\insertgatheredcaptioned}[3][]{%
	\emptyvarerr{\insertgatheredcaptioned}{#2}{Ecuacion no definida}%
	\ifx\hfuzz#3\hfuzz%
		\insertgathered[#1]{#2}%
	\else
		\ifthenelse{\equal{\numberedequation}{true}}{%
			\corevspacevarcm{\marginequationtop}%
			\begin{samepage}%
			\begin{equation}
				\begin{gathered}
					\text{#1} \ensuremath{#2}
				\end{gathered}
			\end{equation}
			\corevspacevarcm{\margingatheredcapttop}%
			\coreinsertequationcaption{#3}%
			\corevspacevarcm{\margingatheredcaptbottom}%
			\end{samepage}
			\coreafterequationfn%
		}{%
			\ifx\hfuzz#1\hfuzz
			\else
				\throwwarning{Label invalido en ecuacion (gathered) sin numero}
			\fi
			\insertgatheredcaptionedanum{#2}{#3}%
		}
		\fi
}

\newcommand{\insertgatheredcaptionedanum}[2]{
	\emptyvarerr{\insertgatheredcaptionedanum}{#1}{Ecuacion no definida}
	\ifx\hfuzz#2\hfuzz%
		\insertgatheredanum{#1}%
	\else
		\corevspacevarcm{\margingatheredtop}%
		\begin{samepage}%
		\begin{gather*}
			\ensuremath{#1}
		\end{gather*}
		\vspace{\dimexpr-0.2cm + \margingatheredcapttop cm}%
		\coreinsertequationcaption{#2}%
		\vspace{\dimexpr-0.05cm + \margingatheredcaptbottom cm}%
		\end{samepage}
		\coreafterequationfn%
	\fi
}

\newcommand{\insertalign}[1]{%
	\emptyvarerr{\insertalign}{#1}{Ecuacion no definida}%
	\ifthenelse{\equal{\numberedequation}{true}}{%
		\corevspacevarcm{\marginaligntop}%
		\begin{samepage}%
		\begin{align}
			\ensuremath{#1}
		\end{align}
		\corevspacevarcm{\marginalignbottom}%
		\end{samepage}
		\coreafterequationfn%
	}{%
		\insertalignanum{#1}%
	}
}

\newcommand{\insertalignanum}[1]{%
	\emptyvarerr{\insertalignanum}{#1}{Ecuacion no definida}%
	\corevspacevarcm{\marginaligntop}%
	\begin{samepage}%
	\begin{align*}
		\ensuremath{#1}
	\end{align*}
	\corevspacevarcm{\marginalignbottom}%
	\end{samepage}
	\coreafterequationfn%
}

\newcommand{\insertaligncaptioned}[2]{%
	\emptyvarerr{\insertaligncaptioned}{#1}{Ecuacion no definida}%
	\ifx\hfuzz#2\hfuzz%
		\insertalign{#1}%
	\else
		\ifthenelse{\equal{\numberedequation}{true}}{%
			\corevspacevarcm{\marginaligntop}%
			\begin{samepage}%
			\begin{align}
				\ensuremath{#1}
			\end{align}
			\corevspacevarcm{\marginaligncapttop}%
			\coreinsertequationcaption{#2}%
			\corevspacevarcm{\marginaligncaptbottom}%
			\end{samepage}
			\coreafterequationfn%
		}{%
			\insertaligncaptionedanum{#1}{#2}%
		}
	\fi
}

\newcommand{\insertaligncaptionedanum}[2]{%
	\emptyvarerr{\insertaligncaptionedanum}{#1}{Ecuacion no definida}%
	\ifx\hfuzz#2\hfuzz%
		\insertalignanum{#1}%
	\else
		\corevspacevarcm{\marginaligntop}%
		\begin{samepage}%
		\begin{align*}
			\ensuremath{#1}
		\end{align*}
		\corevspacevarcm{\marginaligncapttop}%
		\coreinsertequationcaption{#2}%
		\corevspacevarcm{\marginaligncaptbottom}%
		\end{samepage}
		\coreafterequationfn%
	\fi
}

\newcommand{\insertaligned}[2][]{%
	\emptyvarerr{\insertaligned}{#2}{Ecuacion no definida}%
	\ifthenelse{\equal{\numberedequation}{true}}{%
		\corevspacevarcm{\marginequationtop}%
		\begin{samepage}%
		\begin{equation}
			\begin{aligned}
				\text{#1} \ensuremath{#2}
			\end{aligned}
		\end{equation}
		\corevspacevarcm{\marginalignedbottom}%
		\end{samepage}
		\coreafterequationfn%
	}{%
		\ifx\hfuzz#1\hfuzz
		\else
			\throwwarning{Label invalido en ecuacion (aligned) sin numero}
		\fi
		\insertalignedanum{#2}%
	}
}

\newcommand{\insertalignedanum}[1]{%
	\emptyvarerr{\insertalignedanum}{#1}{Ecuacion no definida}%
	\corevspacevarcm{\marginalignedtop}%
	\begin{samepage}%
	\begin{align*}
		\ensuremath{#1}
	\end{align*}
	\vspace{\dimexpr-0.2cm + \marginalignedbottom cm}%
	\end{samepage}
	\coreafterequationfn%
}

\newcommand{\insertalignedcaptioned}[3][]{%
	\emptyvarerr{\insertalignedcaptioned}{#2}{Ecuacion no definida}%
	\ifx\hfuzz#3\hfuzz%
		\insertaligned[#1]{#2}%
	\else
		\ifthenelse{\equal{\numberedequation}{true}}{%
			\corevspacevarcm{\marginequationtop}%
			\begin{samepage}%
			\begin{equation}
				\begin{aligned}
					\text{#1} \ensuremath{#2}
				\end{aligned}
			\end{equation}
			\corevspacevarcm{\marginalignedcapttop}%
			\coreinsertequationcaption{#3}%
			\corevspacevarcm{\marginalignedcaptbottom}%
			\end{samepage}
			\coreafterequationfn%
		}{%
			\ifx\hfuzz#1\hfuzz
			\else
				\throwwarning{Label invalido en ecuacion (aligned) sin numero}
			\fi
			\insertalignedcaptionedanum{#2}{#3}%
		}
	\fi
}

\newcommand{\insertalignedcaptionedanum}[2]{%
	\emptyvarerr{\insertalignedcaptionedanum}{#1}{Ecuacion no definida}%
	\ifx\hfuzz#2\hfuzz%
		\insertalignedanum{#1}%
	\else
		\corevspacevarcm{\marginequationtop}%
		\begin{samepage}%
		\begin{equation}
			\begin{aligned}
				\ensuremath{#1}
			\end{aligned}
		\end{equation}
		\corevspacevarcm{\marginalignedcapttop}%
		\coreinsertequationcaption{#2}%
		\corevspacevarcm{\marginalignedcaptbottom}%
		\end{samepage}
		\coreafterequationfn%
	\fi
}

\newcommand{\addimage}[4][]{%
	\addimageboxed[#1]{#2}{#3}{0}{#4}%
}

\newcommand{\addimageboxed}[5][]{%
	\checkonlyonenvimage%
	\begingroup%
	\setlength{\fboxsep}{0 pt}%
	\setlength{\fboxrule}{#4 pt}%
	\ifthenelse{\equal{\GLOBALenvimageadded}{true}}{%
		\hspace{\marginimagemultright cm}%
		\hspace{-0.125cm}%
	}{}%
	\subfloat[#5]{%
		\fbox{\includegraphics[#3]{#2}}#1}%
	\endgroup%
	\global\def\GLOBALenvimageadded {true}%
}

\newcommand{\addimageanum}[2]{%
	\addimageboxedanum{#1}{#2}{0}%
}

\newcommand{\addimageboxedanum}[3]{%
	\checkonlyonenvimage%
	\begingroup%
	\setlength{\fboxsep}{0 pt}%
	\setlength{\fboxrule}{#3 pt}%
	\ifthenelse{\equal{\GLOBALenvimageadded}{true}}{%
		\hspace{\marginimagemultright cm}%
		\hspace{-0.125cm}%
	}{}%
	\fbox{\includegraphics[#2]{#1}}%
	\endgroup%
	\global\def\GLOBALenvimageadded {true}%
}

\newcommand{\insertimageboxed}[5][]{%
	\emptyvarerr{\insertimageboxed}{#2}{Direccion de la imagen no definida}%
	\emptyvarerr{\insertimageboxed}{#3}{Parametros de la imagen no definidos}%
	\emptyvarerr{\insertimageboxed}{#4}{Ancho de la linea no definido}%
	\checkoutsideenvimage%
	\corevspacevarcm{\marginimagetop}%
	\begin{samepage}%
	\begin{figure}[H]%
		\begingroup%
			\setlength{\fboxsep}{0 pt}%
			\setlength{\fboxrule}{#4 pt}%
			\centering%
			\fbox{\includegraphics[#3]{#2}}%
		\endgroup%
		\ifx\hfuzz#5\hfuzz%
			\corevspacevarcm{\captionlessmarginimage}%
		\else%
			\hspace{0cm}%
			\ifthenelse{\equal{\captionmarginimage}{0}}{}{\corevspacevarcm{\captionmarginimage}}%
			\ifthenelse{\equal{\GLOBALcaptiondefn}{EMPTY-VAR}}{\caption{#5 #1}}{\caption[\GLOBALcaptiondefn]{#5 #1}}%
		\fi
	\end{figure}
	\corevspacevarcm{\marginimagebottom}%
	\end{samepage}
	\resetindexcaption%
}

\newcommand{\insertimageboxedmc}[6][]{%
	\emptyvarerr{\insertimageboxedmc}{#2}{Direccion de la imagen no definida}%
	\emptyvarerr{\insertimageboxedmc}{#3}{Parametros de la imagen no definidos}%
	\emptyvarerr{\insertimageboxedmc}{#4}{Ancho de la linea no definido}%
	\emptyvarerr{\insertimageboxedmc}{#5}{Posicion de la imagen no definida}%
	\checkoutsideenvimage%
	\checkinsidemulticol%
	\checkoutsideappendix%
	\setcaptionmargincm{\captionlrmarginmc}%
	\ifthenelse{\equal{#5}{bottom}}{%
		\begin{samepage}%
		\begin{figure*}[b]
	}{
	\ifthenelse{\equal{#5}{top}}{%
		\begin{samepage}%
		\begin{figure*}[t]
	}{
	\ifthenelse{\equal{#5}{fixed2}}{%
		\end{multicols}
		\begin{samepage}%
		\begin{figure*}[h]
	}{
	\ifthenelse{\equal{#5}{fixed2b}}{%
		\end{multicols}
		\begin{samepage}%
		\begin{figure*}[b]
	}{
	\ifthenelse{\equal{#5}{fixed2t}}{%
		\end{multicols}
		\begin{samepage}%
		\begin{figure*}[t]
	}{
	\ifthenelse{\equal{#5}{fixed3}}{%
		\end{multicols}
		\begin{samepage}%
		\begin{figure*}[h]
	}{
	\ifthenelse{\equal{#5}{fixed3b}}{%
		\end{multicols}
		\begin{samepage}%
		\begin{figure*}[b]
	}{
	\ifthenelse{\equal{#5}{fixed3t}}{%
		\end{multicols}
		\begin{samepage}%
		\begin{figure*}[t]
	}{
	\ifthenelse{\equal{#5}{fixed4}}{%
		\end{multicols}
		\begin{samepage}%
		\begin{figure*}[h]
	}{
	\ifthenelse{\equal{#5}{fixed4b}}{%
		\end{multicols}
		\begin{samepage}%
		\begin{figure*}[h]
	}{
	\ifthenelse{\equal{#5}{fixed4t}}{%
		\end{multicols}
		\begin{samepage}%
		\begin{figure*}[h]
	}{
		\errmessage{LaTeX Warning: Posicion de imagen invalida, valores esperados: bottom,top,fixed2,fixed2b,fixed2t,fixed3,fixed3b,fixed3t,fixed4,fixed4b,fixed4t}
		\stop}}}}}}}}}}
	}
		\begingroup
			\setlength{\fboxsep}{0 pt}
			\setlength{\fboxrule}{#4 pt}
			\centering
			\fbox{\includegraphics[#3]{#2}}%
		\endgroup
		\ifx\hfuzz#6\hfuzz%
			\corevspacevarcm{\captionlessmarginimage}%
		\else
			\hspace{0cm}
			\ifthenelse{\equal{\captionmarginimage}{0}}{}{\corevspacevarcm{\captionmarginimage}}%
			\ifthenelse{\equal{\GLOBALcaptiondefn}{EMPTY-VAR}}{\caption{#6 #1}}{\caption[\GLOBALcaptiondefn]{#6 #1}}
		\fi
	\end{figure*}
	\end{samepage}
	\ifthenelse{\equal{#5}{fixed2}}{%
		\begin{multicols}{2}
	}{
	\ifthenelse{\equal{#5}{fixed2b}}{%
		\begin{multicols}{2}
	}{
	\ifthenelse{\equal{#5}{fixed2t}}{%
		\begin{multicols}{2}
	}{
	\ifthenelse{\equal{#5}{fixed3}}{%
		\begin{multicols}{3}
	}{
	\ifthenelse{\equal{#5}{fixed3b}}{%
		\begin{multicols}{3}
	}{
	\ifthenelse{\equal{#5}{fixed3t}}{%
		\begin{multicols}{3}
	}{
	\ifthenelse{\equal{#5}{fixed4}}{%
		\begin{multicols}{4}
	}{
	\ifthenelse{\equal{#5}{fixed4b}}{%
		\begin{multicols}{4}
	}{
	\ifthenelse{\equal{#5}{fixed4t}}{%
		\begin{multicols}{4}
	}{
	}}}}}}}}}
	\setcaptionmargincm{\captionlrmargin}%
	\resetindexcaption%
}

\newcommand{\inserttableimageboxed}[3]{%
	\emptyvarerr{\inserttableimageboxed}{#1}{Direccion de la imagen no definida}%
	\emptyvarerr{\inserttableimageboxed}{#2}{Parametros de la imagen no definidos}%
	\emptyvarerr{\inserttableimageboxed}{#3}{Ancho de la linea no definido}%
	\checkoutsideenvimage%
	\begingroup%
	\setlength{\fboxsep}{0 pt}%
	\setlength{\fboxrule}{#3 pt}%
	\raisebox{-1\totalheight}{%
		\fbox{\includegraphics[#2]{#1}}}%
	\endgroup%
	\resetindexcaption%
}

\newcommand{\insertimageleftboxed}[5][]{%
	\emptyvarerr{\insertimageleftboxed}{#2}{Direccion de la imagen no definida}%
	\emptyvarerr{\insertimageleftboxed}{#3}{Ancho de la imagen no definido}%
	\emptyvarerr{\insertimageleftboxed}{#4}{Ancho de la linea no definido}%
	\checkoutsideenvimage%
	~%
	\vspace{-\baselineskip}%
	\par%
	\begin{wrapfigure}{l}{#3\linewidth}%
		\setcaptionmargincm{0}%
		\ifthenelse{\equal{\figurecaptiontop}{true}}{}{%
			\vspace{\marginfloatimages pt}%
		}%
		\begingroup%
			\setlength{\fboxsep}{0 pt}%
			\setlength{\fboxrule}{#4 pt}%
			\centering%
			\fbox{\includegraphics[width=\linewidth]{#2}}%
		\endgroup%
		\ifx\hfuzz#5\hfuzz%
			\corevspacevarcm{\captionlessmarginimage}%
		\else%
			\ifthenelse{\equal{\captionmarginimage}{0}}{}{\corevspacevarcm{\captionmarginimage}}%
			\ifthenelse{\equal{\GLOBALcaptiondefn}{EMPTY-VAR}}{\caption{#5 #1}}{\caption[\GLOBALcaptiondefn]{#5 #1}}%
		\fi%
	\end{wrapfigure}
	\setcaptionmargincm{\captionlrmargin}%
	\resetindexcaption%
}

\newcommand{\insertimageleftlineboxed}[6][]{%
	\emptyvarerr{\insertimageleftlineboxed}{#2}{Direccion de la imagen no definida}%
	\emptyvarerr{\insertimageleftlineboxed}{#3}{Ancho de la imagen no definido}%
	\emptyvarerr{\insertimageleftlineboxed}{#4}{Ancho de la linea no definido}%
	\emptyvarerr{\insertimageleftlineboxed}{#5}{Altura en lineas de la imagen flotante izquierda no definida}
	\checkoutsideenvimage%
	~%
	\vspace{-\baselineskip}%
	\par%
	\begin{wrapfigure}[#5]{l}{#3\linewidth}%
		\setcaptionmargincm{0}%
		\ifthenelse{\equal{\figurecaptiontop}{true}}{}{%
			\vspace{\marginfloatimages pt}}%
		\begingroup%
			\setlength{\fboxsep}{0 pt}%
			\setlength{\fboxrule}{#4 pt}%
			\centering%
			\fbox{\includegraphics[width=\linewidth]{#2}}%
		\endgroup%
		\ifx\hfuzz#6\hfuzz%
			\corevspacevarcm{\captionlessmarginimage}%
		\else%
			\ifthenelse{\equal{\captionmarginimage}{0}}{}{\corevspacevarcm{\captionmarginimage}}%
			\ifthenelse{\equal{\GLOBALcaptiondefn}{EMPTY-VAR}}{\caption{#6 #1}}{\caption[\GLOBALcaptiondefn]{#6 #1}}%
		\fi
	\end{wrapfigure}
	\setcaptionmargincm{\captionlrmargin}%
	\resetindexcaption%
}

\newcommand{\insertimagerightboxed}[5][]{%
	\emptyvarerr{\insertimagerightboxed}{#2}{Direccion de la imagen no definida}%
	\emptyvarerr{\insertimagerightboxed}{#3}{Ancho de la imagen no defindo}%
	\emptyvarerr{\insertimagerightboxed}{#4}{Ancho de la linea no definido}%
	\checkoutsideenvimage%
	~%
	\vspace{-\baselineskip}%
	\par%
	\begin{wrapfigure}{r}{#3\linewidth}%
		\setcaptionmargincm{0}%
		\ifthenelse{\equal{\figurecaptiontop}{true}}{}{%
			\vspace{\marginfloatimages pt}%
		}%
		\begingroup%
			\setlength{\fboxsep}{0 pt}%
			\setlength{\fboxrule}{#4 pt}%
			\centering%
			\fbox{\includegraphics[width=\linewidth]{#2}}%
		\endgroup%
		\ifx\hfuzz#5\hfuzz%
			\corevspacevarcm{\captionlessmarginimage}%
		\else%
			\ifthenelse{\equal{\captionmarginimage}{0}}{}{\corevspacevarcm{\captionmarginimage}}%
			\ifthenelse{\equal{\GLOBALcaptiondefn}{EMPTY-VAR}}{\caption{#5 #1}}{\caption[\GLOBALcaptiondefn]{#5 #1}}%
		\fi%
	\end{wrapfigure}
	\setcaptionmargincm{\captionlrmargin}%
	\resetindexcaption%
}

\newcommand{\insertimagerightlineboxed}[6][]{%
	\emptyvarerr{\insertimagerightlineboxed}{#2}{Direccion de la imagen no definida}%
	\emptyvarerr{\insertimagerightlineboxed}{#3}{Ancho de la imagen no defindo}%
	\emptyvarerr{\insertimagerightlineboxed}{#4}{Ancho de la linea no definido}%
	\emptyvarerr{\insertimagerightlineboxed}{#5}{Altura en lineas de la imagen flotante derecha no definida}%
	\checkoutsideenvimage%
	~%
	\vspace{-\baselineskip}%
	\par%
	\begin{wrapfigure}[#5]{r}{#3\linewidth}%
		\setcaptionmargincm{0}%
		\ifthenelse{\equal{\figurecaptiontop}{true}}{}{%
			\vspace{\marginfloatimages pt}%
		}%
		\begingroup%
			\setlength{\fboxsep}{0 pt}%
			\setlength{\fboxrule}{#4 pt}%
			\centering%
			\fbox{\includegraphics[width=\linewidth]{#2}}%
		\endgroup%
		\ifx\hfuzz#6\hfuzz%
			\corevspacevarcm{\captionlessmarginimage}%
		\else%
			\ifthenelse{\equal{\captionmarginimage}{0}}{}{\corevspacevarcm{\captionmarginimage}}%
			\ifthenelse{\equal{\GLOBALcaptiondefn}{EMPTY-VAR}}{\caption{#6 #1}}{\caption[\GLOBALcaptiondefn]{#6 #1}}%
		\fi%
	\end{wrapfigure}
	\setcaptionmargincm{\captionlrmargin}%
	\resetindexcaption%
}

\newcommand{\insertimageleftboxedp}[6][]{%
	\emptyvarerr{\insertimageleftboxedp}{#2}{Direccion de la imagen no definida}%
	\emptyvarerr{\insertimageleftboxedp}{#3}{Ancho del objeto no definido}%
	\emptyvarerr{\insertimageleftboxedp}{#4}{Propiedades de la imagen no defindos}%
	\emptyvarerr{\insertimageleftboxedp}{#5}{Ancho de la linea no definido}%
	\checkoutsideenvimage%
	~%
	\vspace{-\baselineskip}%
	\par%
	\begin{wrapfigure}{l}{#3}%
		\setcaptionmargincm{0}%
		\ifthenelse{\equal{\figurecaptiontop}{true}}{}{%
			\vspace{\marginfloatimages pt}%
		}%
		\begingroup%
			\setlength{\fboxsep}{0 pt}%
			\setlength{\fboxrule}{#5 pt}%
			\centering%
			\fbox{\includegraphics[#4]{#2}}%
		\endgroup%
		\ifx\hfuzz#6\hfuzz%
			\corevspacevarcm{\captionlessmarginimage}%
		\else%
			\ifthenelse{\equal{\captionmarginimage}{0}}{}{\corevspacevarcm{\captionmarginimage}}%
			\ifthenelse{\equal{\GLOBALcaptiondefn}{EMPTY-VAR}}{\caption{#6 #1}}{\caption[\GLOBALcaptiondefn]{#6 #1}}%
		\fi%
	\end{wrapfigure}
	\setcaptionmargincm{\captionlrmargin}%
	\resetindexcaption%
}

\newcommand{\insertimageleftlineboxedp}[7][]{%
	\emptyvarerr{\insertimageleftlineboxedp}{#2}{Direccion de la imagen no definida}%
	\emptyvarerr{\insertimageleftlineboxedp}{#3}{Ancho del objeto no definido}%
	\emptyvarerr{\insertimageleftlineboxedp}{#4}{Propiedades de la imagen no definidos}%
	\emptyvarerr{\insertimageleftlineboxedp}{#5}{Ancho de la linea no definido}%
	\emptyvarerr{\insertimageleftlineboxedp}{#6}{Altura en lineas de la imagen flotante izquierda no definida}%
	\checkoutsideenvimage%
	~%
	\vspace{-\baselineskip}%
	\par%
	\begin{wrapfigure}[#6]{l}{#3}%
		\setcaptionmargincm{0}%
		\ifthenelse{\equal{\figurecaptiontop}{true}}{}{%
			\vspace{\marginfloatimages pt}%
		}%
		\begingroup%
			\setlength{\fboxsep}{0 pt}%
			\setlength{\fboxrule}{#5 pt}%
			\centering%
			\fbox{\includegraphics[#4]{#2}}%
		\endgroup%
		\ifx\hfuzz#7\hfuzz%
			\corevspacevarcm{\captionlessmarginimage}%
		\else%
			\ifthenelse{\equal{\captionmarginimage}{0}}{}{\corevspacevarcm{\captionmarginimage}}%
			\ifthenelse{\equal{\GLOBALcaptiondefn}{EMPTY-VAR}}{\caption{#7 #1}}{\caption[\GLOBALcaptiondefn]{#7 #1}}%
		\fi%
	\end{wrapfigure}
	\setcaptionmargincm{\captionlrmargin}%
	\resetindexcaption%
}

\newcommand{\insertimagerightboxedp}[6][]{%
	\emptyvarerr{\insertimagerightboxedp}{#2}{Direccion de la imagen no definida}%
	\emptyvarerr{\insertimagerightboxedp}{#3}{Ancho del objeto no definido}%
	\emptyvarerr{\insertimagerightboxedp}{#4}{Propiedades de la imagen no definidos}%
	\emptyvarerr{\insertimagerightboxedp}{#5}{Ancho de la linea no definido}%
	\checkoutsideenvimage%
	~%
	\vspace{-\baselineskip}%
	\par%
	\begin{wrapfigure}{r}{#3}%
		\setcaptionmargincm{0}%
		\ifthenelse{\equal{\figurecaptiontop}{true}}{}{%
			\vspace{\marginfloatimages pt}%
		}%
		\begingroup%
			\setlength{\fboxsep}{0 pt}%
			\setlength{\fboxrule}{#5 pt}%
			\centering%
			\fbox{\includegraphics[#4]{#2}}%
		\endgroup%
		\ifx\hfuzz#6\hfuzz%
			\corevspacevarcm{\captionlessmarginimage}%
		\else%
			\ifthenelse{\equal{\captionmarginimage}{0}}{}{\corevspacevarcm{\captionmarginimage}}%
			\ifthenelse{\equal{\GLOBALcaptiondefn}{EMPTY-VAR}}{\caption{#6 #1}}{\caption[\GLOBALcaptiondefn]{#6 #1}}%
		\fi%
	\end{wrapfigure}
	\setcaptionmargincm{\captionlrmargin}%
	\resetindexcaption%
}

\newcommand{\insertimagerightlineboxedp}[7][]{%
	\emptyvarerr{\insertimagerightlineboxedp}{#2}{Direccion de la imagen no definida}%
	\emptyvarerr{\insertimagerightlineboxedp}{#3}{Ancho del objeto no definido}%
	\emptyvarerr{\insertimagerightlineboxedp}{#4}{Propiedades de la imagen no definidos}%
	\emptyvarerr{\insertimagerightlineboxedp}{#5}{Ancho de la linea no definido}%
	\emptyvarerr{\insertimagerightlineboxedp}{#6}{Altura en lineas de la imagen flotante derecha no definida}%
	\checkoutsideenvimage%
	~%
	\vspace{-\baselineskip}%
	\par%
	\begin{wrapfigure}[#6]{r}{#3}%
		\setcaptionmargincm{0}%
		\ifthenelse{\equal{\figurecaptiontop}{true}}{}{%
			\vspace{\marginfloatimages pt}%
		}%
		\begingroup%
			\setlength{\fboxsep}{0 pt}%
			\setlength{\fboxrule}{#5 pt}%
			\centering%
			\fbox{\includegraphics[#4]{#2}}%
		\endgroup%
		\ifx\hfuzz#7\hfuzz%
			\corevspacevarcm{\captionlessmarginimage}%
		\else%
			\ifthenelse{\equal{\captionmarginimage}{0}}{}{\corevspacevarcm{\captionmarginimage}}%
			\ifthenelse{\equal{\GLOBALcaptiondefn}{EMPTY-VAR}}{\caption{#7 #1}}{\caption[\GLOBALcaptiondefn]{#7 #1}}%
		\fi%
	\end{wrapfigure}
	\setcaptionmargincm{\captionlrmargin}%
	\resetindexcaption%
}

\newcommand{\coreinsertkeyimage}[2]{%
	\expandafter\includegraphics\expandafter[#1]{\expandafter#2}%
}

\makeatletter
\define@key{Gin}{resolution}{\pdfimageresolution=#1\relax}
\makeatother

\def\coreintializetitlenumbering {%
	\ifthenelse{\equal{\GLOBALchapternumenabled}{false}}{%
		\ifthenelse{\equal{\GLOBALsectionalph}{true}}{%
			\renewcommand{\thesection}{\Alph{section}}%
		}{%
			\renewcommand{\thesection}{\arabic{section}}%
		}%
	}{%
		\ifthenelse{\equal{\GLOBALsectionalph}{true}}{%
			\renewcommand{\thesection}{\thechapter.\Alph{section}}%
		}{%
			\renewcommand{\thesection}{\thechapter.\arabic{section}}%
		}%
	}%
	\ifthenelse{\equal{\GLOBALsectionanumenabled}{true}}{%
		\renewcommand{\thesubsection}{\arabic{subsection}}%
	}{%
		\renewcommand{\thesubsection}{\thesection.\arabic{subsection}}%
	}%
	\ifthenelse{\equal{\GLOBALsubsectionanumenabled}{true}}{%
		\renewcommand{\thesubsubsection}{\arabic{subsubsection}}%
	}{%
		\renewcommand{\thesubsubsection}{\thesubsection.\arabic{subsubsection}}%
	}%
	\ifthenelse{\equal{\GLOBALsubsectionanumenabled}{true}}{%
		\renewcommand{\thesubsubsubsection}{\arabic{subsubsection}.\arabic{subsubsubsection}}%
	}{%
		\renewcommand{\thesubsubsubsection}{\thesubsubsection.\arabic{subsubsubsection}}%
	}%
	\hbadness=10000%
}

\pretocmd{\chapter}{
	\ifthenelse{\equal{\showsectioncaptioncode}{chap}}{ 
		\addtocounter{templateListings}{\value{lstlisting}}%
		\setcounter{lstlisting}{0}%
	}{}
	\ifthenelse{\equal{\showsectioncaptioneqn}{chap}}{ 
		\addtocounter{templateEquations}{\value{equation}}%
		\setcounter{equation}{0}%
	}{}
	\ifthenelse{\equal{\equationrestart}{chap}}{ 
		\addtocounter{templateEquations}{\value{equation}}%
		\setcounter{equation}{0}%
	}{}
	\ifthenelse{\equal{\showsectioncaptionfig}{chap}}{ 
		\addtocounter{templateFigures}{\value{figure}}%
		\setcounter{figure}{0}%
	}{}
	\ifthenelse{\equal{\showsectioncaptiontab}{chap}}{ 
		\addtocounter{templateTables}{\value{table}}%
		\setcounter{table}{0}%
	}{}
	\global\def\GLOBALchapternumenabled {true}%
	\coreintializetitlenumbering%
}{}{}

\pretocmd{\section}{
	\ifthenelse{\equal{\showsectioncaptioncode}{sec}}{ 
		\addtocounter{templateListings}{\value{lstlisting}}%
		\setcounter{lstlisting}{0}%
	}{}
	\ifthenelse{\equal{\showsectioncaptioneqn}{sec}}{ 
		\addtocounter{templateEquations}{\value{equation}}%
		\setcounter{equation}{0}%
	}{}
	\ifthenelse{\equal{\equationrestart}{sec}}{ 
		\addtocounter{templateEquations}{\value{equation}}%
		\setcounter{equation}{0}%
	}{}
	\ifthenelse{\equal{\showsectioncaptionfig}{sec}}{ 
		\addtocounter{templateFigures}{\value{figure}}%
		\setcounter{figure}{0}%
	}{}
	\ifthenelse{\equal{\showsectioncaptiontab}{sec}}{ 
		\addtocounter{templateTables}{\value{table}}%
		\setcounter{table}{0}%
	}{}
	\global\def\GLOBALsectionanumenabled {false}%
	\global\def\GLOBALsubsectionanumenabled {false}%
	\coreintializetitlenumbering%
}{}{}

\pretocmd{\subsection}{
	\ifthenelse{\equal{\showsectioncaptioncode}{ssec}}{ 
		\addtocounter{templateListings}{\value{lstlisting}}%
		\setcounter{lstlisting}{0}%
	}{}
	\ifthenelse{\equal{\showsectioncaptioneqn}{ssec}}{ 
		\addtocounter{templateEquations}{\value{equation}}%
		\setcounter{equation}{0}%
	}{}
	\ifthenelse{\equal{\equationrestart}{ssec}}{ 
		\addtocounter{templateEquations}{\value{equation}}%
		\setcounter{equation}{0}%
	}{}
	\ifthenelse{\equal{\showsectioncaptionfig}{ssec}}{ 
		\addtocounter{templateFigures}{\value{figure}}%
		\setcounter{figure}{0}%
	}{}
	\ifthenelse{\equal{\showsectioncaptiontab}{ssec}}{ 
		\addtocounter{templateTables}{\value{table}}%
		\setcounter{table}{0}%
	}{}
	\global\def\GLOBALsubsectionanumenabled {false}%
	\coreintializetitlenumbering%
}{}{}

\pretocmd{\subsubsection}{
	\ifthenelse{\equal{\showsectioncaptioncode}{sssec}}{ 
		\addtocounter{templateListings}{\value{lstlisting}}%
		\setcounter{lstlisting}{0}%
	}{}
	\ifthenelse{\equal{\showsectioncaptioneqn}{sssec}}{ 
		\addtocounter{templateEquations}{\value{equation}}%
		\setcounter{equation}{0}%
	}{}
	\ifthenelse{\equal{\equationrestart}{sssec}}{ 
		\addtocounter{templateEquations}{\value{equation}}%
		\setcounter{equation}{0}%
	}{}
	\ifthenelse{\equal{\showsectioncaptionfig}{sssec}}{ 
		\addtocounter{templateFigures}{\value{figure}}%
		\setcounter{figure}{0}%
	}{}
	\ifthenelse{\equal{\showsectioncaptiontab}{sssec}}{ 
		\addtocounter{templateTables}{\value{table}}%
		\setcounter{table}{0}%
	}{}
	\coreintializetitlenumbering%
}{}{}

\pretocmd{\subsubsubsection}{
	\ifthenelse{\equal{\showsectioncaptioncode}{ssssec}}{ 
		\addtocounter{templateListings}{\value{lstlisting}}%
		\setcounter{lstlisting}{0}%
	}{}
	\ifthenelse{\equal{\showsectioncaptioneqn}{ssssec}}{ 
		\addtocounter{templateEquations}{\value{equation}}%
		\setcounter{equation}{0}%
	}{}
	\ifthenelse{\equal{\equationrestart}{ssssec}}{ 
		\addtocounter{templateEquations}{\value{equation}}%
		\setcounter{equation}{0}%
	}{}
	\ifthenelse{\equal{\showsectioncaptionfig}{ssssec}}{ 
		\addtocounter{templateFigures}{\value{figure}}%
		\setcounter{figure}{0}%
	}{}
	\ifthenelse{\equal{\showsectioncaptiontab}{ssssec}}{ 
		\addtocounter{templateTables}{\value{table}}%
		\setcounter{table}{0}%
	}{}
}{}{}

\newcommand{\sectionanum}[1]{%
	\emptyvarerr{\sectionanum}{#1}{Titulo no definido}%
	\phantomsection%
	\needspace{3\baselineskip}%
	\section*{#1}%
	\addcontentsline{toc}{section}{#1}%
	\ifthenelse{\equal{\anumsecaddtocounter}{true}}{\stepcounter{section}}{}%
	\changeheadertitle{#1}%
	\setcounter{subsection}{0}%
	\renewcommand{\thesubsection}{\arabic{subsection}}%
	\global\def\GLOBALsectionanumenabled {true}%
}

\newcommand{\sectionanumheadless}[1]{%
	\emptyvarerr{\sectionanumnoheadless}{#1}{Titulo no definido}%
	\section*{#1}%
	\addcontentsline{toc}{section}{#1}%
	\ifthenelse{\equal{\anumsecaddtocounter}{true}}{\stepcounter{section}}{}%
	\setcounter{subsection}{0}%
	\renewcommand{\thesubsection}{\arabic{subsection}}%
	\global\def\GLOBALsectionanumenabled {true}%
}

\newcommand{\subsectionanum}[1]{%
	\emptyvarerr{\subsectionanum}{#1}{Subtitulo no definido}%
	\subsection*{#1}%
	\addcontentsline{toc}{subsection}{#1}
	\ifthenelse{\equal{\anumsecaddtocounter}{true}}{\stepcounter{subsection}}{}%
	\setcounter{subsubsection}{0}%
	\renewcommand{\thesubsubsection}{\arabic{subsubsection}}%
	\global\def\GLOBALsubsectionanumenabled {true}%
}

\newcommand{\subsubsectionanum}[1]{%
	\emptyvarerr{\subsubsectionanum}{#1}{Sub-subtitulo no definido}%
	\subsubsection*{#1}%
	\addcontentsline{toc}{subsubsection}{#1}%
	\ifthenelse{\equal{\anumsecaddtocounter}{true}}{\stepcounter{subsubsection}}{}%
	\setcounter{subsubsubsection}{0}%
	\renewcommand{\thesubsubsubsection}{\arabic{subsubsubsection}}%
}

\newcommand{\subsubsubsectionanum}[1]{%
	\emptyvarerr{\subsubsubsectionanum}{#1}{Sub-sub-subtitulo no definido}%
	\subsubsubsection*{#1}%
	\addcontentsline{toc}{subsubsubsection}{#1}%
	\ifthenelse{\equal{\anumsecaddtocounter}{true}}{\stepcounter{subsubsubsection}}{}%
}

\newcommand{\changeheadertitle}[1]{%
	\emptyvarerr{\changeheadertitle}{#1}{Titulo no definido}%
	\markboth{#1}{}%
}

\newcommand{\newchapter}[1]{%
	\emptyvarerr{\newchapter}{#1}{Titulo no definido}%
	\clearpage%
	\stepcounter{section}%
	\phantomsection%
	\needspace{3\baselineskip}%
	\vspace* {3cm}%
	\noindent {\huge{\textbf{\namechapter\ \thesection}}} \\%
	\vspace* {0.5cm} \\%
	\noindent {\Huge{\textbf{#1}}} \\%
	\vspace {0.5cm} \\%
	\addcontentsline{toc}{section}{\protect\numberline{\thesection}#1}%
	\markboth{#1}{}%
}

\newcommand{\itemresize}[2]{%
	\emptyvarerr{\itemresize}{#1}{Tamano del nuevo objeto no definido}%
	\emptyvarerr{\itemresize}{#2}{Objeto a redimensionar no definido}%
	\resizebox{#1\linewidth}{!}{#2}%
}

\newcommand{\hreftext}[1]{%
	\ifthenelse{\equal{\fonturl}{same}}{%
		#1%
	}{%
	\ifthenelse{\equal{\fonturl}{tt}}{%
		\texttt{#1}%
	}{%
	\ifthenelse{\equal{\fonturl}{rm}}{%
		\textrm{#1}%
	}{%
	\ifthenelse{\equal{\fonturl}{sf}}{%
		\textsf{#1}%
	}{}}}}%
}

\newcommand{\insertemail}[1]{%
	\href{mailto:#1}{\hreftext{#1}}%
}

\newcommand{\settablerowcolors}[1]{%
	\emptyvarerr{\settablerowcolors}{#1}{Posicion de fila no definida}
	
	\ifthenelse{\equal{\GLOBALtablerowcolorswitch}{false}}{
		\ifthenelse{\equal{\tablerowfirstcolor}{none}}{
			\ifthenelse{\equal{\tablerowsecondcolor}{none}}{
				\rowcolors{#1}{}{}
			}{
				\rowcolors{#1}{\tablerowsecondcolor}{}
			}
		}{
			\ifthenelse{\equal{\tablerowsecondcolor}{none}}{
				\rowcolors{#1}{}{\tablerowfirstcolor}
			}{
				\rowcolors{#1}{\tablerowsecondcolor}{\tablerowfirstcolor}
			}
		}
	}{
		\ifthenelse{\equal{\tablerowfirstcolor}{none}}{
			\ifthenelse{\equal{\tablerowsecondcolor}{none}}{
				\rowcolors{#1}{}{}
			}{
				\rowcolors{#1}{}{\tablerowsecondcolor}
			}
		}{
			\ifthenelse{\equal{\tablerowsecondcolor}{none}}{
				\rowcolors{#1}{\tablerowfirstcolor}{}
			}{
				\rowcolors{#1}{\tablerowfirstcolor}{\tablerowsecondcolor}
			}
		}
	}
	
	\global\def\GLOBALtablerowcolorindex {#1}
}

\newcommand{\settablerowcolorslast}{
	\ifthenelse{\equal{\GLOBALtablerowcolorswitch}{false}}{
		\ifthenelse{\equal{\tablerowfirstcolor}{none}}{
			\ifthenelse{\equal{\tablerowsecondcolor}{none}}{
				\rowcolors{\GLOBALtablerowcolorindex}{}{}
			}{
				\rowcolors{\GLOBALtablerowcolorindex}{\tablerowsecondcolor}{}
			}
		}{
			\ifthenelse{\equal{\tablerowsecondcolor}{none}}{
				\rowcolors{\GLOBALtablerowcolorindex}{}{\tablerowfirstcolor}
			}{
				\rowcolors{\GLOBALtablerowcolorindex}{\tablerowsecondcolor}{\tablerowfirstcolor}
			}
		}
	}{
		\ifthenelse{\equal{\tablerowfirstcolor}{none}}{
			\ifthenelse{\equal{\tablerowsecondcolor}{none}}{
				\rowcolors{\GLOBALtablerowcolorindex}{}{}
			}{
				\rowcolors{\GLOBALtablerowcolorindex}{}{\tablerowsecondcolor}
			}
		}{
			\ifthenelse{\equal{\tablerowsecondcolor}{none}}{
				\rowcolors{\GLOBALtablerowcolorindex}{\tablerowfirstcolor}{}
			}{
				\rowcolors{\GLOBALtablerowcolorindex}{\tablerowfirstcolor}{\tablerowsecondcolor}
			}
		}
	}
}

\newcommand{\enabletablerowcolor}[1][]{
	\ifx\hfuzz#1\hfuzz
		\settablerowcolors{2}
	\else
		\settablerowcolors{#1}
	\fi
}

\newcommand{\disabletablerowcolor}{\rowcolors{2}{}{}}

\newcommand{\switchtablerowcolors}{
	\ifthenelse{\equal{\GLOBALtablerowcolorswitch}{false}}{
		\global\def\GLOBALtablerowcolorswitch {true}
	}{
		\global\def\GLOBALtablerowcolorswitch {false}
	}
	\settablerowcolorslast
}

\newcommand{\settablecellpadding}[2]{
	\emptyvarerr{\settablecellpadding}{#1}{Padding horizontal no definido}
	\emptyvarerr{\settablecellpadding}{#2}{Padding vertical no definido}
	\setlength{\tabcolsep}{#1 em} 
	\def\arraystretch {#2} 
}

\newcommand{\resettablecellpadding}{
	\settablecellpadding{\tablepaddingh}{\tablepaddingv}
}

\newcommand{\changepagesize}[3][]{
	\emptyvarerr{\changepagesize}{#2}{Ancho de la pagina no definida}
	\emptyvarerr{\changepagesize}{#3}{Altura de la pagina no definida}
	\ifthenelse{\equal{\compilertype}{lualatex}}{
		\throwwarning{Funcion no valida en compilador lualatex}
	}{
		\clearpage
		\ifthenelse{\equal{#1}{}}{
			\newgeometry{left=\pagemarginleft cm, top=\pagemargintop cm, right=\pagemarginright cm, bottom=\pagemarginbottom cm, paperwidth=#2 cm, paperheight=#3 cm}
		}{
		\ifthenelse{\equal{#1}{0}}{
			\newgeometry{left=\pagemarginleft cm, top=\pagemargintop cm, right=\pagemarginright cm, bottom=\pagemarginbottom cm, paperwidth=#2 cm, paperheight=#3 cm}
		}{
		\ifthenelse{\equal{#1}{90}}{
			\newgeometry{left=\pagemarginleft cm, top=\pagemargintop cm, right=\pagemarginright cm, bottom=\pagemarginbottom cm, paperwidth=#3 cm, paperheight=#2 cm}
		}{
			\throwbadconfig{Orientacion de pagina no valido}{\changepagesize}{0,90}}}
		}
	}
}

\newcommand{\changepagesizeformat}[2][]{%
	\emptyvarerr{\changepagesizeformat}{#2}{Formato de pagina no definido}
	\ifthenelse{\equal{#2}{4A0}}{
		\changepagesize[#1]{168.2}{237.8}
	}{
	\ifthenelse{\equal{#2}{2A0}}{
		\changepagesize[#1]{118.9}{168.2}
	}{
	\ifthenelse{\equal{#2}{A0}}{
		\changepagesize[#1]{84.1}{118.9}
	}{
	\ifthenelse{\equal{#2}{A1}}{
		\changepagesize[#1]{59.4}{84.1}
	}{
	\ifthenelse{\equal{#2}{A2}}{
		\changepagesize[#1]{42.0}{84.1}
	}{
	\ifthenelse{\equal{#2}{A3}}{
		\changepagesize[#1]{29.7}{42.0}
	}{
	\ifthenelse{\equal{#2}{A4}}{
		\changepagesize[#1]{21.0}{29.7}
	}{
	\ifthenelse{\equal{#2}{A5}}{
		\changepagesize[#1]{14.8}{21.0}
	}{
	\ifthenelse{\equal{#2}{A6}}{
		\changepagesize[#1]{10.5}{14.8}
	}{
	\ifthenelse{\equal{#2}{letter}}{
		\changepagesize[#1]{21.59}{27.94}
	}{
	\ifthenelse{\equal{#2}{legal}}{
		\changepagesize[#1]{21.59}{35.6}
	}{
	\ifthenelse{\equal{#2}{foolscap}}{
		\changepagesize[#1]{20.3}{33.0}
	}{
	\ifthenelse{\equal{#2}{executive}}{
		\changepagesize[#1]{18.41}{26.67}
	}{
	\ifthenelse{\equal{#2}{ledger}}{
		\changepagesize[#1]{27.94}{43.18}
	}{
	\ifthenelse{\equal{#2}{tabloid}}{
		\changepagesize[#1]{43.18}{27.94}
	}{
	\ifthenelse{\equal{#2}{ANSIC}}{
		\changepagesize[#1]{55.9}{43.2}
	}{
	\ifthenelse{\equal{#2}{ANSID}}{
		\changepagesize[#1]{86.4}{55.9}
	}{
	\ifthenelse{\equal{#2}{ANSIE}}{
		\changepagesize[#1]{111.8}{86.4}
	}{
	\ifthenelse{\equal{#2}{B0}}{
		\changepagesize[#1]{100}{141.4}
	}{
	\ifthenelse{\equal{#2}{B1}}{
		\changepagesize[#1]{70.7}{100}
	}{
	\ifthenelse{\equal{#2}{B2}}{
		\changepagesize[#1]{50}{70.7}
	}{
	\ifthenelse{\equal{#2}{B3}}{
		\changepagesize[#1]{35.3}{50}
	}{
	\ifthenelse{\equal{#2}{B4}}{
		\changepagesize[#1]{25}{35.3}
	}{
	\ifthenelse{\equal{#2}{B5}}{
		\changepagesize[#1]{17.6}{25}
	}{
	\ifthenelse{\equal{#2}{B6}}{
		\changepagesize[#1]{12.5}{17.6}
	}{
		\throwbadconfig{Estilo de pagina no valido}{\changepagesizeformat}{4A0,2A0,A0,A1,A2,A3,A4,A5,A6,letter,legal,foolscap,executive,ledger,tabloid,ANSIC,ANSID,ANSIE,B0,B1,B2,B3,B4,B5,B6}}}}}}}}}}}}}}}}}}}}}}}}}
	}
}

\global\def\GLOBALtwocolumnap {l}
\global\def\GLOBALtwocolumnav {t}
\global\def\GLOBALtwocolumnbp {l}
\global\def\GLOBALtwocolumnbv {t}
\global\def\GLOBALthreecolumnap {l}
\global\def\GLOBALthreecolumnav {t}
\global\def\GLOBALthreecolumnbp {l}
\global\def\GLOBALthreecolumnbv {t}
\global\def\GLOBALthreecolumncp {l}
\global\def\GLOBALthreecolumncv {t}

\newcommand{\corecheckcolumnvvalue}[1]{%
	\ifthenelse{\equal{#1}{c}}{}{%
	\ifthenelse{\equal{#1}{t}}{}{%
	\ifthenelse{\equal{#1}{b}}{}{%
		\errmessage{LaTeX Warning: Posicion vertical columna invalido, valores esperados: c,t,b}%
	}}}%
}

\newcommand{\corecheckcolumnpvalue}[1]{%
	\ifthenelse{\equal{#1}{c}}{}{%
	\ifthenelse{\equal{#1}{l}}{}{%
	\ifthenelse{\equal{#1}{r}}{}{%
		\errmessage{LaTeX Warning: Alineacion columna invalida, valores esperados: c,l,r}%
	}}}%
}

\newcommand{\createtwocolumn}[6][]{%
	\setcaptionmargincm{0}%
	\begin{samepage}%
	\begin{flushleft}%
		\vspace{-0.5\baselineskip}%
		\begin{minipage}{1\linewidth}%
			\begin{minipage}[t][#1][\GLOBALtwocolumnav]{#2\linewidth}%
				\ifthenelse{\equal{\GLOBALtwocolumnap}{c}}{%
					\begin{center}#5\end{center}%
				}{%
				\ifthenelse{\equal{\GLOBALtwocolumnap}{l}}{%
					\begin{raggedright}#5\end{raggedright}%
				}{%
				\ifthenelse{\equal{\GLOBALtwocolumnap}{r}}{%
					\hfill\begin{raggedleft}#5\end{raggedleft}%
				}{%
					\errmessage{LaTeX Warning: Alineacion columna izquierda incorrecta, valores esperados: c,l,r}
				}}}
			\end{minipage}%
			\hspace{#4 cm}%
			\begin{minipage}[t][#1][\GLOBALtwocolumnbv]{#3\linewidth}%
				\ifthenelse{\equal{\GLOBALtwocolumnbp}{c}}{%
					\begin{center}#6\end{center}%
				}{%
				\ifthenelse{\equal{\GLOBALtwocolumnbp}{l}}{%
					\begin{raggedright}#6\end{raggedright}%
				}{%
				\ifthenelse{\equal{\GLOBALtwocolumnbp}{r}}{%
					\hfill\begin{raggedleft}#6\end{raggedleft}%
				}{%
					\errmessage{LaTeX Warning: Alineacion columna derecha incorrecta, valores esperados: c,l,r}
				}}}
			\end{minipage}%
		\end{minipage}%
	\end{flushleft}%
	~ \vspace{-0.5\baselineskip}%
	\end{samepage}
	\setcaptionmargincm{\captionlrmargin}%
}

\newcommand{\createthreecolumn}[9][]{%
	\setcaptionmargincm{0}%
	\begin{samepage}%
	\begin{flushleft}%
		\vspace{-0.5\baselineskip}%
		\begin{minipage}{1\linewidth}%
			\begin{minipage}[t][#1][\GLOBALthreecolumnav]{#2\linewidth}%
				\ifthenelse{\equal{\GLOBALthreecolumnap}{c}}{%
					\begin{center}#7\end{center}%
				}{%
				\ifthenelse{\equal{\GLOBALthreecolumnap}{l}}{%
					\begin{raggedright}#7\end{raggedright}%
				}{%
				\ifthenelse{\equal{\GLOBALthreecolumnap}{r}}{%
					\hfill\begin{raggedleft}#7\end{raggedleft}%
				}{%
					\errmessage{LaTeX Warning: Alineacion columna izquierda incorrecta, valores esperados: c,l,r}
				}}}
			\end{minipage}%
			\hspace{#5 cm}%
			\begin{minipage}[t][#1][\GLOBALthreecolumnbv]{#3\linewidth}%
				\ifthenelse{\equal{\GLOBALthreecolumnbp}{c}}{%
					\begin{center}#8\end{center}%
				}{%
				\ifthenelse{\equal{\GLOBALthreecolumnbp}{l}}{%
					\begin{raggedright}#8\end{raggedright}%
				}{%
				\ifthenelse{\equal{\GLOBALthreecolumnbp}{r}}{%
					\hfill\begin{raggedleft}#8\end{raggedleft}%
				}{%
					\errmessage{LaTeX Warning: Alineacion columna central incorrecta, valores esperados: c,l,r}
				}}}
			\end{minipage}%
			\hspace{#6 cm}%
			\begin{minipage}[t][#1][\GLOBALthreecolumncv]{#4\linewidth}%
				\ifthenelse{\equal{\GLOBALthreecolumncp}{c}}{%
					\begin{center}#9\end{center}%
				}{%
				\ifthenelse{\equal{\GLOBALthreecolumncp}{l}}{%
					\begin{raggedright}#9\end{raggedright}%
				}{%
				\ifthenelse{\equal{\GLOBALthreecolumncp}{r}}{%
					\hfill\begin{raggedleft}#9\end{raggedleft}%
				}{%
					\errmessage{LaTeX Warning: Alineacion columna derecha incorrecta, valores esperados: c,l,r}
				}}}
			\end{minipage}%
		\end{minipage}%
	\end{flushleft}%
	~ \vspace{-0.5\baselineskip}%
	\end{samepage}
	\setcaptionmargincm{\captionlrmargin}%
}

\def\predocpageromannumber {false}
\def\predocresetpagenumber {false}

\newwrite\fileauthornames
\newwrite\fileauthordata
\newwrite\fileauthornamesdata

\immediate\openout\fileauthornames=\jobname.authn
\immediate\openout\fileauthordata=\jobname.authd
\immediate\openout\fileauthornamesdata=\jobname.authnd
\AtBeginDocument{%
	\immediate\closeout\fileauthornames
	\immediate\closeout\fileauthordata
	\immediate\closeout\fileauthornamesdata
}

\newcounter{authornumber}
\newcommand{\addauthor}[4][]{
	\immediate\write\fileauthornames{\unexpanded{\titleauthorname}[\theauthornumber]\unexpanded{{#2}}\LOCALpercentchar}
	\immediate\write\fileauthordata{\unexpanded{\titleauthordata}[\theauthornumber]\unexpanded{{#3}{#4}{#1}}\LOCALpercentchar}%
	\immediate\write\fileauthornamesdata{\unexpanded{\titleauthornamedata}[\theauthornumber]\unexpanded{{#2}{#3}{#4}{#1}}\LOCALpercentchar}%
	\stepcounter{authornumber}
}

\ifthenelse{\equal{\titlestyle}{style1}}{
	\def\LOCALhasname {false}
	\newcommand{\titleauthorname}[2][]{%
		\ifthenelse{\equal{\LOCALhasname}{true}}{\hspace{\titleauthorspacing cm}}{}%
		\normalsize{#2} \textsuperscript{(#1)}%
		\def\LOCALhasname {true}%
	}
	\newcommand{\titleauthordata}[4][]{%
		\ifthenelse{\equal{\LOCALhasname}{true}}{\vspace{0.025cm}\\}{}%
		\indent \footnotesize{\textsuperscript{(#1)} #2%
			\ifthenelse{\equal{#3}{}}{}{, \insertemail{#3}}%
			\ifthenelse{\equal{#4}{}}{}{, ORCID \href{https://orcid.org/#4}{#4}}}%
		\def\LOCALhasname {true}%
	}
	\newcommand{\inserttitle}{%
		\begin{center}%
			\ifthenelse{\equal{\titlebold}{true}}{%
				\textbf{\Large{\documenttitle}} \\%
			}{%
				\Large{\documenttitle} \\%
			}
			\vspace{0.6cm}%
			\begin{minipage}[t]{\titleauthormaxwidth\textwidth}
				\centering
				\input{\jobname.authn}%
			\end{minipage}
			\vspace{0.3cm}%
			\def\LOCALhasname {false}%
			~ \\%
			\begin{minipage}[t]{\titleauthormaxwidth\textwidth}%
				\centering%
				\input{\jobname.authd}%
			\end{minipage}
		\end{center}%
		\vspace{0.05cm}%
		\normalsize%
	}
}{
\ifthenelse{\equal{\titlestyle}{style2}}{
	\def\LOCALhasname {false}
	\newcommand{\titleauthorname}[2][]{%
		\ifthenelse{\equal{\LOCALhasname}{true}}{\hspace{\titleauthorspacing cm}}{}%
		\normalsize{#2} \textsuperscript{(#1)}%
		\def\LOCALhasname {true}%
	}
	\newcommand{\titleauthordata}[4][]{%
		\ifthenelse{\equal{\LOCALhasname}{true}}{\vspace{0.025cm}\\}{}%
		\indent \footnotesize{\textsuperscript{(#1)} #2%
		\ifthenelse{\equal{#3}{}}{}{, \insertemail{#3}}%
		\ifthenelse{\equal{#4}{}}{}{, ORCID \href{https://orcid.org/#4}{#4}}}%
		\def\LOCALhasname {true}%
		\vspace{-0.05cm}%
	}
	\newcommand{\inserttitle}{%
		\begin{center}%
			\ifthenelse{\equal{\titlebold}{true}}{%
				\textbf{\Large{\documenttitle}} \\%
			}{%
				\Large{\documenttitle} \\%
			}
			\vspace{0.7cm}%
			\input{\jobname.authn}%
			\vspace{0.2cm}%
		\end{center}%
		\par%
		\def\LOCALhasname {false}%
		\begin{minipage}[t]{\titleauthormaxwidth\textwidth}%
			\input{\jobname.authd}%
		\end{minipage}
		\vspace{0.3cm}%
		\normalsize%
	}
}{
\ifthenelse{\equal{\titlestyle}{style3}}{
	\def\LOCALhasname {false}
	\newcommand{\titleauthorname}[2][]{%
		\ifthenelse{\equal{\LOCALhasname}{true}}{,\ }{}%
		\normalsize{#2}%
		\def\LOCALhasname {true}%
	}
	\newcommand{\inserttitle}{%
		\begin{center}%
			\ifthenelse{\equal{\titlebold}{true}}{%
				\textbf{\Large{\documenttitle}} \\%
			}{%
				\Large{\documenttitle} \\%
			}
			\vspace{0.7cm}%
			\begin{minipage}[t]{\titleauthormaxwidth\textwidth}%
				\centering%
				\input{\jobname.authn}%
			\end{minipage}
			\vspace{0.3cm}%
			\def\LOCALhasname {false}%
		\end{center}%
		\normalsize%
	}
}{
\ifthenelse{\equal{\titlestyle}{style4}}{
	\def\LOCALhasname {false}
	\newcommand{\titleauthorname}[2][]{%
		\ifthenelse{\equal{\LOCALhasname}{true}}{,\ }{}%
		\normalsize{#2}%
		\def\LOCALhasname {true}%
	}
	\newcommand{\inserttitle}{%
		\begin{center}%
			\ifthenelse{\equal{\titlebold}{true}}{%
				\textbf{\Large{\documenttitle}} \\%
			}{%
				\Large{\documenttitle} \\%
			}
			\vspace{0.7cm}%
			\begin{minipage}[t]{\titleauthormaxwidth\textwidth}%
				\centering%
				\input{\jobname.authn}%
			\end{minipage}
			\vspace{0.3cm}%
			\def\LOCALhasname {false}%
			\vspace{0.1cm}%
			~ \\%
			\small{\documentdate} \\
			\vspace{0.3cm}%
		\end{center}%
		\normalsize%
	}
}{
\ifthenelse{\equal{\titlestyle}{style5}}{
	\def\LOCALhasname {false}
	\newcommand{\titleauthornamedata}[5][]{%
		\ifthenelse{\equal{\LOCALhasname}{true}}{\hspace{\titleauthorspacing cm}}{}%
		\begin{tabular}[t]{C{4cm}}
			\normalsize{#2} \\
			\small{#3} \\
			\ifthenelse{\equal{#4}{}}{}{\small{\insertemail{#4}} \\}
			\ifthenelse{\equal{#5}{}}{}{\small{ORCID \href{https://orcid.org/#5}{#5}}}
		\end{tabular}
		\def\LOCALhasname {true}%
	}
	\newcommand{\inserttitle}{%
		\begin{center}%
			\ifthenelse{\equal{\titlebold}{true}}{%
				\textbf{\Large{\documenttitle}} \\%
			}{%
				\Large{\documenttitle} \\%
			}
			\vspace{0.9cm}%
			\noindent%
			\settablecellpadding{0}{1.5}%
			\linespread{0.6}\selectfont{%
				\begin{minipage}[t]{\titleauthormaxwidth\textwidth}%
					\centering%
					\input{\jobname.authnd}%
				\end{minipage}
			}
			\resettablecellpadding%
			\def\LOCALhasname {false}%
		\end{center}%
		\normalsize%
	}
}{
	\throwbadconfig{Valor configuracion incorrecto}{\titlestyle}{style1 .. style5}}}}}
}

\newenvironment{appendixd}{
	\global\def\GLOBALenvappendix {true}%
	\begingroup 
	\phantomsection
	\changeheadertitle{\nameltappendixsection} 
	\global\def\GLOBALsectionalph {true} 
	\bookmarksetup{
		numbered={true},
		openlevel={\thetemplateBookmarksLevelPrev}
	}
	\appendixtitleon
	\appendixtitletocon
	\bookmarksetupnext{level=part}
	\begin{appendices} 
		\ifthenelse{\equal{\showappendixsecindex}{true}}{}{
			\pdfbookmark{\nameappendixsection}{appendix} 
		}
		\ifthenelse{\equal{\appendixindepobjnum}{true}}{
			\counterwithin{equation}{section}
			\counterwithin{figure}{section}
			\counterwithin{lstlisting}{section}
			\counterwithin{table}{section}}{
		}
	}{
	\end{appendices}
	\global\def\GLOBALsectionalph {false} 
	\bookmarksetupnext{level={\thetemplateBookmarksLevelPrev}} 
	\bookmarksetup{
		numbered={\cfgpdfsecnumbookmarks},
		openlevel={\cfgbookmarksopenlevel}
	}
	\endgroup
}

\newcommand{\coreinitsourcecodep}[4]{
	\emptyvarerr{\coreinitsourcecodep}{#2}{Estilo de codigo no definido}
	\checkvalidsourcecodestyle{#2}
	\ifthenelse{\equal{\showlinenumbers}{true}}{
		\rightlinenumbers}{
	}
	\lstset{
		backgroundcolor=\color{\sourcecodebgcolor}
	}
	\ifthenelse{\equal{\codecaptiontop}{true}}{
		\ifx\hfuzz#4\hfuzz
			\ifx\hfuzz#3\hfuzz
				\lstset{
					escapeinside={(*@}{@*)},
					style=#2
				}
			\else
				\lstset{
					escapeinside={(*@}{@*)},
					style=#2,
					#3
				}
			\fi
		\else
			\ifx\hfuzz#3\hfuzz
				\lstset{
					caption={#4 #1},
					captionpos=t,
					escapeinside={(*@}{@*)},
					style=#2
				}
			\else
				\lstset{
					caption={#4 #1},
					captionpos=t,
					escapeinside={(*@}{@*)},
					style=#2,
					#3
				}
			\fi
		\fi
	}{
		\ifx\hfuzz#4\hfuzz
			\ifx\hfuzz#3\hfuzz
				\lstset{
					escapeinside={(*@}{@*)},
					style=#2
				}
			\else
				\lstset{
					escapeinside={(*@}{@*)},
					style=#2,
					#3
				}
			\fi
		\else
			\ifx\hfuzz#3\hfuzz
				\lstset{
					caption={#4 #1},
					captionpos=b,
					style=#2
				}
			\else
				\lstset{
					caption={#4 #1},
					captionpos=b,
					escapeinside={(*@}{@*)},
					style=#2,
					#3
				}
			\fi
		\fi	
	}
}

\lstnewenvironment{sourcecodep}[4][]{%
	\coreinitsourcecodep{#1}{#2}{#3}{#4}%
}{%
	\ifthenelse{\equal{\showlinenumbers}{true}}{%
		\leftlinenumbers}{%
	}%
}

\newcommand{\importsourcecodep}[5][]{%
	\coreinitsourcecodep{#1}{#2}{#3}{#5}%
	\inputlisting{#4}%
	\ifthenelse{\equal{\showlinenumbers}{true}}{%
		\leftlinenumbers}{%
	}%
}

\newcommand{\coreinitsourcecode}[3]{
	\emptyvarerr{\coreinitsourcecode}{#2}{Estilo de codigo no definido}
	\checkvalidsourcecodestyle{#2}
	\ifthenelse{\equal{\showlinenumbers}{true}}{
		\rightlinenumbers}{
	}
	\lstset{
		backgroundcolor=\color{\sourcecodebgcolor}
	}
	\ifthenelse{\equal{\codecaptiontop}{true}}{
		\ifx\hfuzz#3\hfuzz
			\lstset{
				escapeinside={(*@}{@*)},
				style=#2
			}
		\else
			\lstset{
				escapeinside={(*@}{@*)},
				caption={#3 #1},
				captionpos=t,
				style=#2
			}
		\fi
	}{
		\ifx\hfuzz#3\hfuzz
			\lstset{
				escapeinside={(*@}{@*)},
				style=#2
			}
		\else
			\lstset{
				escapeinside={(*@}{@*)},
				caption={#3 #1},
				captionpos=b,
				style=#2
			}
		\fi
	}
}

\lstnewenvironment{sourcecode}[3][]{%
	\coreinitsourcecode{#1}{#2}{#3}%
}{%
	\ifthenelse{\equal{\showlinenumbers}{true}}{%
		\leftlinenumbers}{%
	}%
}

\newcommand{\importsourcecode}[4][]{%
	\coreinitsourcecode{#1}{#2}{#4}%
	\lstinputlisting{#3}%
	\ifthenelse{\equal{\showlinenumbers}{true}}{%
		\leftlinenumbers}{%
	}
}

\newenvironment{images}[2][]{%
	\def\envimageslabelvar {#1}
	\def\envimagescaptionvar {#2}
	\global\def\GLOBALenvimageinitialized {true}
	\global\def\GLOBALenvimageadded {false}
	
	\corevspacevarcm{\marginimagetop}%
	\setcaptionmargincm{\captionmarginmultimg} 
	
	\begin{samepage}%
	\begin{figure}[H] \centering%
		\corevspacevarcm{\marginimagemulttop}%
		}{%
		\setcaptionmargincm{\captionlrmargin}%
		\ifthenelse{\equal{\envimagescaptionvar}{}}{ 
			\corevspacevarcm{\captionlessmarginimage}%
		}{%
			\ifthenelse{\equal{\captionmarginimages}{0}}{}{\corevspacevarcm{\captionmarginimages}}%
			\caption{\envimagescaptionvar\envimageslabelvar}%
		}%
	\end{figure}%

	\setcaptionmargincm{\captionlrmargin}%
	\corevspacevarcm{\marginimagebottom}%
	\end{samepage}
	
	\global\def\GLOBALenvimageinitialized {false}
}

\colorlet{numb}{magenta!60!black}
\colorlet{punct}{red!60!black}
\definecolor{delim}{RGB}{20,105,176}
\definecolor{dkcyan}{RGB}{0,123,167}
\definecolor{dkgray}{RGB}{90,90,90}
\definecolor{dkgreen}{RGB}{0,150,0}
\definecolor{gray}{RGB}{127,127,127}
\definecolor{lbrown}{RGB}{255,252,249}
\definecolor{lgray}{RGB}{240,240,240}
\definecolor{mauve}{RGB}{150,0,210}
\definecolor{ocre}{RGB}{243,102,25}

\lstdefinestyle{abap}{
	language=ABAP
}

\lstdefinestyle{ada}{
	language=[2005]Ada
}

\lstdefinelanguage[x64]{Assembler}[x86masm]{Assembler}{
	morekeywords={
		CDQE,CMPSQ,CMPXCHG16B,CQO,IRETQ,JRCXZ,LODSQ,MOVSXD,POPFQ,PUSHFQ,r8,r8b,r8d,r8w,r9,r9b,r9d,r9w,r10,r10b,r10d,r10w,r11,r11b,r11d,r11w,r12,r12b,r12d,r12w,r13,r13b,r13d,r13w,r14,r14b,r14d,r14w,r15,r15b,r15d,r15w,rax,rbp,rbx,rcx,rdi,RDTSCP,rdx,rsi,rsp,SCASQ,STOSQ,SWAPGS
	}
}
\lstdefinestyle{assemblerx64}{
	language=[x64]Assembler
}
\lstdefinestyle{assemblerx86}{
	language=[x86masm]Assembler
}

\lstdefinestyle{awk}{
	language=[gnu]Awk
}

\lstdefinestyle{bash}{
	language=bash,
	breakatwhitespace=false,
	morecomment=[l]{rem},
	morecomment=[s]{::}{::},
	morekeywords={
		call,cp,dig,gcc,git,grep,ls,mv,python,rm,sudo,vim
	},
	sensitive=false
}

\lstdefinestyle{basic}{
	language=[Visual]Basic
}

\lstdefinestyle{c}{
	language=C,
	breakatwhitespace=false,
	keepspaces=true
}

\lstdefinestyle{caml}{
	language=[light]Caml
}

\lstdefinestyle{cmake}{
	language=[gnu] make,
	keywordstyle=[2]\color{dkcyan},
	morekeywords=[1]{
		add_custom_command,add_custom_target,add_definitions,add_executable,add_library,add_subdirectory,cmake_minimum_required,cmake_policy,configure_file,cuda_add_library,cuda_include_directories,else,elseif,endforeach,endfunction,endif,endmacro,execute_process,file,find_package,find_path,find_program,foreach,function,get_directory_property,get_filename_component,get_filename_component,get_source_file_property,if,include,include_directories,install,link_directories,list,macro,mark_as_advanced,message,option,project,set,set_property,set_target_properties,string,target_compile_options,target_link_libraries,unset
	},
	morekeywords=[2]{
		AND,APPEND,ARCHIVE,CACHE,CMAKE_MODULE_PATH,COMMAND,COMMENT,COMPILE_DEFINITIONS,CONFIG,DEFINED,DEPENDS,DESTINATION,DIRECTORY,ENDIF,ENV,EQUAL,ERROR_QUIET,EXISTS,FATAL_ERROR,FILES,FILES_MATCHING,FIND,FIND,FIND_LIBRARY,FORCE,GLOB,GREATER,IF,INCLUDE_DIRECTORIES,IS_ABSOLUTE,LESS,LIBRARY,LINK_PRIVATE,LIST,MAIN_DEPENDENCY,MAKE_DIRECTORY,MARK_AS_ADVANCED,MATCHALL,MATCHES,NOT,OFF,ON,OPTIONAL,OR,OUTPUT,OUTPUT_STRIP_TRAILING_WHITESPACE,OUTPUT_VARIABLE,PARENT_SCOPE,PATTERN,PRE_BUILD_COMMAND,PRE_LINK,PRIVATE,PROPERTIES,PUBLIC,REGEX,RELEASE,RENAME,REQUIRED,RUNTIME,SET,STATIC,STREQUAL,TARGET,TARGETS,TOUPPER,UNIX,VERSION,VERSION_EQUAL,VERSION_LESS,WIN32,WORKING_DIRECTORY
	}
}

\lstdefinestyle{cobol}{
	language=Cobol
}

\lstdefinestyle{cpp}{
	language=C++,
	breakatwhitespace=false
}

\lstdefinestyle{csharp}{
	language=csh,
	morecomment=[l]{//},
	morecomment=[s]{/*}{*/},
	morekeywords={
		abstract,as,base,bool,break,byte,case,catch,char,checked,class,const,continue,decimal,default,delegate,do,double,else,enum,event,explicit,extern,false,finally,fixed,float,for,foreach,goto,if,implicit,in,int,interface,internal,is,lock,long,namespace,new,null,object,operator,out,override,params,private,protected,public,readonly,ref,return,sbyte,sealed,short,sizeof,stackalloc,static,string,struct,switch,this,throw,true,try,typeof,uint,ulong,unchecked,unsafe,ushort,using,virtual,void,volatile,while
	}
}

\lstdefinelanguage{CSS}{
	morecomment=[s]{/*}{*/},
	morekeywords={
		-moz-binding,-moz-border-bottom-colors,-moz-border-left-colors,-moz-border-radius,-moz-border-radius-bottomleft,-moz-border-radius-bottomright,-moz-border-radius-topleft,-moz-border-radius-topright,-moz-border-right-colors,-moz-border-top-colors,-moz-opacity,-moz-outline,-moz-outline-color,-moz-outline-style,-moz-outline-width,-moz-user-focus,-moz-user-input,-moz-user-modify,-moz-user-select,-replace,-set-link-source,-use-link-source,accelerator,azimuth,background,background-attachment,background-color,background-image,background-position,background-position-x,background-position-y,background-repeat,behavior,border,border-bottom,border-bottom-color,border-bottom-style,border-bottom-width,border-collapse,border-color,border-left,border-left-color,border-left-style,border-left-width,border-right,border-right-color,border-right-style,border-right-width,border-spacing,border-style,border-top,border-top-color,border-top-style,border-top-width,border-width,bottom,caption-side,clear,clip,color,content,counter-increment,counter-reset,cue,cue-after,cue-before,cursor,direction,display,elevation,empty-cells,filter,float,font,font-family,font-size,font-size-adjust,font-stretch,font-style,font-variant,font-weight,height,ime-mode,include-source,layer-background-color,layer-background-image,layout-flow,layout-grid,layout-grid-char,layout-grid-char-spacing,layout-grid-line,layout-grid-mode,layout-grid-type,left,letter-spacing,line-break,line-height,list-style,list-style-image,list-style-position,list-style-type,margin,margin-bottom,margin-left,margin-right,margin-top,marker-offset,marks,max-height,max-width,min-height,min-width,orphans,outline,outline-color,outline-style,outline-width,overflow,overflow-X,overflow-Y,padding,padding-bottom,padding-left,padding-right,padding-top,page,page-break-after,page-break-before,page-break-inside,pause,pause-after,pause-before,pitch,pitch-range,play-during,position,quotes,richness,right,ruby-align,ruby-overhang,ruby-position,scrollbar-3d-light-color,scrollbar-arrow-color,scrollbar-base-color,scrollbar-dark-shadow-color,scrollbar-face-color,scrollbar-highlight-color,scrollbar-shadow-color,scrollbar-track-color,size,speak,speak-header,speak-numeral,speak-punctuation,speech-rate,stress,table-layout,text-align,text-align-last,text-autospace,text-decoration,text-indent,text-justify,text-kashida-space,text-overflow,text-shadow,text-transform,text-underline-position,top,unicode-bidi,vertical-align,visibility,voice-family,volume,white-space,widows,width,word-break,word-spacing,word-wrap,writing-mode,z-index,zoom
	},
	morestring=[s]{:}{;},
	sensitive=true
}
\lstdefinestyle{css}{
	language=CSS,
	breakatwhitespace=true
}

\lstdefinestyle{csv}{
	language={}
}

\lstdefinestyle{cuda}{
	language=C++,
	breakatwhitespace=false,
	emph={
		cudaFree,cudaMalloc,__device__,__global__,__host__,__shared__,__syncthreads
	},
	emphstyle=\color{dkcyan}\ttfamily,
	morecomment=[l][\color{magenta}]{\#},
	moredelim=[s][\ttfamily]{<<<}{>>>}
}

\lstdefinestyle{dart}{
	language=Java,
	emph=[2]{
		findAllElements,findElements
	},
	morekeywords={
		*,get,library,List,num,set,String,var
	}
}

\lstdefinelanguage{docker}{
	comment=[l]{\#},
	keywords={
		ADD,CMD,COPY,ENTRYPOINT,ENV,EXPOSE,FROM,LABEL,MAINTAINER,ONBUILD,RUN,STOPSIGNAL,USER,VOLUME,WORKDIR
	},
	morestring=[b]',
	morestring=[b]"
}
\lstdefinestyle{docker}{
	language=docker,
	breakatwhitespace=true
}

\lstdefinestyle{elisp}{
	language=elisp
}

\lstdefinestyle{elixir}{
	morekeywords={
		case,catch,def,do,else,false,use,alias,receive,timeout,defmacro,defp,for,if,import,defmodule,defprotocol,nil,defmacrop,defoverridable,defimpl,super,fn,raise,true,try,end,with,unless
	},
	otherkeywords={
		<-,->, |>, \%\{, \}, \{, \, (, )
	},
	morecomment=[l]{\#},
	morecomment=[n]{/*}{*/},
	morecomment=[s][\color{purple}]{:}{\ },
	morestring=[s][\color{mauve}]"",
	sensitive=true
}

\lstdefinestyle{erlang}{
	language=erlang
}

\lstdefinestyle{fortran}{
	language=[95]Fortran,
	breakatwhitespace=false
}

\lstdefinestyle{fsharp}{
	morecomment=[l][\color{dkgreen}]{///},
	morecomment=[l][\color{dkgreen}]{//},
	morecomment=[s][\color{dkgreen}]{{(*}{*)}},
	morestring=[b]",
	morekeywords={
		abstract,and,Application,Array,Async,async,begin,cloud,do,else,end,false,finally,for,fun,function,if,in,inherit,interface,let,List,match,member,module,mutable,namespace,new,of,open,rec,return,Seq,static,System,then,true,try,type,use,while,with,yield
	},
	otherkeywords={
		by,do!,from,let!,order,return!,select,use!,var,where,yield!
	},
	sensitive=true
}

\lstdefinelanguage{GLSL}{
	alsoletter={\#},
	morekeywords=[1]{
		attribute,bool,break,bvec2,bvec3,bvec4,case,centroid,const,continue,default,discard,do,else,false,flat,float,for,highp,if,in,inout,int,invariant,isampler1D,isampler1DArray,isampler2D,isampler2DArray,isampler2DMS,isampler2DMSArray,isampler2DRect,isampler3D,isamplerBuffer,isamplerCube,ivec2,ivec3,ivec4,layout,lowp,mat2,mat2x2,mat2x3,mat2x4,mat3,mat3x2,mat3x3,mat3x4,mat4,mat4x2,mat4x3,mat4x4,mediump,noperspective,out,precision,return,sampler1D,sampler1DArray,sampler1DArrayShadow,sampler1DShadow,sampler2D,sampler2DArray,sampler2DArrayShadow,sampler2DMS,sampler2DMSArray,sampler2DRect,sampler2DRectShadow,sampler2DShadow,sampler3D,samplerBuffer,samplerCube,samplerCubeShadow,smooth,struct,switch,true,uint,uniform,usampler1D,usampler1DArray,usampler2D,usampler2DArray,usampler2DMS,usampler2DMSArray,usampler2DRect,usampler3D,usamplerBuffer,usamplerCube,uvec2,uvec3,uvec4,varying,vec2,vec3,vec4,void,while
	},
	morekeywords=[2]{
		abs,acos,acosh,all,any,asin,asinh,atan,atan,atanh,ceil,clamp,cos,cosh,cross,degrees,determinant,dFdx,dFdy,distance,dot,EmitVertex,EndPrimitive,equal,exp,exp2,faceforward,floatBitsToInt,floatBitsToUint,floor,fract,fwidth,greaterThan,greaterThanEqual,intBitsToFloat,inverse,inversesqrt,isinf,isnan,length,lessThan,lessThanEqual,log,log2,matrixCompMult,max,min,mix,mod,modf,noise1,noise2,noise3,noise4,normalize,not,notEqual,outerProduct,pow,radians,reflect,refract,round,roundEven,shadow1D,shadow1DLod,shadow1DProj,shadow1DProjLod,shadow2D,shadow2DLod,shadow2DProj,shadow2DProjLod,sign,sin,sinh,smoothstep,sqrt,step,tan,tanh,texelFetch,texelFetchOffset,texture,texture1D,texture1DProj,texture1DProjLod,texture2D,texture2DLod,texture2DProj,texture2DProjLod,texture3D,texture3DLod,texture3DProj,texture3DProjLod,textureCube,textureCubeLod,textureGrad,textureGradOffset,textureLod,textureLodOffset,textureOffset,textureProj,textureProjGrad,textureProjGradOffset,textureProjLod,textureProjLodOffset,textureProjOffset,textureSize,transpose,trunc,uintBitsToFloat
	},
	morekeywords=[3]{
		\#version,core,gl_ClipDistance,gl_ClipDistance,gl_ClipVertex,gl_DepthRange,gl_FragColor,gl_FragCoord,gl_FragData,gl_FragDepth,gl_FrontFacing,gl_InstanceID,gl_Layer,gl_MaxClipDistances,gl_MaxCombinedTextureImageUnits,gl_MaxDrawBuffers,gl_MaxDrawBuffers,gl_MaxFragmentInputComponents,gl_MaxFragmentUniformComponents,gl_MaxGeometryInputComponents,gl_MaxGeometryOutputComponents,gl_MaxGeometryOutputVertices,gl_MaxGeometryOutputVertices,gl_MaxGeometryTextureImageUnits,gl_MaxGeometryTotalOutputComponents,gl_MaxGeometryUniformComponents,gl_MaxGeometryVaryingComponents,gl_MaxTextureImageUnits,gl_MaxVaryingComponents,gl_MaxVaryingFloats,gl_MaxVertexAttribs,gl_MaxVertexOutputComponents,gl_MaxVertexTextureImageUnits,gl_MaxVertexUniformComponents,gl_PerVertex,gl_PointCoord,gl_PointSize,gl_Position,gl_PrimitiveID,gl_VertexID
	},
	morecomment=[l]{//},
	morecomment=[s]{/*}{*/}
}
\lstdefinestyle{glsl}{
	language=GLSL,
	keywordstyle=[3]\color{dkcyan}\ttfamily,
	prebreak=\raisebox{0ex}[0ex][0ex]{\ensuremath{\hookleftarrow}},
	sensitive=true,
	upquote=true
}

\lstdefinestyle{gnuplot}{
	language=Gnuplot
}

\lstdefinestyle{go}{
	language=Go
}

\lstdefinestyle{haskell}{
	language=haskell,
	morecomment=[l]\%
}

\lstdefinelanguage{HTML5}{
	language=html,
	alsoletter={<>=-},
	morecomment=[s]{<!--}{-->},
	ndkeywords={
		=,
		accept-charset=,accept=,accesskey=,action=,align=,alt=,async=,autocomplete=,autofocus=,autoplay=,autosave=,bgcolor=,border=,buffered=,challenge=,charset=,checked=,cite=,class=,code=,codebase=,color=,cols=,colspan=,content=,contenteditable=,contextmenu=,controls=,coords=,data=,datetime=,default=,defer=,dir=,dirname=,disabled=,download=,draggable=,dropzone=,enctype=,for=,form=,formaction=,headers=,height=,hidden=,high=,href=,hreflang=,http-equiv=,icon=,id=,ismap=,itemprop=,keytype=,kind=,label=,lang=,language=,list=,loop=,low=,manifest=,max=,maxlength=,media=,method=,min=,multiple=,name=,novalidate=,open=,optimum=,pattern=,ping=,placeholder=,poster=,preload=,pubdate=,radiogroup=,readonly=,rel=,required=,reversed=,rows=,rowspan=,sandbox=,scope=,scoped=,seamless=,selected=,shape=,size=,sizes=,span=,spellcheck=,src=,srcdoc=,srclang=,start=,step=,style=,summary=,tabindex=,target=,title=,type=,usemap=,value=,width=,wrap=,
		-moz-binding:,-moz-border-bottom-colors:,-moz-border-left-colors:,-moz-border-radius-bottomleft:,-moz-border-radius-bottomright:,-moz-border-radius-topleft:,-moz-border-radius-topright:,-moz-border-radius:,-moz-border-right-colors:,-moz-border-top-colors:,-moz-opacity:,-moz-outline-color:,-moz-outline-style:,-moz-outline-width:,-moz-outline:,-moz-transform:,-moz-user-focus:,-moz-user-input:,-moz-user-modify:,-moz-user-select:,-replace:,-set-link-source:,-use-link-source:,accelerator:,azimuth:,background-attachment:,background-color:,background-image:,background-position-x:,background-position-y:,background-position:,background-repeat:,background:,behavior:,border-bottom-color:,border-bottom-style:,border-bottom-width:,border-bottom:,border-collapse:,border-color:,border-left-color:,border-left-style:,border-left-width:,border-left:,border-right-color:,border-right-style:,border-right-width:,border-right:,border-spacing:,border-style:,border-top-color:,border-top-style:,border-top-width:,border-top:,border-width:,border:,bottom:,caption-side:,clear:,clip:,color:,content:,counter-increment:,counter-reset:,cue-after:,cue-before:,cue:,cursor:,direction:,display:,elevation:,empty-cells:,filter:,float:,font-family:,font-size-adjust:,font-size:,font-stretch:,font-style:,font-variant:,font-weight:,font:,height:,ime-mode:,include-source:,layer-background-color:,layer-background-image:,layout-flow:,layout-grid-char-spacing:,layout-grid-char:,layout-grid-line:,layout-grid-mode:,layout-grid-type:,layout-grid:,left:,letter-spacing:,line-break:,line-height:,list-style-image:,list-style-position:,list-style-type:,list-style:,margin-bottom:,margin-left:,margin-right:,margin-top:,margin:,marker-offset:,marks:,max-height:,max-width:,min-height:,min-width:,orphans:,outline-color:,outline-style:,outline-width:,outline:,overflow-X:,overflow-Y:,overflow:,padding-bottom:,padding-left:,padding-right:,padding-top:,padding:,page-break-after:,page-break-before:,page-break-inside:,page:,pause-after:,pause-before:,pause:,pitch-range:,pitch:,play-during:,position:,quotes:,richness:,right:,ruby-align:,ruby-overhang:,ruby-position:,scrollbar-3d-light-color:,scrollbar-arrow-color:,scrollbar-base-color:,scrollbar-dark-shadow-color:,scrollbar-face-color:,scrollbar-highlight-color:,scrollbar-shadow-color:,scrollbar-track-color:,size:,speak-header:,speak-numeral:,speak-punctuation:,speak:,speech-rate:,stress:,table-layout:,text-align-last:,text-align:,text-autospace:,text-decoration:,text-indent:,text-justify:,text-kashida-space:,text-overflow:,text-shadow:,text-transform:,text-underline-position:,top:,transform:,transition-duration:,transition-property:,transition-timing-function:,unicode-bidi:,vertical-align:,visibility:,voice-family:,volume:,white-space:,widows:,width:,word-break:,word-spacing:,word-wrap:,writing-mode:,z-index:,zoom:
	},
	otherkeywords={
		<,</,>,</a,<a,</a>,</abbr,<abbr,</abbr>,</address,<address,</address>,</area,<area,</area>,</area,<area,</area>,</article,<article,</article>,</aside,<aside,</aside>,</audio,<audio,</audio>,</audio,<audio,</audio>,</b,<b,</b>,</base,<base,</base>,</bdi,<bdi,</bdi>,</bdo,<bdo,</bdo>,</blockquote,<blockquote,</blockquote>,</body,<body,</body>,</br,<br,</br>,</button,<button,</button>,</canvas,<canvas,</canvas>,</caption,<caption,</caption>,</cite,<cite,</cite>,</code,<code,</code>,</col,<col,</col>,</colgroup,<colgroup,</colgroup>,</data,<data,</data>,</datalist,<datalist,</datalist>,</dd,<dd,</dd>,</del,<del,</del>,</details,<details,</details>,</dfn,<dfn,</dfn>,</div,<div,</div>,</dl,<dl,</dl>,</dt,<dt,</dt>,</em,<em,</em>,</embed,<embed,</embed>,</fieldset,<fieldset,</fieldset>,</figcaption,<figcaption,</figcaption>,</figure,<figure,</figure>,</footer,<footer,</footer>,</form,<form,</form>,</h1,<h1,</h1>,</h2,<h2,</h2>,</h3,<h3,</h3>,</h4,<h4,</h4>,</h5,<h5,</h5>,</h6,<h6,</h6>,</head,<head,</head>,</header,<header,</header>,</hr,<hr,</hr>,</html,<html,</html>,</i,<i,</i>,</iframe,<iframe,</iframe>,</img,<img,</img>,</input,<input,</input>,</ins,<ins,</ins>,</kbd,<kbd,</kbd>,</keygen,<keygen,</keygen>,</label,<label,</label>,</legend,<legend,</legend>,</li,<li,</li>,</link,<link,</link>,</main,<main,</main>,</map,<map,</map>,</mark,<mark,</mark>,</math,<math,</math>,</menu,<menu,</menu>,</menuitem,<menuitem,</menuitem>,</meta,<meta,</meta>,</meter,<meter,</meter>,</nav,<nav,</nav>,</noscript,<noscript,</noscript>,</object,<object,</object>,</ol,<ol,</ol>,</optgroup,<optgroup,</optgroup>,</option,<option,</option>,</output,<output,</output>,</p,<p,</p>,</param,<param,</param>,</pre,<pre,</pre>,</progress,<progress,</progress>,</q,<q,</q>,</rp,<rp,</rp>,</rt,<rt,</rt>,</ruby,<ruby,</ruby>,</s,<s,</s>,</samp,<samp,</samp>,</script,<script,</script>,</section,<section,</section>,</select,<select,</select>,</small,<small,</small>,</source,<source,</source>,</span,<span,</span>,</strong,<strong,</strong>,</style,<style,</style>,</summary,<summary,</summary>,</sup,<sup,</sup>,</svg,<svg,</svg>,</table,<table,</table>,</tbody,<tbody,</tbody>,</td,<td,</td>,</template,<template,</template>,</textarea,<textarea,</textarea>,</tfoot,<tfoot,</tfoot>,</th,<th,</th>,</thead,<thead,</thead>,</time,<time,</time>,</title,<title,</title>,</tr,<tr,</tr>,</track,<track,</track>,</u,<u,</u>,</ul,<ul,</ul>,</var,<var,</var>,</video,<video,</video>,</wbr,<wbr,</wbr>,/>,<!
	},
	sensitive=true,
	tag=[s]
}
\lstdefinestyle{html}{
	language=HTML5,
	alsodigit={.:;},
	alsolanguage=JavaScript,
	firstnumber=1,
	ndkeywordstyle=\color{dkgreen}\bfseries,
	numberfirstline=true
}

\lstdefinestyle{ini}{
	language={},
	commentstyle=\color{gray}\ttfamily,
	keywordstyle={\color{black}\bfseries},
	morecomment=[l]{;},
	morecomment=[l]{\#},
	morecomment=[s][\color{dkgreen}\bfseries]{[}{]},
	morekeywords={},
	otherkeywords={=,:}
}

\lstdefinestyle{java}{
	language=Java,
	breakatwhitespace=true,
	keepspaces=true
}

\lstdefinelanguage{JavaScript}{
	comment=[l]{//},
	keepspaces=true,
	keywords={
		break,else,false,for,function,if,in,new,null,return,true,typeof,var,while
	},
	morecomment=[s]{/*}{*/},
	morestring=[b]',
	morestring=[b]",
	morestring=[b]`,
	ndkeywords={
		await,async,case,catch,class,const,default,do,enum,export,extends,finally,from,implements,import,instanceof,let,static,super,switch,then,this,throw,try
	},
	ndkeywordstyle=\color{blue}\bfseries,
	sensitive=false
}
\lstdefinestyle{javascript}{
	language=JavaScript
}

\lstdefinestyle{json}{
	literate=*{0}{{{\color{numb}0}}}{1}{1}{{{\color{numb}1}}}{1}{2}
	{{{\color{numb}2}}}{1}{3}{{{\color{numb}3}}}{1}{4}{{{\color{numb}4}}}
	{1}{5}{{{\color{numb}5}}}{1}{6}{{{\color{numb}6}}}{1}{7}{{{\color{numb}7}}}
	{1}{8}{{{\color{numb}8}}}{1}{9}{{{\color{numb}9}}}{1}{:}
	{{{\color{punct}{:}}}}{1}{,}{{{\color{punct}{,}}}}{1}{\{}
	{{{\color{delim}{\{}}}}{1}{\}}{{{\color{delim}{\}}}}}
	{1}{[}{{{\color{delim}{[}}}}{1}{]}{{{\color{delim}{]}}}}{1},
	tabsize=2
}

\lstdefinestyle{julia}{
	keywordsprefix=\@,
	morecomment=[l]{\#},
	morekeywords={
		abstract,Any,applicable,assert,baremodule,begin,bitstype,Bool,break,catch,ccall,Complex64,Complex128,const,continue,convert,dlopen,dlsym,do,edit,else,elseif,end,eps,error,exit,export,finalizer,Float32,Float64,for,function,global,hash,if,im,immutable,import,importall,in,Inf,Int,Int8,Int16,Int32,Int64,invoke,is,isa,isequal,let,load,local,macro,method_exists,module,Nan,new,None,Nothing,ntuple,pi,promote,promote_type,quote,realmax,realmin,return,sizeof,subtype,system,throw,try,tuple,type,typealias,typemax,typemin,typeof,uid,Uint,Uint8,Uint16,Uint32,Uint64,using,while,whos
	},
	morestring=[b]',
	morestring=[b]",
	sensitive=true
}

\lstdefinestyle{kotlin}{
	comment=[l]{//},
	emph={delegate,filter,first,firstOrNull,forEach,lazy,map,mapNotNull,println,
		return@},
	emphstyle={\color{blue}},
	keywords={
		abstract,actual,as,as?,break,by,class,companion,continue,data,do,dynamic,else,enum,expect,false,final,for,fun,get,if,import,in,interface,internal,is,null,object,override,package,private,public,return,set,super,suspend,this,throw,true,try,typealias,val,var,vararg,when,where,while
	},
	morecomment=[s]{/*}{*/},
	morestring=[b]",
	morestring=[s]{"""*}{*"""},
	ndkeywords={
		@Deprecated,@JvmField,@JvmName,@JvmOverloads,@JvmStatic,@JvmSynthetic,Array,Byte,Double,Float,Int,Integer,Iterable,Long,Runnable,Short,String
	},
	ndkeywordstyle=\color{BurntOrange}\bfseries,
	sensitive=true
}

\lstdefinestyle{latex}{
	language=TeX,
	morekeywords={
		aacos,aasin,aatan,acos,addimage,addimageanum,addimageboxed,align,asin,atan,begin,bibitem,bibliography,bigstrut,boldmath,bookmarksetup,boxed,cancelto,caption,changeheadertitle,checkmark,checkvardefined,cite,clearpage,dd,degree,eqref,equal,frac,fracnpartial,fullcite,hline,href,ifthenelse,imageshspace,imagesnewline,imagesvspace,includefullhfpdf,includehfpdf,insertalign,insertalignanum,insertaligncaptioned,insertaligncaptioned,insertaligncaptionedanum,insertaligned,insertalignedanum,insertalignedcaptioned,insertalignedcaptionedanum,insertemail,insertemptypage,inserteqimage,insertequation,insertequationanum,insertequationcaptioned,insertequationcaptionedanum,insertgather,insertgatheranum,insertgathercaptioned,insertgathercaptionedanum,insertgathered,insertgatheredanum,insertgatheredcaptioned,insertgatheredcaptionedanum,insertimage,insertimageleft,insertimageright,insertindextitle,insertindextitlepage,insertphone,isundefined,itemresize,label,LaTeX,lipsum,lpow,makeatletter,makeatother,newcommand,newcounter,newp,newpage,pow,quotes,ref,renewcommand,section,sectionanum,setcounter,setlength,shortcite,sourcecode,sourcecodep,subsection,subsectionanum,subsubsection,subsubsectionanum,subsubsubsection,subsubsubsection,subsubsubsectionanum,textbf,textit,textregistered,textsuperscript,texttt,throwbadconfig,unboldmath,url,xspace
	}
}

\lstdefinestyle{lisp}{
	language=Lisp,
	morekeywords={if}
}

\lstdefinestyle{llvm}{
	language=LLVM
}

\lstdefinestyle{lua}{
	language={[5.3]Lua}
}

\lstdefinestyle{make}{
	language=[gnu] make
}

\lstdefinelanguage{Maple}{
	morecomment=[l]\#,
	morekeywords={
		and,assuming,break,by,catch,description,do,done,elif,else,end,error,export,fi,finally,for,from,global,if,implies,in,intersect,local,minus,mod,module,next,not,o,option,options,or,proc,quit,read,restart,return,save,stop,subset,then,to,try,union,use,uses,with,while,xor
	},
	morestring=[b]",
	morestring=[d],
	sensitive=true
} 
\lstdefinestyle{maple}{
	language=Maple
}

\lstdefinestyle{mathematica}{
	language=Mathematica
}

\lstdefinestyle{matlab}{
	language=Matlab,
	deletekeywords={fft},
	keepspaces=true,
	morecomment=[l]\%,
	morecomment=[n]{\%\{\^^M}{\%\}\^^M},
	morekeywords={
		addOptional,box,break,catch,cell,classdef,continue,deal,double,end,factorial,for,gradient,hessian,if,isa,ltitr,matlab2tikz,methods,minor,movegui,normcdf,normpdf,on,ones,parse,persistent,poissrnd,properties,repmat,solve,strcat,subs,syms,try,var,warning,xlim,ylim
	}
}

\lstdefinestyle{mercury}{
	language=Mercury
}

\lstdefinestyle{modula2}{
	language=Modula-2
}

\lstdefinestyle{objectivec}{
	language=[Objective]C,
	breakatwhitespace=false,
	keepspaces=true,
	moredirectives={
		import
	},
	morekeywords={
		@catch,@class,@dynamic,@encode,@end,@finally,@implementation,@interface,@package,@private,@property,@protected,@protocol,@public,@selector,@synchronized,@synthesize,@throw,@try,assign,BOOL,bycopy,byref,Class,copy,id,IMP,in,inout,Nil,nil,NO,nonatomic,oneway,out,readonly,readwrite,retain,SEL,self,super,YES,_cmd
	}
}

\lstdefinestyle{octave}{
	language=Octave,
	keepspaces=true,
	morecomment=[l]\%,
	morecomment=[n]{\%\{\^^M}{\%\}\^^M}
}

\lstdefinestyle{opencl}{
	language=C++,
	breakatwhitespace=false,
	emph={
		bool2,bool3,bool4,bool8,bool16,char2,char3,char4,char8,char16,complex,constant,event_t,float2,float3,float4,float8,float16,global,half2,half3,half4,half8,half16,image2d_t,image3d_t,imaginary,int2,int3,int4,int8,int16,kernel,local,long2,long3,long4,long8,long16,private,quad,quad2,quad3,quad4,quad8,quad16,sampler_t,short2,short3,short4,short8,short16,uchar2,uchar3,uchar4,uchar8,uchar16,uint2,uint3,uint4,uint8,uint16,ulong2,ulong3,ulong4,ulong8,ulong16,ushort2,ushort3,ushort4,ushort8,ushort16,__constant,__global,__kernel,__local,__private
	},
	emphstyle=\color{dkcyan}\ttfamily,
	morecomment=[l][\color{magenta}]{\#}
}

\lstdefinestyle{opensees}{
	language=tcl,
	breakatwhitespace=false,
	emph=[1]{
		-accel,-beamUniform,-dir,-dof,-ele,-eleRange,-file,-height,-increment,-initial,-iNode,-integration,-iterate,-jNode,-kNode,-mass,-mat,-matConcrete,-matShear,-matSteel,-max,-maxDim,-maxEta,-maxIter,-min,-minEta,-ndf,-ndm,-node,-nodeRange,-numSublevels,-numSubSteps,-perpDirn,-region,-rho,-sections,-thick,-time,-tol,-type,-width
	},
	emphstyle=[1]\color{black}\bfseries\em,
	keepspaces=true,
	morecomment=[l]{\#},
	morekeywords={
		algorithm,analysis,analyze,constraints,deformation,disp,eleLoad,element,equalDOF,fix,fixX,fixY,fixZmodel,geomTransf,initialize,integrator,layer,loadConst,mass,model,node,numberer,patch,pattern,printA,PySimple1Gen,reaction,recorder,region,rigidDiaphragm,section,system,test,uniaxialMaterial,wipe,wipeAnalysis
	},
	ndkeywords={
		9_4_QuadUP,20_8_BrickUP,AC3D8,Aggregator,ArcLength,ASI3D8,AV3D4,AxialSp,AxialSpHD,BandGeneral,BARSLIP,BasicBuilder,bbarBrick,bbarBrickUP,bbarQuad,bbarQuadUP,BeamColumnJoint,BeamContact2D,BeamContact3D,BeamEndContact3D,BFGS,Bilin,BilinearOilDamper,Bond_SP01,BoucWen,Brick20N,brickUP,Broyden,BWBN,Cast,CatenaryCable,CentralDifference,CFSSSWP,CFSWSWP,Concrete01,Concrete01WithSITC,Concrete02,Concrete03,Concrete04,Concrete06,Concrete07,ConcreteCM,ConcreteD,ConfinedConcrete01,constraintsTypeGravity,Corotational,corotTruss,corotTrussSection,CoupledZeroLength,DeformedShape,dispBeamColumn,dispBeamColumnInt,DisplacementControl,Dodd_Restrepo,Drift,ECC01,Elastic,elasticBeamColumn,ElasticBilin,ElasticMultiLinear,ElasticPP,ElasticPPGap,ElasticTimoshenkoBeam,ElasticTubularJoint,elastomericBearingBoucWen,elastomericBearingPlasticity,ElastomericX,Element,EnergyIncr,enhancedQuad,ENT,Explicitdifference,Fatigue,flatSliderBearing,forceBeamColumn,forceBeamColumn,FourNodeTetrahedron,FPBearingPTV,FRPConfinedConcrete,GeneralizedAlpha,Hardening,HDR,HHT,HyperbolicGapMaterial,Hysteretic,ImpactMaterial,InitStrainMaterial,InitStressMaterial,Joint2D,KikuchiAikenHDR,KikuchiAikenLRB,KikuchiBearing,KrylovNewton,Lagrange,LeadRubberX,LimitState,Linear,LoadControl,LoadControl,MinMax,MinUnbalDispNorm,mkdir,ModElasticBeam2d,ModifiedNewton,ModIMKPeakOriented,ModIMKPinching,MultiLinear,multipleShearSpring,MVLEM,Newmark,Newton,NewtonLineSearch,Node,NodeNumbers,nonlinearBeamColumn,NormDispIncr,numberer,Parallel,PathIndependentMaterial,pattern,PDelta,Pinching4,PinchingLimitStateMaterial,Plain,PyLiq1,PySimple1,quad,quadr,quadUP,QzSimple1,RambergOsgoodSteel,rayleigh,RCM,rect,ReinforcingSteel,RJWatsonEqsBearing,SAWS,SecantNewton,SelfCentering,Series,SFI_MVLEM,ShallowFoundationGen,ShellDKGQ,ShellDKGT,ShellMITC4,ShellNL,ShellNLDKGQ,ShellNLDKGT,SimpleContact2D,SimpleContact3D,singleFPBearing,SparseGeneral,SSPbrick,SSPbrickUP,SSPquad,SSPquadUP,Static,stdBrick,Steel01,Steel01,Steel02,Steel4,SteelMPF,straight,SurfaceLoad,TFP,Transient,TRBDF2,tri31,TripleFrictionPendulum,truss,trussSection,twoNodeLink,TzLiq1,TzSimple1,UniformExcitation,ViewScale,Viscous,ViscousDamper,VS3D4,YamamotoBiaxialHDR,zeroLength,zeroLengthContact,zeroLengthContactNTS2D,zeroLengthImpact3D,zeroLengthImpact3D,zeroLengthInterface2D,zeroLengthND,zeroLengthSection
	},
	ndkeywordstyle=\color{dkcyan}\ttfamily
}

\lstdefinestyle{pascal}{
	language=Pascal,
	morecomment=[l]{//},
	sensitive=false
}

\lstdefinestyle{perl}{
	language=Perl,
	alsoletter={\%},
	breakatwhitespace=false,
	keepspaces=true
}

\lstdefinestyle{php}{
	language=php,
	emph=[1]{
		php
	},
	emph=[2]{
		if,and,or,else
	},
	emph=[3]{
		abstract,as,const,else,elseif,endfor,endforeach,endif,extends,final,for,foreach,global,if,implements,private,protected,public,static,var
	},
	emphstyle=[1]\color{black},
	emphstyle=[2]\color{blue},
	keywords={
		abstract,and,array,as,break,callable,case,catch,class,clone,const,continue,declare,default,die,do,echo,else,elseif,empty,enddeclare,endfor,endforeach,endif,endswitch,endwhile,eval,exit,extends,final,finally,for,foreach,function,global,goto,if,implements,include,include_once,instanceof,insteadof,interface,isset,list,namespace,new,or,print,private,protected,public,require,require_once,return,static,switch,throw,trait,try,unset,use,var,while,xor,yield,__halt_compiler
	},
	showlines=true,
	upquote=true
}

\lstdefinestyle{plaintext}{
	language={},
	keepspaces=true,
	postbreak={},
	tabsize=4
}

\lstdefinestyle{postscript}{
	language=PostScript,
	keepspaces=true
}

\lstdefinestyle{powershell}{
	alsodigit={-},
	morecomment=[l]{\#},
	morecomment=[n]{<\#}{\#>},
	morekeywords={
		Add-Content,Add-PSSnapin,Clear-Content,Clear-History,Clear-Host,Clear-Item,Clear-ItemProperty,Clear-Variable,Compare-Object,Connect-PSSession,Convert-Path,ConvertFrom-String,Copy-Item,Copy-ItemProperty,Disable-PSBreakpoint,Disconnect-PSSession,Enable-PSBreakpoint,Enter-PSSession,Exit-PSSession,Export-Alias,Export-Csv,Export-PSSession,ForEach-Object,Format-Custom,Format-Hex,Format-List,Format-Table,Format-Wide,Get-Alias,Get-ChildItem,Get-Clipboard,Get-Command,Get-ComputerInfo,Get-Content,Get-History,Get-Item,Get-ItemProperty,Get-ItemPropertyValue,Get-Job,Get-Location,Get-Member,Get-Module,Get-Process,Get-PSBreakpoint,Get-PSCallStack,Get-PSDrive,Get-PSSession,Get-PSSnapin,Get-Service,Get-TimeZone,Get-Unique,Get-Variable,Get-WmiObject,Group-Object,help,Import-Alias,Import-Csv,Import-Module,Import-PSSession,Invoke-Command,Invoke-Expression,Invoke-History,Invoke-Item,Invoke-RestMethod,Invoke-WebRequest,Invoke-WmiMethod,Measure-Object,mkdir,Move-Item,Move-ItemProperty,New-Alias,New-Item,New-Module,New-PSDrive,New-PSSession,New-PSSessionConfigurationFile,New-Variable,Out-GridView,Out-Host,Out-Printer,Pop-Location,powershell_ise.exe,Push-Location,Receive-Job,Receive-PSSession,Remove-Item,Remove-ItemProperty,Remove-Job,Remove-Module,Remove-PSBreakpoint,Remove-PSDrive,Remove-PSSession,Remove-PSSnapin,Remove-Variable,Remove-WmiObject,Rename-Item,Rename-ItemProperty,Resolve-Path,Resume-Job,Select-Object,Select-String,Set-Alias,Set-Clipboard,Set-Content,Set-Item,Set-ItemProperty,Set-Location,Set-PSBreakpoint,Set-TimeZone,Set-Variable,Set-WmiInstance,Show-Command,Sort-Object,Start-Job,Start-Process,Start-Service,Start-Sleep,Stop-Job,Stop-Process,Stop-Service,Suspend-Job,Tee-Object,Trace-Command,Wait-Job,Where-Object,Write-Output
	},
	morekeywords={
		Do,Else,For,ForEach,Function,If,In,Until,While
	},
	morestring=[b]{"},
	morestring=[b]{'},
	morestring=[s]{@'}{'@},
	morestring=[s]{@"}{"@},
	sensitive=false
}

\lstdefinestyle{prolog}{
	language=Prolog
}

\lstdefinestyle{promela}{
	language=Promela
}

\lstdefinestyle{pseudocode}{
	language={},
	backgroundcolor=\color{white},
	breakatwhitespace=false,
	commentstyle=\color{gray}\upshape,
	frame=tb,
	keepspaces=true,
	keywords={
		and,be,begin,break,datatype,do,elif,else,end,for,foreach,fun,function,if,in,input,let,not,null,or,output,pop,procedure,push,repeat,return,swap,until,while,xor
	},
	keywordstyle=\color{black}\bfseries,
	mathescape=true,
	morecomment=[l]{//},
	morecomment=[l]{\#},
	morecomment=[s]{/*}{*/},
	morecomment=[s]{/**}{*/},
	sensitive=false,
	stringstyle=\color{dkgray}\bfseries\em
}

\lstdefinelanguage{pythonEXTENDED}{
	language=Python,
	breakatwhitespace=false,
	emph={
		AbstractSet,Any,AsyncContextManager,AsyncGenerator,AsyncIterable,AsyncIterator,Awaitable,AwaitableGenerator,BinaryIO,ByteString,Callable,Collection,Container,ContextManager,Coroutine,Dict,False,ForwardRef,Generator,GenericMeta,Hashable,IO,ItemsView,Iterable,Iterator,KeysView,List,Mapping,MappingView,Match,Meta,MutableMapping,MutableSequence,MutableSet,NamedTuple,None,Pattern,Reversible,Sequence,Sized,SupportInts,SupportsAbs,SupportsBytes,SupportsComplex,SupportsFloat,SupportsIndex,SupportsRound,TextIO,True,Tuple,TypeAlias,TYPE_CHECKING,Union,ValuesView,__add__,__and__,__eq__,__floordiv__,__ge__,__gt__,__init__,__le__,__lt__,__main__,__mod__,__mul__,__name__,__ne__,__or__,__pow__,__repr__,__str__,__sub__,__truediv__,__xor__
	},
	emphstyle=\color{dkcyan}\ttfamily,
	keepspaces=true,
	morecomment=[s][\color{BurntOrange}]{@}{\ },
	morekeywords={
		as,assert,close,listdir,self,sorted,split,strip,with
	}
}
\lstdefinestyle{python}{
	language=pythonEXTENDED
}

\lstdefinestyle{qsharp}{
	mathescape=true,
	morecomment=[l]{//},
	morecomment=[l][\color{dkgreen}]{///},
	morekeywords={
		Adj,Adjoint,adjoint,and,apply,as,auto,BigInt,body,Bool,borrowing,Controlled,controlled,Ctl,distribute,Double,elif,else,fail,false,fixup,for,function,if,in,Int,intrinsic,invert,is,let,mutable,namespace,new,newtype,not,One,open,operation,or,Pauli,PauliI,PauliX,PauliY,PauliZ,Qubit,Range,repeat,Result,return,self,set,String,true,Unit,until,using,while,within,Zero
	},
	morekeywords=[2]{
		Assert,AssertProb,CCNOT,CNOT,Exp,ExpFrac,H,I,M,Measure,Message,R,R1,R1Frac,Random,Reset,ResetAll,RFrac,Rx,Ry,Rz,S,SWAP,T,X,Y,Z
	},
	sensitive=true
}

\lstdefinestyle{r}{
	language=R,
	alsoletter={.<-},
	alsoother={._$},
	deletekeywords={
		df,data,frame,length,as,character
	},
	morecomment=[l]\#,
	morestring=[d]',
	morestring=[d]",
	otherkeywords={
		!,!=,~,$,*,\&,\%/\%,\%*\%,\%\%,<-,<<-,/
	}
}

\lstdefinestyle{racket}{
	alsoletter={',`,-,/,>,<,\#,\%},
	morekeywords=[1]{
		define,define-macro,define-stream,define-syntax,lambda,stream-lambda
	},
	morekeywords=[2]{
		->,always_publish,and,\#',\#\%module-begin,\#lang,\#`,begin,begin-for-syntax,Boolean,call-with-current-continuation,call-with-input-file,call-with-output-file,callback,call/cc,case,cond,define-context,define-controller,define-struct/contract,define/contract,delay,do,else,environment,eval,fold,for,for-each,force,get,if,implement,in-range,Integer,label,let,let*,let*-values,let-syntax,let-values,letrec,letrec-syntax,map,maybe_publish,message-box,module,new,not,or,or/c,parent,provide,quasiquote,query,quote,rename-out,require,send,submod,syntax,syntax-case,syntax-rules,unquote,unquote-splicing,when,when-provided,when-required,with-syntax
	},
	morekeywords=[3]{
		export,import
	},
	morecomment=[l]{;},
	moredelim=**[is][\color{lgray}]{<<@<<}{>>@>>},
	moredelim=**[is][\itshape\color{mauve}]{<<;<<}{>>;>>},
	morecomment=[s]{\#|}{|\#},
	morestring=[s]{"}{"},
	sensitive=true
}

\lstdefinestyle{reil}{
	comment=[l]{;},
	keywords=[1]{
		ADD,add,and,AND,BISZ,bisz,bsh,BSH,div,DIV,jcc,JCC,LDM,ldm,MOD,mod,mul,MUL,nop,NOP,or,OR,stm,STM,STR,str,sub,SUB,undef,UNDEF,unkn,UNKN,XOR,xor
	},
	keywords=[3]{
		ah,AH,al,AL,AX,ax,bh,BH,BL,bl,bp,BP,bpl,BPL,BX,bx,ch,CH,cl,CL,cx,CX,DH,dh,di,DI,dil,DIL,dl,DL,DX,dx,EAX,eax,EBP,ebp,ebx,EBX,ECX,ecx,EDI,edi,edx,EDX,esi,ESI,esp,ESP,r8,R8,r8b,R8B,r8d,R8D,r8w,R8W,r9,R9,R9B,r9b,R9D,r9d,r9w,R9W,r10,R10,R10B,r10b,R10D,r10d,r10w,R10W,r11,R11,r11b,R11B,r11d,R11D,R11W,r11w,R12,r12,R12B,r12b,r12d,R12D,r12w,R12W,R13,r13,r13b,R13B,R13D,r13d,R13W,r13w,r14,R14,R14B,r14b,r14d,R14D,R14W,r14w,r15,R15,r15b,R15B,r15d,R15D,R15W,r15w,RAX,rax,rbp,RBP,rbx,RBX,RCX,rcx,RDI,rdi,rdx,RDX,RSI,rsi,RSP,rsp,SI,si,SIL,sil,SP,sp,spl,SPL
	},
	sensitive=true
}

\lstdefinestyle{ruby}{
	language=Ruby,
	breakatwhitespace=true,
	morestring=[s][]{\#\{}{\}},
	morestring=*[d]{"},
	sensitive=true
}

\lstdefinelanguage{Rust}{
	sensitive,
	alsodigit={},
	alsoletter={!},
	alsoother={},
	morecomment=[l]{//},
	morecomment=[s]{/*}{*/},
	moredelim=[s][{\itshape\color[rgb]{0,0,0.75}}]{\#[}{]},
	morekeywords=[2]{ 
		Add,AddAssign,Any,AsciiExt,AsInner,AsInnerMut,AsMut,AsRawFd,AsRawHandle,AsRawSocket,AsRef,Binary,BitAnd,BitAndAssign,Bitor,BitOr,BitOrAssign,BitXor,BitXorAssign,Borrow,BorrowMut,Boxed,BoxPlace,BufRead,BuildHasher,CastInto,CharExt,Clone,CoerceUnsized,CommandExt,Copy,Debug,DecodableFloat,Default,Deref,DerefMut,DirBuilderExt,DirEntryExt,Display,Div,DivAssign,DoubleEndedIterator,DoubleEndedSearcher,Drop,EnvKey,Eq,Error,ExactSizeIterator,ExitStatusExt,Extend,FileExt,FileTypeExt,Float,Fn,FnBox,FnMut,FnOnce,Freeze,From,FromInner,FromIterator,FromRawFd,FromRawHandle,FromRawSocket,FromStr,FullOps,FusedIterator,Generator,Hash,Hasher,Index,IndexMut,InPlace,Int,Into,IntoCow,IntoInner,IntoIterator,IntoRawFd,IntoRawHandle,IntoRawSocket,IsMinusOne,IsZero,Iterator,JoinHandleExt,LargeInt,LowerExp,LowerHex,MetadataExt,Mul,MulAssign,Neg,Not,Octal,OpenOptionsExt,Ord,OsStrExt,OsStringExt,Packet,PartialEq,PartialOrd,Pattern,PermissionsExt,Place,Placer,Pointer,Product,Put,RangeArgument,RawFloat,Read,Rem,RemAssign,Seek,Shl,ShlAssign,Shr,ShrAssign,Sized,SliceConcatExt,SliceExt,SliceIndex,Stats,Step,StrExt,Sub,SubAssign,Sum,Sync,TDynBenchFn,Terminal,Termination,ToOwned,ToSocketAddrs,ToString,Try,TryFrom,TryInto,UnicodeStr,Unsize,UpperExp,UpperHex,WideInt,Write
	},
	morekeywords=[2]{
		Send
	},
	morekeywords=[3]{ 
		bool,char,f32,f64,i8,i16,i32,i64,isize,str,u8,u16,u32,u64,unit,usize,i128,u128
	},
	morekeywords=[4]{ 
		Err,false,None,Ok,Some,true
	},
	morekeywords=[5]{ 
		assert!,assert_eq!,assert_ne!,cfg!,column!,compile_error!,concat!,concat_idents!,debug_assert!,debug_assert_eq!,debug_assert_ne!,env!,eprint!,eprintln!,file!,format!,format_args!,include!,include_bytes!,include_str!,line!,module_path!,option_env!,panic!,print!,println!,select!,stringify!,thread_local!,try!,unimplemented!,unreachable!,vec!,write!,writeln!
	},
	morekeywords={ 
		abstract,alignof,become,box,do,final,macro,offsetof,override,priv, proc,pure,sizeof,typeof,unsized,virtual,yield
	},
	morekeywords={
		as,const,let,move,mut,ref,static
	},
	morekeywords={
		break,continue,else,for,if,in,loop,match,return,while
	},
	morekeywords={
		crate,extern,mod,pub,super
	},
	morekeywords={
		dyn,enum,fn,impl,Self,self,struct,trait,type,union,use,where
	},
	morekeywords={
		unsafe
	},
	morestring=[b]{"}
}
\lstdefinestyle{rust}{
	language=Rust,
	keywordstyle=[2]\color[rgb]{0.75,0,0}, 
	keywordstyle=[3]\color[rgb]{0,0.5,0}, 
	keywordstyle=[4]\color[rgb]{0,0.5,0}, 
	keywordstyle=[5]\color[rgb]{0,0,0.75} 
}

\lstdefinestyle{scala}{
	language=scala,
	breakatwhitespace=true,
	morecomment=[l]{//},
	morecomment=[n]{/*}{*/},
	morekeywords={
		abstract,case,catch,class,def,do,else,extends,false,final,finally,for,if,implicit,import,match,mixin,new,null,object,override,package,private,protected,requires,return,sealed,super,this,throw,trait,true,try,type,val,var,while,with,yield
	},
	morestring=[b]',
	morestring=[b]",
	morestring=[b]""",
	otherkeywords={
		=>,<-,<\%,<:,>:,\#,@
	}
}

\lstdefinestyle{scheme}{
	language=Lisp,
	morecomment=[l]{;},
	morekeywords={
		and,begin,case,case-lambda,cond,cond-expand,define,delay,delay-force,do,else,force,guard,if,lambda,let,let*,let*-values,let-syntax,let-values,letrec,letrec*,letrec-syntax,make-parameter,make-promise,map,or,parameterize,promise?,quasiquote,quote,set!,syntax-rules,unless,when
	},
	morestring=[b]"
}

\lstdefinestyle{scilab}{
	language=Scilab
}

\lstdefinestyle{simula}{
	language=Simula
}

\lstdefinestyle{sparql}{
	language=SPARQL
}

\lstdefinestyle{sql}{
	language=SQL,
	breakatwhitespace=true
}

\lstdefinestyle{swift}{
	language=Swift
}

\lstdefinestyle{tcl}{
	language=tcl,
	breakatwhitespace=false,
	keepspaces=true,
	morecomment=[l]{\#}
}

\lstdefinestyle{vbscript}{
	language=[Visual]Basic,
	extendedchars=true
}

\lstdefinestyle{verilog}{
	language=Verilog
}

\lstdefinelanguage{VHDL}{
	morekeywords=[1]{
		ALL,all,and,architecture,begin,downto,end,entity,in,is,library,Not,of,or,out,port,use
	},
	morekeywords=[2]{
		IEEE,NUMERIC_STD,STD_LOGIC,std_logic,STD_LOGIC_1164,STD_LOGIC_ARITH,STD_LOGIC_UNSIGNED,STD_LOGIC_VECTOR,std_logic_vector
	},
	morecomment=[l]--
}
\lstdefinestyle{vhdl}{
	language=VHDL
}

\lstdefinelanguage{XML}{
	morecomment=[s]{<?}{?>},
	morekeywords={
		encoding,type,version,xmlns
	},
	morestring=[b]",
	morestring=[s]{>}{<}
}
\lstdefinestyle{xml}{
	language=XML,
	tabsize=2
}

\newcommand{\checkvalidsourcecodestyle}[1]{%
	\ifthenelse{\equal{#1}{abap}}{}{%
	\ifthenelse{\equal{#1}{ada}}{}{%
	\ifthenelse{\equal{#1}{assemblerx64}}{}{%
	\ifthenelse{\equal{#1}{assemblerx86}}{}{%
	\ifthenelse{\equal{#1}{awk}}{}{%
	\ifthenelse{\equal{#1}{bash}}{}{%
	\ifthenelse{\equal{#1}{basic}}{}{%
	\ifthenelse{\equal{#1}{c}}{}{%
	\ifthenelse{\equal{#1}{caml}}{}{%
	\ifthenelse{\equal{#1}{cmake}}{}{%
	\ifthenelse{\equal{#1}{cobol}}{}{%
	\ifthenelse{\equal{#1}{cpp}}{}{%
	\ifthenelse{\equal{#1}{csharp}}{}{%
	\ifthenelse{\equal{#1}{css}}{}{%
	\ifthenelse{\equal{#1}{csv}}{}{%
	\ifthenelse{\equal{#1}{cuda}}{}{%
	\ifthenelse{\equal{#1}{dart}}{}{%
	\ifthenelse{\equal{#1}{docker}}{}{%
	\ifthenelse{\equal{#1}{elisp}}{}{%
	\ifthenelse{\equal{#1}{elixir}}{}{%
	\ifthenelse{\equal{#1}{erlang}}{}{%
	\ifthenelse{\equal{#1}{fortran}}{}{%
	\ifthenelse{\equal{#1}{fsharp}}{}{%
	\ifthenelse{\equal{#1}{glsl}}{}{%
	\ifthenelse{\equal{#1}{gnuplot}}{}{%
	\ifthenelse{\equal{#1}{go}}{}{%
	\ifthenelse{\equal{#1}{haskell}}{}{%
	\ifthenelse{\equal{#1}{html}}{}{%
	\ifthenelse{\equal{#1}{ini}}{}{%
	\ifthenelse{\equal{#1}{java}}{}{%
	\ifthenelse{\equal{#1}{javascript}}{}{%
	\ifthenelse{\equal{#1}{json}}{}{%
	\ifthenelse{\equal{#1}{julia}}{}{%
	\ifthenelse{\equal{#1}{kotlin}}{}{%
	\ifthenelse{\equal{#1}{latex}}{}{%
	\ifthenelse{\equal{#1}{lisp}}{}{%
	\ifthenelse{\equal{#1}{llvm}}{}{%
	\ifthenelse{\equal{#1}{lua}}{}{%
	\ifthenelse{\equal{#1}{make}}{}{%
	\ifthenelse{\equal{#1}{maple}}{}{%
	\ifthenelse{\equal{#1}{mathematica}}{}{%
	\ifthenelse{\equal{#1}{matlab}}{}{%
	\ifthenelse{\equal{#1}{mercury}}{}{%
	\ifthenelse{\equal{#1}{modula2}}{}{%
	\ifthenelse{\equal{#1}{objectivec}}{}{%
	\ifthenelse{\equal{#1}{octave}}{}{%
	\ifthenelse{\equal{#1}{opencl}}{}{%
	\ifthenelse{\equal{#1}{opensees}}{}{%
	\ifthenelse{\equal{#1}{pascal}}{}{%
	\ifthenelse{\equal{#1}{perl}}{}{%
	\ifthenelse{\equal{#1}{php}}{}{%
	\ifthenelse{\equal{#1}{plaintext}}{}{%
	\ifthenelse{\equal{#1}{postscript}}{}{%
	\ifthenelse{\equal{#1}{powershell}}{}{%
	\ifthenelse{\equal{#1}{prolog}}{}{%
	\ifthenelse{\equal{#1}{promela}}{}{%
	\ifthenelse{\equal{#1}{pseudocode}}{}{%
	\ifthenelse{\equal{#1}{python}}{}{%
	\ifthenelse{\equal{#1}{qsharp}}{}{%
	\ifthenelse{\equal{#1}{r}}{}{%
	\ifthenelse{\equal{#1}{racket}}{}{%
	\ifthenelse{\equal{#1}{reil}}{}{%
	\ifthenelse{\equal{#1}{ruby}}{}{%
	\ifthenelse{\equal{#1}{rust}}{}{%
	\ifthenelse{\equal{#1}{scala}}{}{%
	\ifthenelse{\equal{#1}{scheme}}{}{%
	\ifthenelse{\equal{#1}{scilab}}{}{%
	\ifthenelse{\equal{#1}{simula}}{}{%
	\ifthenelse{\equal{#1}{sparql}}{}{%
	\ifthenelse{\equal{#1}{sql}}{}{%
	\ifthenelse{\equal{#1}{swift}}{}{%
	\ifthenelse{\equal{#1}{tcl}}{}{%
	\ifthenelse{\equal{#1}{vbscript}}{}{%
	\ifthenelse{\equal{#1}{verilog}}{}{%
	\ifthenelse{\equal{#1}{vhdl}}{}{%
	\ifthenelse{\equal{#1}{xml}}{}{%
		\errmessage{LaTeX Warning: Estilo de codigo desconocido. Valores esperados: abap,ada,assemblerx64,assemblerx86,awk,bash,basic,c,caml,cmake,cobol,cpp,csharp,css,csv,cuda,dart,docker,elisp,elixir,erlang,fortran,fsharp,glsl,gnuplot,go,haskell,html,ini,java,javascript,json,julia,kotlin,latex,lisp,llvm,lua,make,maple,mathematica,matlab,mercury,modula2,objectivec,octave,opencl,opensees,pascal,perl,php,plaintext,postscript,powershell,prolog,promela,pseudocode,python,qsharp,r,racket,reil,ruby,rust,scala,scheme,scilab,simula,sparql,sql,swift,tcl,vbscript,verilog,vhdl,xml}%
		\stop%
	}}}}}}}}}}}}}}}}}}}}}}}}}}}}}}}}}}}}}}}}}}}}}}}}}}}}}}}}}}}}}}}}}}}}}}}}}}}}%
}

\newcommand{\inlinesourcecodeboxed}[3][]{%
	\emptyvarerr{\inlinesourcecodeboxed}{#2}{Estilo de codigo no definido}%
	\emptyvarerr{\inlinesourcecodeboxed}{#3}{Codigo no definido}%
	\lstset{%
		basicstyle={\sourcecodeilfonts\sourcecodeilfontf\color{\maintextcolor}}%
	}
	\checkvalidsourcecodestyle{#2}%
	\ifthenelse{\equal{#1}{}}{%
		\Colorbox{\sourcecodebgcolor}{\lstinline[style=#2]!#3!}%
	}{%
	\ifthenelse{\equal{#1}{NOCOLOR}}{%
		\lstinline[style=#2]!#3!%
	}{%
		\Colorbox{#1}{\lstinline[style=#2]!#3!}%
	}}%
	\lstset{%
		basicstyle={\sourcecodefonts\sourcecodefontf\color{\maintextcolor}}%
	}
}

\RequirePackage{enumitem}
\makeatletter
\def\greek#1{\expandafter\@greek\csname c@#1\endcsname}
\def\Greek#1{\expandafter\@Greek\csname c@#1\endcsname}
\def\@greek#1{%
	\ifcase#1%
		\or $\alpha$%
		\or $\beta$%
		\or $\gamma$%
		\or $\delta$%
		\or $\epsilon$%
		\or $\zeta$%
		\or $\eta$%
		\or $\theta$%
		\or $\iota$%
		\or $\kappa$%
		\or $\lambda$%
		\or $\mu$%
		\or $\nu$%
		\or $\xi$%
		\or $o$%
		\or $\pi$%
		\or $\rho$%
		\or $\sigma$%
		\or $\tau$%
		\or $\upsilon$%
		\or $\phi$%
		\or $\chi$%
		\or $\psi$%
		\or $\omega$%
	\fi%
}
\def\@Greek#1{%
	\ifcase#1%
		\or $\mathrm{A}$%
		\or $\mathrm{B}$%
		\or $\Gamma$%
		\or $\Delta$%
		\or $\mathrm{E}$%
		\or $\mathrm{Z}$%
		\or $\mathrm{H}$%
		\or $\Theta$%
		\or $\mathrm{I}$%
		\or $\mathrm{K}$%
		\or $\Lambda$%
		\or $\mathrm{M}$%
		\or $\mathrm{N}$%
		\or $\Xi$%
		\or $\mathrm{O}$%
		\or $\Pi$%
		\or $\mathrm{P}$%
		\or $\Sigma$%
		\or $\mathrm{T}$%
		\or $\mathrm{Y}$%
		\or $\Phi$%
		\or $\mathrm{X}$%
		\or $\Psi$%
		\or $\Omega$%
	\fi%
}
\makeatother
\AddEnumerateCounter{\greek}{\@greek}{24}
\AddEnumerateCounter{\Greek}{\@Greek}{12}

\checkvardefined{\documenttitle}

\makeatletter
	\g@addto@macro\documenttitle\xspace
\makeatother

\ifthenelse{\equal{\documentsubtitle}{}}{
	\def\documenttitlehf {\documenttitle}
}{
	\makeatletter
		\g@addto@macro\documentsubtitle\xspace
	\makeatother
	\def\documenttitlehf {\documentsubtitle}
}

\ifthenelse{\equal{\cfgpdfsecnumbookmarks}{true}}{
	\bookmarksetup{numbered}}{
}

\ifthenelse{\equal{\cfgshowbookmarkmenu}{true}}{
	\def\cfgpdfpagemode {UseOutlines}
	}{
	\def\cfgpdfpagemode {UseNone}
}
\ifthenelse{\equal{\usepdfmetadata}{true}}{
	\def\pdfmetainfotitle {\documenttitle}
}{
	\def\pdfmetainfotitle {}
}
\hypersetup{
	keeppdfinfo,
	bookmarksopen={\cfgpdfbookmarkopen},
	bookmarksopenlevel={\cfgbookmarksopenlevel},
	bookmarkstype={toc},
	pdfcenterwindow={\cfgpdfcenterwindow},
	pdfcopyright={\cfgpdfcopyright},
	pdfcreator={LaTeX},
	pdfdisplaydoctitle={\cfgpdfdisplaydoctitle},
	pdfencoding={unicode},
	pdffitwindow={\cfgpdffitwindow},
	pdfinfo={
		Document.Title={\pdfmetainfotitle},
		Template.Author.Alias={ppizarror},
		Template.Author.Email={pablo@ppizarror.com},
		Template.Author.Web={https://ppizarror.com},
		Template.Author={Pablo Pizarro R.},
		Template.Date={02/10/2021},
		Template.Encoding={UTF-8},
		Template.Latex.Compiler={pdflatex},
		Template.License.Type={MIT},
		Template.License.Web={https://opensource.org/licenses/MIT},
		Template.Name={Template-Articulo},
		Template.Type={Normal},
		Template.Version.Dev={1.0.0-5-ARTC},
		Template.Version.Hash={2A6CDEF6F6D329FC88B6AAB372C6CCA4},
		Template.Version.Release={1.0.0},
		Template.Web.Dev={https://github.com/Template-Latex/Template-Articulo},
		Template.Web.Manual={https://latex.ppizarror.com/articulo},
	},
	pdfkeywords={\cfgpdfkeywords},
	pdfmenubar={\cfgpdfmenubar},
	pdfpagelayout={\cfgpdflayout},
	pdfpagemode={\cfgpdfpagemode},
	pdfproducer={Template-Articulo v1.0.0 | (Pablo Pizarro R.) ppizarror.com},
	pdfremotestartview={Fit},
	pdfstartpage={1},
	pdfstartview={\cfgpdfpageview},
	pdftitle={\pdfmetainfotitle},
	pdftoolbar={\cfgpdftoolbar}
}

\graphicspath{{./\defaultimagefolder}}

\makeatletter
\ifthenelse{\equal{\fpremovetopbottomcenter}{true}}{
	\setlength{\@fptop}{0pt}
	\setlength{\@fpbot}{0pt}
}{}
\makeatother

\ifthenelse{\equal{\showlinenumbers}{true}}{
	\setlength{\linenumbersep}{0.50cm}
	
	}{
}

\color{\maintextcolor} 
\arrayrulecolor{\tablelinecolor} 
\sethlcolor{\highlightcolor} 
\ifthenelse{\equal{\showborderonlinks}{true}}{
	\hypersetup{
		citebordercolor=\numcitecolor,
		linkbordercolor=\linkcolor,
		urlbordercolor=\urlcolor
	}
}{
	\hypersetup{ 
		hidelinks,
		colorlinks=true,
		citecolor=\numcitecolor,
		linkcolor=\linkcolor,
		urlcolor=\urlcolor
	}
}
\ifthenelse{\equal{\pagescolor}{white}}{}{
	\pagecolor{\pagescolor}
}

\setcaptionmargincm{\captionlrmargin}
\ifthenelse{\equal{\captiontextbold}{true}}{ 
	\renewcommand{\captiontextbold}{bf}}{
	\renewcommand{\captiontextbold}{}
}
\ifthenelse{\equal{\captiontextsubnumbold}{true}}{ 
	\renewcommand{\captiontextsubnumbold}{bf}}{
	\renewcommand{\captiontextsubnumbold}{}
}

\corecheckfontsize{\captionfontsize}
\corecheckfontsize{\subcaptionfsize}
\makeatletter
\renewcommand\p@subfigure{\thefigure\captionsubchar}
\renewcommand\p@subtable{\thetable\captionsubchar}
\makeatother

\ifthenelse{\equal{\figurecaptiontop}{true}}{
	\floatsetup[figure]{position=above}}{
}

\ifthenelse{\equal{\tablecaptiontop}{true}}{
	\floatsetup[table]{position=top}
	}{
	\floatsetup[table]{position=bottom}
}

\ifthenelse{\equal{\captionalignment}{justified}}{ 
	\captionsetup{
		format=plain,
		justification=justified
	}
}{
\ifthenelse{\equal{\captionalignment}{centered}}{ 
	\captionsetup{
		justification=centering
	}
}{
\ifthenelse{\equal{\captionalignment}{left}}{ 
	\captionsetup{
		justification=raggedright,
		singlelinecheck=false
	}
}{
\ifthenelse{\equal{\captionalignment}{right}}{ 
	\captionsetup{
		justification=raggedleft,
		singlelinecheck=false
	}
}{
	\throwbadconfig{Posicion de leyendas desconocida}{\captionalignment}{justified,centered,left,right}}}}
}

\ifthenelse{\equal{\stylecitereferences}{natbib}}{
	\def\twocolumnreferencesmargin{-0.35cm}
	\bibliographystyle{\natbibrefstyle}
	\setlength{\bibsep}{\natbibrefsep pt}
	\newcommand{\shortcite}[1]{\citep{#1}}
	\newcommand{\fullcite}[1]{\citet{#1}}
	\setcitestyle{open={\natbibrefcitecharopen},close={\natbibrefcitecharclose}}
	\ifthenelse{\equal{\natbibrefcitesepcomma}{true}}{
		\setcitestyle{comma}
	}{
		\setcitestyle{semicolon}
	}
	\ifthenelse{\equal{\natbibrefcitetype}{numbers}}{
		\setcitestyle{numbers}
	}{
	\ifthenelse{\equal{\natbibrefcitetype}{authoryear}}{
		\setcitestyle{authoryear}
	}{
	\ifthenelse{\equal{\natbibrefcitetype}{super}}{
		\setcitestyle{super}
	}{
		\throwbadconfig{Tipo cita natbib desconocido}{\natbibrefcitetype}{numbers,authoryear,super}}}
	}
}{
\ifthenelse{\equal{\stylecitereferences}{apacite}}{
	\def\twocolumnreferencesmargin{-0.39cm}
	\bibliographystyle{\apacitestyle}
	\setlength{\bibitemsep}{\apaciterefsep pt}
	\newcommand{\citep}[1]{\fullcite{#1}}
	\newcommand{\citet}[1]{\shortcite{#1}}
}{
\ifthenelse{\equal{\stylecitereferences}{bibtex}}{
	\def\twocolumnreferencesmargin{-0.35cm}
	\bibliographystyle{\bibtexstyle}
	\newlength{\bibitemsep}
	\setlength{\bibitemsep}{.2\baselineskip plus .05\baselineskip minus .05\baselineskip}
	\newlength{\bibparskip}\setlength{\bibparskip}{0pt}
	\let\oldthebibliography\thebibliography
	\renewcommand\thebibliography[1]{
		\oldthebibliography{#1}
		\setlength{\parskip}{\bibitemsep}
		\setlength{\itemsep}{\bibparskip}
	}
	\setlength{\bibitemsep}{\bibtexrefsep pt}
}{
\ifthenelse{\equal{\stylecitereferences}{custom}}{
	\coretemplatemessage{Usando estilo citas referencias custom, importar librerias y configuraciones posterior al llamado de template.tex en archivo principal}
}{
	\throwbadconfig{Estilo citas desconocido}{\stylecitereferences}{bibtex,apacite,natbib,custom}}}}
}

\makeatletter
\ifthenelse{\equal{\stylecitereferences}{apacite}}{
	\ifthenelse{\equal{\apaciterefnumber}{true}}{
		\newcounter{apaciteNumberCounter}
		\renewcommand{\theapaciteNumberCounter}{
			\apaciterefcitecharopen\arabic{apaciteNumberCounter}\apaciterefcitecharclose
		}
		\patchcmd{\@lbibitem}{\item[}{\item[\stepcounter{apaciteNumberCounter}{\hss\llap{\theapaciteNumberCounter}\quad}}{}{}
		\setlength{\bibleftmargin}{2.54em}
		\setlength{\bibindent}{-0.54em}
	}{}
}{}
\makeatother

\ifthenelse{\equal{\stylecitereferences}{apacite}}{
	\ifthenelse{\equal{\apaciteshowurl}{false}}{
		
		\AtBeginEnvironment{APACrefURL}{\renewcommand{\url}[1]{}}
		\renewcommand{\doiprefix}{doi:~\kern-1pt}
	}{}
}{}

\makeatletter
\ifthenelse{\equal{\twocolumnreferences}{true}}{
	\renewenvironment{thebibliography}[1]
	{\begin{multicols}{2}[\section*{\refname}\vspace{\twocolumnreferencesmargin}]
		\@mkboth{\MakeUppercase\refname}{\MakeUppercase\refname}
		\list{\@biblabel{\@arabic\c@enumiv}}
		{\settowidth\labelwidth{\@biblabel{#1}}
			\leftmargin\labelwidth
			\advance\leftmargin\labelsep
			\@openbib@code
			\usecounter{enumiv}
			\let\p@enumiv\@empty
			\renewcommand\theenumiv{\@arabic\c@enumiv}}
		\sloppy
		\clubpenalty 4000
		\@clubpenalty \clubpenalty
		\widowpenalty 4000
		\sfcode`\.\@m}
		{\def\@noitemerr
		{\@latex@warning{Ambiente `thebibliography' no definido}}
		\endlist\end{multicols}}}{}
\makeatother

\patchcmd{\appendices}{\quad}{\charappendixsection\spacingaftersection}{}{}

\begingroup
	\makeatletter
	\let\newcounter\@gobble\let\setcounter\@gobbletwo
	\globaldefs\@ne\let\c@loldepth\@ne
	\newlistof{listings}{lol}{\lstlistlistingname}
	\newlistentry{lstlisting}{lol}{0}
	\makeatother
\endgroup

\newcommand{\listindexequationsname}{\namelteqn}
\newlistof{myindexequations}{equ}{\listindexequationsname}
\newcommand{\myindexequations}[1]{
	\addcontentsline{equ}{myindexequations}{\protect\numberline{\theequation}#1}
}
\DeclareTotalCounter{templateIndexEquations}

\makeatletter
	\def\ifGm@preamble#1{\@firstofone}
	\appto\restoregeometry{
		\pdfpagewidth=\paperwidth
		\pdfpageheight=\paperheight}
	\apptocmd\newgeometry{
		\pdfpagewidth=\paperwidth
		\pdfpageheight=\paperheight}{}{}
\makeatother

\hbadness=\maxdimen
\vbadness=\maxdimen

\makeatletter
\def\Hv@scale {.95}
\makeatother

\makeatletter 
\preto\tabular{\global\rownum=\z@}
\preto\tabularx{\global\rownum=\z@}
\makeatother

\strictpagecheck

\ttfamily \hyphenchar\the\font=`\-

\ifthenelse{\equal{\compilertype}{pdf2latex}}{
	\pdfcompresslevel=\pdfcompilecompression
	
	\pdfdecimaldigits=2
	
	\pdfinclusionerrorlevel=0
	
	\pdfminorversion=\pdfcompileversion
	
	\pdfobjcompresslevel=\pdfcompileobjcompression
}{
\ifthenelse{\equal{\compilertype}{xelatex}}{
}{
\ifthenelse{\equal{\compilertype}{lualatex}}{
}{
	\throwbadconfig{Compilador desconocido}{\compilertype}{pdf2latex,xelatex,lualatex}}}
}

\newcounter{subsubsubsection}[subsubsection]

\makeatletter
	\def\toclevel@subsubsubsection {4}
	\def\toclevel@paragraph {5}
	\def\toclevel@subparagraph {6}
	\ifthenelse{\equal{\charaftersectionnum}{}}{ 
		\def\l@subsubsubsection {\@dottedtocline{4}{6.97em}{4em}}
		\def\l@paragraph {\@dottedtocline{5}{10.97em}{5em}}
		\def\l@subparagraph {\@dottedtocline{6}{14em}{6em}}
	}{ 
		\def\l@subsubsubsection {\@dottedtocline{4}{7.83em}{4.15em}}
		\def\l@paragraph {\@dottedtocline{5}{11.98em}{4.92em}}
		\def\l@subparagraph {\@dottedtocline{6}{14.65em}{5.69em}}
	}
\makeatother

\makeatletter
\newcommand\sectionpunct[2]{%
	\expandafter\def\csname @seccntfmt@#1\endcsname##1{%
		\csname the##1\endcsname#2%
	}%
}
\def\@seccntformat#1{\@ifundefined{#1@cntformat}%
	{\csname the#1\endcsname} 
	{\csname #1@cntformat\endcsname} 
}
\newcommand\section@cntformat{\thesection\charaftersectionnum\spacingaftersection}
\newcommand\subsection@cntformat{\thesubsection\charaftersectionnum\spacingaftersection}
\newcommand\subsubsection@cntformat{\thesubsubsection\charaftersectionnum\spacingaftersection}
\makeatother

\titlespacing{\section}{\sectionspacingleft pt}{\sectionspacingtop pt plus 0pt minus 4pt}{\sectionspacingbottom pt plus 0pt minus 2pt}
\titlespacing{\subsection}{\ssectionspacingleft pt}{\ssectionspacingtop pt plus 0pt minus 2pt}{\ssectionspacingbottom pt plus 0pt minus 2pt}
\titlespacing{\subsubsection}{\sssectionspacingleft pt}{\sssectionspacingtop pt plus 0pt minus 2pt}{\sssectionspacingbottom pt plus 0pt minus 2pt}
\titlespacing*{\subsubsubsection}{\ssssectionspacingleft pt}{\ssssectionspacingtop pt plus 0pt minus 2pt}{\ssssectionspacingbottom pt plus 0pt minus 2pt}
\makeatletter
\renewcommand\paragraph{\@startsection{paragraph}{5}{\paragspacingleft pt}
	{\paragspacingtop pt \@plus 0pt \@minus 2pt}
	{\paragspacingbottom pt \@plus 0pt \@minus 2pt}
	{\color{\paragcolor}\normalfont\paragfontsize\paragfontstyle}}
\renewcommand\subparagraph{\@startsection{subparagraph}{6}{\paragsubspacingleft pt}
	{\paragsubspacingtop pt \@plus 0pt \@minus 2pt}
	{\paragsubspacingbottom pt \@plus 0pt \@minus 2pt}
	{\color{\paragsubcolor}\normalfont\paragsubfontsize\paragsubfontstyle}}
\makeatother

\ifthenelse{\equal{\footnoterestart}{none}}{
}{
\ifthenelse{\equal{\footnoterestart}{sec}}{
	\counterwithin*{footnote}{section}
}{
\ifthenelse{\equal{\footnoterestart}{ssec}}{
	\counterwithin*{footnote}{subsection}
}{
\ifthenelse{\equal{\footnoterestart}{sssec}}{
	\counterwithin*{footnote}{subsubsection}
}{
\ifthenelse{\equal{\footnoterestart}{ssssec}}{
	\counterwithin*{footnote}{subsubsubsection}
}{
\ifthenelse{\equal{\footnoterestart}{page}}{
	\counterwithin*{footnote}{page}
}{
\ifthenelse{\equal{\footnoterestart}{chap}}{
	\counterwithin*{footnote}{chapter}
}{
	\throwbadconfig{Formato reinicio numero footnote desconocido}{\footnoterestart}{none,chap,page,sec,ssec,sssec,ssssec}}}}}}}
}

\ifthenelse{\equal{\footnoterulefigure}{false}}{
	\floatsetup[figure]{footnoterule=none}}{
}
\ifthenelse{\equal{\footnoteruletable}{false}}{
	\floatsetup[table]{footnoterule=none}}{
}

\makeatletter
\def\@xfootnote[#1]{%
  \protected@xdef\@thefnmark{#1}%
  \@footnotemark\@footnotetext}
\makeatother

\ifthenelse{\equal{\equationrestart}{none}}{
}{
\ifthenelse{\equal{\equationrestart}{chap}}{
}{
\ifthenelse{\equal{\equationrestart}{sec}}{
}{
\ifthenelse{\equal{\equationrestart}{ssec}}{
}{
\ifthenelse{\equal{\equationrestart}{sssec}}{
}{
\ifthenelse{\equal{\equationrestart}{ssssec}}{
}{
	\throwbadconfig{Formato reinicio numero ecuacion desconocido}{\equationrestart}{none,chap,sec,ssec,sssec,ssssec}}}}}}
}

\newtheoremstyle{miestilo}{\baselineskip}{3pt}{\itshape}{}{\bfseries}{}{.5em}{}
\newtheoremstyle{miobs}{\baselineskip}{3pt}{}{}{\bfseries}{}{.5em}{}
\theoremstyle{miestilo}

\ifthenelse{\equal{\showsectioncaptionmat}{none}}{
	\newtheorem{defn}{\namemathdefn}
	\newtheorem{teo}{\namemaththeorem}
	\newtheorem{cor}{\namemathcol}
	\newtheorem{lema}{\namemathlem}
	\newtheorem{prop}{\namemathprp}
}{
\ifthenelse{\equal{\showsectioncaptionmat}{chap}}{
	\newtheorem{defn}{\namemathdefn}[chapter]
	\newtheorem{teo}{\namemaththeorem}[chapter]
	\newtheorem{cor}{\namemathcol}[chapter]
	\newtheorem{lema}{\namemathlem}[chapter]
	\newtheorem{prop}{\namemathprp}[chapter]
}{
\ifthenelse{\equal{\showsectioncaptionmat}{sec}}{
	\newtheorem{defn}{\namemathdefn}[section]
	\newtheorem{teo}{\namemaththeorem}[section]
	\newtheorem{cor}{\namemathcol}[section]
	\newtheorem{lema}{\namemathlem}[section]
	\newtheorem{prop}{\namemathprp}[section]
}{
\ifthenelse{\equal{\showsectioncaptionmat}{ssec}}{
	\newtheorem{defn}{\namemathdefn}[subsection]
	\newtheorem{teo}{\namemaththeorem}[subsection]
	\newtheorem{cor}{\namemathcol}[subsection]
	\newtheorem{lema}{\namemathlem}[subsection]
	\newtheorem{prop}{\namemathprp}[subsection]
}{
\ifthenelse{\equal{\showsectioncaptionmat}{sssec}}{
	\newtheorem{defn}{\namemathdefn}[subsubsection]
	\newtheorem{teo}{\namemaththeorem}[subsubsection]
	\newtheorem{cor}{\namemathcol}[subsubsection]
	\newtheorem{lema}{\namemathlem}[subsubsection]
	\newtheorem{prop}{\namemathprp}[subsubsection]
}{
\ifthenelse{\equal{\showsectioncaptionmat}{ssssec}}{

}{
	\throwbadconfig{Valor configuracion incorrecto}{\showsectioncaptionmat}{none,chap,sec,ssec,sssec,ssssec}}}}}}
}
\theoremstyle{miobs}

\unaccentedoperators

\AtEndDocument{
	\addtocounter{equation}{\value{templateEquations}}
	\addtocounter{figure}{\value{templateFigures}}
	\addtocounter{lstlisting}{\value{templateListings}}
	\addtocounter{table}{\value{templateTables}}
}

\newcolumntype{C}[1]{>{\centering\let\newline\\\arraybackslash\hspace{0pt}}m{#1}}
\newcolumntype{\CColor}[2]{>{\columncolor{#1}\centering\let\newline\\\arraybackslash\hspace{0pt}}m{#2}}

\newcolumntype{P}[1]{>{\centering\let\newline\\\arraybackslash\hspace{0pt}}p{#1}}
\newcolumntype{\PColor}[2]{>{\columncolor{#1}\centering\let\newline\\\arraybackslash\hspace{0pt}}p{#2}}

\newcolumntype{B}[1]{>{\centering\let\newline\\\arraybackslash\hspace{0pt}}b{#1}}
\newcolumntype{\BColor}[2]{>{\columncolor{#1}\centering\let\newline\\\arraybackslash\hspace{0pt}}b{#2}}

\newcolumntype{L}[1]{>{\raggedright\let\newline\\\arraybackslash\hspace{0pt}}m{#1}}
\newcolumntype{\LColor}[2]{>{\columncolor{#1}\raggedright\let\newline\\\arraybackslash\hspace{0pt}}m{#2}}
\newcolumntype{T}[1]{>{\raggedright\let\newline\\\arraybackslash\hspace{0pt}}p{#1}}
\newcolumntype{\TColor}[2]{>{\columncolor{#1}\raggedright\let\newline\\\arraybackslash\hspace{0pt}}p{#2}}
\newcolumntype{F}[1]{>{\raggedright\let\newline\\\arraybackslash\hspace{0pt}}b{#1}}
\newcolumntype{\FColor}[2]{>{\columncolor{#1}\raggedright\let\newline\\\arraybackslash\hspace{0pt}}b{#2}}

\newcolumntype{R}[1]{>{\raggedleft\let\newline\\\arraybackslash\hspace{0pt}}m{#1}}
\newcolumntype{\RColor}[2]{>{\columncolor{#1}\raggedleft\let\newline\\\arraybackslash\hspace{0pt}}m{#2}}
\newcolumntype{H}[1]{>{\raggedleft\let\newline\\\arraybackslash\hspace{0pt}}p{#1}}
\newcolumntype{\HColor}[2]{>{\columncolor{#1}\raggedleft\let\newline\\\arraybackslash\hspace{0pt}}p{#2}}
\newcolumntype{G}[1]{>{\raggedleft\let\newline\\\arraybackslash\hspace{0pt}}b{#1}}
\newcolumntype{\GColor}[2]{>{\columncolor{#1}\raggedleft\let\newline\\\arraybackslash\hspace{0pt}}b{#2}}

\BeforeBeginEnvironment{tablenotes}{%
	\tablenotesfontsize\selectfont%
}
\AfterEndEnvironment{tablenotes}{%
	\normalsize\selectfont%
}

\let\SOURCEcaptionlrmargin\captionlrmargin
\newcounter{multicoldepth}
\BeforeBeginEnvironment{multicols}{%
	\def\captionlrmargin {\captionlrmarginmc}%
	\global\def\GLOBALenvmulticol {true}%
	\setcaptionmargincm{\captionlrmargin}%
	\addtocounter{multicoldepth}{1}%
}
\AfterEndEnvironment{multicols}{%
	\def\captionlrmargin {\SOURCEcaptionlrmargin}%
	\setcaptionmargincm{\captionlrmargin}%
	\addtocounter{multicoldepth}{-1}
	\ifnumequal{\number\value{multicoldepth}}{0}{%
		\global\def\GLOBALenvmulticol {false}
	}{}
}

\AtBeginDocument{%
	\normalfont%
	\setlength{\parindent}{\documentparindent pt}%
	\setlength{\parskip}{\documentparskip pt}%
}

\newcommand{\templatePagecfg}{
	
	\clearpage
	\ifthenelse{\equal{\predocpageromannumber}{true}}{ 
		\ifthenelse{\equal{\predocpageromanupper}{true}}{
			\pagenumbering{Roman}
		}{
			\pagenumbering{roman}
		}}{
		\pagenumbering{arabic}
	}
	\setcounter{page}{1}
	\setcounter{footnote}{0}
	
	\setpagemargincm{\pagemarginleft}{\pagemargintop}{\pagemarginright}{\pagemarginbottom}
	\resettablecellpadding
	
	\ifthenelse{\equal{\pointdecimal}{true}}{
		\decimalpoint}{
	}
	
	\renewcommand{\abstractname}{\nameabstract} 
	\renewcommand{\appendixname}{\nameltappendixsection} 
	\renewcommand{\appendixpagename}{\nameappendixsection} 
	\renewcommand{\appendixtocname}{\nameappendixsection} 
	\renewcommand{\contentsname}{\nameltcont} 
	\renewcommand{\figurename}{\nameltwfigure} 
	\renewcommand{\listfigurename}{\nameltfigure} 
	\renewcommand{\listtablename}{\namelttable} 
	\renewcommand{\lstlistingname}{\nameltwsrc} 
	\renewcommand{\lstlistlistingname}{\nameltsrc} 
	\renewcommand{\refname}{\namereferences} 
	\renewcommand{\bibname}{\namereferences} 
	\renewcommand{\tablename}{\nameltwtable} 
	
	\sectionfont{%
		\color{\sectioncolor} \sectionfontsize \sectionfontstyle \selectfont%
	}
	\subsectionfont{%
		\color{\ssectioncolor} \ssectionfontsize \ssectionfontstyle \selectfont%
	}
	\subsubsectionfont{%
		\color{\sssectioncolor} \sssectionfontsize \sssectionfontstyle \selectfont%
	}
	\titleformat{\subsubsubsection}{%
		\color{\ssssectioncolor} \ssssectionfontsz \ssssectionfontstyle
		}{%
		\thesubsubsubsection\charaftersectionnum\spacingaftersection}{0em}{%
	}
	\def\bibfont {\fontsizerefbibl} 
	
	\ifthenelse{\equal{\stylecitereferences}{apacite}}{
		\renewcommand{\BOthers}[1]{\apacitebothers\hbox{}}
	}{}
	
	\ifthenelse{\isundefined{\authorshf}}{%
		\def\authorshf {}}{%
	}%
	\fancyheadoffset{0pt} 
	\def\hfheaderimageparamsA {height=\baselineskip} 
	\ifthenelse{\equal{\hfstyle}{style1}}{
		\pagestyle{fancy}
		\newcommand{\COREstyledefinition}{
			\fancyhf{}
			\fancyfoot[C]{\thepage}
			\renewcommand{\headrulewidth}{0pt}
			\renewcommand{\footrulewidth}{0pt}
		}
		\setlength{\headheight}{49pt}
		\COREstyledefinition
	}{
	\ifthenelse{\equal{\hfstyle}{style2}}{
		\pagestyle{fancy}
		\newcommand{\COREstyledefinition}{
			\fancyhf{}
			\fancyfoot[R]{\thepage}
			\renewcommand{\headrulewidth}{0pt}
			\renewcommand{\footrulewidth}{0pt}
		}
		\setlength{\headheight}{49pt}
		\COREstyledefinition
	}{
	\ifthenelse{\equal{\hfstyle}{style3}}{
		\pagestyle{fancy}
		\newcommand{\COREstyledefinition}{
			\fancyhf{}
			\fancyhead[LE,RO]{\authorshf: \documenttitlehf}
			\fancyhead[RE,LO]{\thepage}
			\renewcommand{\headrulewidth}{0pt}
			\renewcommand{\footrulewidth}{0pt}
		}
		\fancypagestyle{portraitstyle}{
			\fancyhf{}
			\fancyhead[L]{\journalname}
			\renewcommand{\headrulewidth}{0pt}
			\renewcommand{\footrulewidth}{0pt}
		}
		\thispagestyle{portraitstyle}
		\COREstyledefinition
	}{
	\ifthenelse{\equal{\hfstyle}{style4}}{
		\pagestyle{fancy}
		\newcommand{\COREstyledefinition}{
			\fancyhf{}
			\fancyhead[LE,RO]{\small \textit{\authorshf}}
			\fancyhead[RE,LO]{\small \textit{\journalname}}
			\fancyfoot[C]{\small \thepage}
			\renewcommand{\headrulewidth}{0pt}
			\renewcommand{\footrulewidth}{0pt}
		}
		\fancypagestyle{portraitstyle}{
			\fancyhf{}
			\fancyhead[C]{\small \journalname}
			\renewcommand{\headrulewidth}{0pt}
			\renewcommand{\footrulewidth}{0pt}
		}
		\thispagestyle{portraitstyle}
		\COREstyledefinition
	}{
	\ifthenelse{\equal{\hfstyle}{style5}}{
		\pagestyle{fancy}
		\newcommand{\COREstyledefinition}{
			\fancyhf{}
			\fancyhead[LE,RO]{\thepage}
			\fancyhead[RE]{\authorshf}
			\fancyhead[LO]{\documenttitlehf}
			\renewcommand{\headrulewidth}{0.5pt}
			\renewcommand{\footrulewidth}{0pt}
		}
		\fancypagestyle{portraitstyle}{
			\fancyhf{}
			\fancyhead[L]{\journalname}
			\renewcommand{\headrulewidth}{0.5pt}
			\renewcommand{\footrulewidth}{0pt}
		}
		\thispagestyle{portraitstyle}
		\COREstyledefinition
	}{
	\ifthenelse{\equal{\hfstyle}{style6}}{
		\pagestyle{fancy}
		\newcommand{\COREstyledefinition}{
			\fancyhf{}
			\fancyhead[R]{\journalname}
			\fancyfoot[C]{\thepage}
			\renewcommand{\headrulewidth}{0pt}
			\renewcommand{\footrulewidth}{0pt}
		}
		\COREstyledefinition
	}{
	\ifthenelse{\equal{\hfstyle}{style7}}{
		\pagestyle{fancy}
		\newcommand{\COREstyledefinition}{
			\fancyhf{}
			\fancyhead[LE,RO]{\textbf{\thepage}}
			\fancyhead[RE]{\textbf{\authorshf}}
			\fancyhead[LO]{\textbf{\documenttitlehf}}
			\renewcommand{\headrulewidth}{0pt}
			\renewcommand{\footrulewidth}{0pt}
		}
		\fancypagestyle{portraitstyle}{
			\fancyhf{}
			\renewcommand{\headrulewidth}{0pt}
			\renewcommand{\footrulewidth}{0pt}
		}
		\thispagestyle{portraitstyle}
		\COREstyledefinition
	}{
	\ifthenelse{\equal{\hfstyle}{style8}}{
		\pagestyle{fancy}
		\newcommand{\COREstyledefinition}{
			\fancyhf{}
			\fancyfoot[C]{\thepage}
			\renewcommand{\headrulewidth}{0pt}
			\renewcommand{\footrulewidth}{0pt}
		}
		\fancypagestyle{portraitstyle}{
			\fancyhf{}
			\fancyhead[C]{\journalname}
			\fancyfoot[C]{\thepage}
			\renewcommand{\headrulewidth}{0pt}
			\renewcommand{\footrulewidth}{0pt}
		}
		\thispagestyle{portraitstyle}
		\COREstyledefinition
	}{
	\ifthenelse{\equal{\hfstyle}{style9}}{
		\pagestyle{fancy}
		\newcommand{\COREstyledefinition}{
			\fancyhf{}
			\fancyhead[LE]{\thepage\quad\quad\authorshf}
			\fancyhead[RO]{\documenttitlehf\quad\quad\thepage}
			\renewcommand{\headrulewidth}{0pt}
			\renewcommand{\footrulewidth}{0pt}
		}
		\fancypagestyle{portraitstyle}{
			\fancyhf{}
			\fancyfoot[L]{\journalname}
			\renewcommand{\headrulewidth}{0pt}
			\renewcommand{\footrulewidth}{0pt}
		}
		\thispagestyle{portraitstyle}
		\COREstyledefinition
	}{
	\ifthenelse{\equal{\hfstyle}{style10}}{
		\pagestyle{fancy}
		\newcommand{\COREstyledefinition}{
			\fancyhf{}
			\fancyhead[C]{\journalname}
			\fancyfoot[R]{\thepage}
			\renewcommand{\headrulewidth}{0pt}
			\renewcommand{\footrulewidth}{0pt}
		}
		\COREstyledefinition
	}{
	\ifthenelse{\equal{\hfstyle}{style11}}{
		\pagestyle{fancy}
		\newcommand{\COREstyledefinition}{
			\fancyhf{}
			\fancyhead[LE,RO]{\small \journalname}
			\fancyhead[LO,RE]{\small \authorshf}
			\renewcommand{\headrulewidth}{0pt}
			\renewcommand{\footrulewidth}{0pt}
		}
		\fancypagestyle{portraitstyle}{
			\fancyhf{}
			\renewcommand{\headrulewidth}{0pt}
			\renewcommand{\footrulewidth}{0pt}
		}
		\thispagestyle{portraitstyle}
		\COREstyledefinition
	}{
	\ifthenelse{\equal{\hfstyle}{style12}}{
		\pagestyle{fancy}
		\newcommand{\COREstyledefinition}{
			\fancyhf{}
			\fancyhead[LE,RO]{\small \thepage}
			\fancyhead[C]{\small \textit{\authorshf / \journalname}}
			\renewcommand{\headrulewidth}{0pt}
			\renewcommand{\footrulewidth}{0pt}
		}
		\fancypagestyle{portraitstyle}{
			\fancyhf{}
			\fancyhead[C]{\small \journalname}
			\renewcommand{\headrulewidth}{0pt}
			\renewcommand{\footrulewidth}{0pt}
		}
		\thispagestyle{portraitstyle}
		\COREstyledefinition
	}{
	\ifthenelse{\equal{\hfstyle}{style13}}{
		\pagestyle{fancy}
		\newcommand{\COREstyledefinition}{
			\fancyhf{}
			\fancyhead[R]{\small \textit{\documenttitlehf}}
			\fancyfoot[L]{\small \journalname}
			\fancyfoot[R]{\small \thepage\ / \pageref{TotPages}}
			\renewcommand{\headrulewidth}{0.5pt}
			\renewcommand{\footrulewidth}{0.5pt}
		}
		\fancypagestyle{portraitstyle}{
			\fancyhf{}
			\fancyfoot[L]{\small \journalname}
			\renewcommand{\headrulewidth}{0pt}
			\renewcommand{\footrulewidth}{0.5pt}
		}
		\thispagestyle{portraitstyle}
		\COREstyledefinition
	}{
	\ifthenelse{\equal{\hfstyle}{style14}}{
		\pagestyle{fancy}
		\newcommand{\COREstyledefinition}{
			\fancyhf{}
			\fancyhead[C]{\small \journalname}
			\fancyfoot[R]{\small \thepage\ / \pageref{TotPages}}
			\renewcommand{\headrulewidth}{0pt}
			\renewcommand{\footrulewidth}{0pt}
		}
		\fancypagestyle{portraitstyle}{
			\fancyhf{}
			\fancyhead[C]{\small \journalname}
			\renewcommand{\headrulewidth}{0pt}
			\renewcommand{\footrulewidth}{0pt}
		}
		\thispagestyle{portraitstyle}
		\COREstyledefinition
	}{
	\ifthenelse{\equal{\hfstyle}{style15}}{
		\pagestyle{fancy}
		\newcommand{\COREstyledefinition}{
			\fancyhf{}
			\fancyhead[L]{\small \textit{\journalname}}
			\fancyhead[R]{\small \thepage\ \namepageof \pageref{TotPages}}
			\renewcommand{\headrulewidth}{0.5pt}
			\renewcommand{\footrulewidth}{0pt}
		}
		\fancypagestyle{portraitstyle}{
			\fancyhf{}
			\fancyfoot[C]{\small \journalname}
			\renewcommand{\headrulewidth}{0pt}
			\renewcommand{\footrulewidth}{0.5pt}
		}
		\thispagestyle{portraitstyle}
		\COREstyledefinition
	}{
	\ifthenelse{\equal{\hfstyle}{style16}}{
		\pagestyle{fancy}
		\newcommand{\COREstyledefinition}{
			\fancyhf{}
			\fancyhead[L]{\small \textit{\journalname}}
			\fancyhead[R]{\small \thepage\ \namepageof \pageref{TotPages}}
			\renewcommand{\headrulewidth}{0pt}
			\renewcommand{\footrulewidth}{0pt}
		}
		\fancypagestyle{portraitstyle}{
			\fancyhf{}
			\fancyfoot[C]{\small \journalname}
			\renewcommand{\headrulewidth}{0pt}
			\renewcommand{\footrulewidth}{0pt}
		}
		\thispagestyle{portraitstyle}
		\COREstyledefinition
	}{
	\ifthenelse{\equal{\hfstyle}{style17}}{
		\pagestyle{fancy}
		\newcommand{\COREstyledefinition}{
			\fancyhf{}
			\renewcommand{\headrulewidth}{0pt}
			\renewcommand{\footrulewidth}{0pt}
		}
		\renewcommand{\sectionmark}[1]{\markboth{##1}{}}
		\COREstyledefinition
	}{
		\throwbadconfigondoc{Estilo de header-footer incorrecto}{\hfstyle}{style1 .. style17}}}}}}}}}}}}}}}}}
	}
	\fancypagestyle{plain}{
		\fancyheadoffset{0pt}
		\COREstyledefinition
	}
	\floatpagestyle{plain}
	\rotfloatpagestyle{plain}

	\ifthenelse{\equal{\showlinenumbers}{true}}{
		\linenumbers}{
	}
	
}

\newcommand{\templateFinalcfg}{
	
	\markboth{}{}
	\clearpage
	
	\ifthenelse{\equal{\showsectioncaptioncode}{none}}{
		\def\sectionobjectnumcode {}
	}{
	\ifthenelse{\equal{\showsectioncaptioncode}{sec}}{
		\def\sectionobjectnumcode {\thesection\sectioncaptiondelimiter}
	}{
	\ifthenelse{\equal{\showsectioncaptioncode}{ssec}}{
		\def\sectionobjectnumcode {\thesubsection\sectioncaptiondelimiter}
	}{
	\ifthenelse{\equal{\showsectioncaptioncode}{sssec}}{
		\def\sectionobjectnumcode {\thesubsubsection\sectioncaptiondelimiter}
	}{
	\ifthenelse{\equal{\showsectioncaptioncode}{ssssec}}{
		\def\sectionobjectnumcode {\thesubsubsubsection\sectioncaptiondelimiter}
	}{
	\ifthenelse{\equal{\showsectioncaptioncode}{chap}}{
		\def\sectionobjectnumcode {\thechapter\sectioncaptiondelimiter}
	}{
		\throwbadconfig{Valor configuracion incorrecto}{\showsectioncaptioncode}{none,chap,sec,ssec,sssec,ssssec}}}}}}
	}
	
	\ifthenelse{\equal{\showsectioncaptioneqn}{none}}{
		\def\sectionobjectnumeqn {}
	}{
	\ifthenelse{\equal{\showsectioncaptioneqn}{sec}}{
		\def\sectionobjectnumeqn {\thesection\sectioncaptiondelimiter}
	}{
	\ifthenelse{\equal{\showsectioncaptioneqn}{ssec}}{
		\def\sectionobjectnumeqn {\thesubsection\sectioncaptiondelimiter}
	}{
	\ifthenelse{\equal{\showsectioncaptioneqn}{sssec}}{
		\def\sectionobjectnumeqn {\thesubsubsection\sectioncaptiondelimiter}
	}{
	\ifthenelse{\equal{\showsectioncaptioneqn}{ssssec}}{
		\def\sectionobjectnumeqn {\thesubsubsubsection\sectioncaptiondelimiter}
	}{
	\ifthenelse{\equal{\showsectioncaptioneqn}{chap}}{
		\def\sectionobjectnumeqn {\thechapter\sectioncaptiondelimiter}
	}{
		\throwbadconfig{Valor configuracion incorrecto}{\showsectioncaptioneqn}{none,chap,sec,ssec,sssec,ssssec}}}}}}
	}
	
	\ifthenelse{\equal{\showsectioncaptionfig}{none}}{
		\def\sectionobjectnumfig {}
	}{
	\ifthenelse{\equal{\showsectioncaptionfig}{sec}}{
		\def\sectionobjectnumfig {\thesection\sectioncaptiondelimiter}
	}{
	\ifthenelse{\equal{\showsectioncaptionfig}{ssec}}{
		\def\sectionobjectnumfig {\thesubsection\sectioncaptiondelimiter}
	}{
	\ifthenelse{\equal{\showsectioncaptionfig}{sssec}}{
		\def\sectionobjectnumfig {\thesubsubsection\sectioncaptiondelimiter}
	}{
	\ifthenelse{\equal{\showsectioncaptionfig}{ssssec}}{
		\def\sectionobjectnumfig {\thesubsubsubsection\sectioncaptiondelimiter}
	}{
	\ifthenelse{\equal{\showsectioncaptionfig}{chap}}{
		\def\sectionobjectnumfig {\thechapter\sectioncaptiondelimiter}
	}{
		\throwbadconfig{Valor configuracion incorrecto}{\showsectioncaptionfig}{none,chap,sec,ssec,sssec,ssssec}}}}}}
	}
	
	\ifthenelse{\equal{\showsectioncaptiontab}{none}}{
		\def\sectionobjectnumtab {}
	}{
	\ifthenelse{\equal{\showsectioncaptiontab}{sec}}{
		\def\sectionobjectnumtab {\thesection\sectioncaptiondelimiter}
	}{
	\ifthenelse{\equal{\showsectioncaptiontab}{ssec}}{
		\def\sectionobjectnumtab {\thesubsection\sectioncaptiondelimiter}
	}{
	\ifthenelse{\equal{\showsectioncaptiontab}{sssec}}{
		\def\sectionobjectnumtab {\thesubsubsection\sectioncaptiondelimiter}
	}{
	\ifthenelse{\equal{\showsectioncaptiontab}{ssssec}}{
		\def\sectionobjectnumtab {\thesubsubsubsection\sectioncaptiondelimiter}
	}{
	\ifthenelse{\equal{\showsectioncaptiontab}{chap}}{
		\def\sectionobjectnumtab {\thechapter\sectioncaptiondelimiter}
	}{
		\throwbadconfig{Valor configuracion incorrecto}{\showsectioncaptiontab}{none,chap,sec,ssec,sssec,ssssec}}}}}}
	}
	
	\ifthenelse{\equal{\captionnumcode}{arabic}}{
		\renewcommand{\thelstlisting}{\sectionobjectnumcode\arabic{lstlisting}}
	}{
	\ifthenelse{\equal{\captionnumcode}{alph}}{
		\renewcommand{\thelstlisting}{\sectionobjectnumcode\alph{lstlisting}}
	}{
	\ifthenelse{\equal{\captionnumcode}{Alph}}{
		\renewcommand{\thelstlisting}{\sectionobjectnumcode\Alph{lstlisting}}
	}{
	\ifthenelse{\equal{\captionnumcode}{roman}}{
		\renewcommand{\thelstlisting}{\sectionobjectnumcode\roman{lstlisting}}
	}{
	\ifthenelse{\equal{\captionnumcode}{Roman}}{
		\renewcommand{\thelstlisting}{\sectionobjectnumcode\Roman{lstlisting}}
	}{
		\throwbadconfig{Tipo numero codigo fuente desconocido}{\captionnumcode}{arabic,alph,Alph,roman,Roman}}}}}
	}
	
	\ifthenelse{\equal{\captionnumequation}{arabic}}{
		\renewcommand{\theequation}{\sectionobjectnumeqn\arabic{equation}}
	}{
	\ifthenelse{\equal{\captionnumequation}{alph}}{
		\renewcommand{\theequation}{\sectionobjectnumeqn\alph{equation}}
	}{
	\ifthenelse{\equal{\captionnumequation}{Alph}}{
		\renewcommand{\theequation}{\sectionobjectnumeqn\Alph{equation}}
	}{
	\ifthenelse{\equal{\captionnumequation}{roman}}{
		\renewcommand{\theequation}{\sectionobjectnumeqn\roman{equation}}
	}{
	\ifthenelse{\equal{\captionnumequation}{Roman}}{
		\renewcommand{\theequation}{\sectionobjectnumeqn\Roman{equation}}
	}{
		\throwbadconfig{Tipo numero ecuacion desconocido}{\captionnumequation}{arabic,alph,Alph,roman,Roman}}}}}
	}
	
	\ifthenelse{\equal{\captionnumfigure}{arabic}}{
		\renewcommand{\thefigure}{\sectionobjectnumfig\arabic{figure}}
	}{
	\ifthenelse{\equal{\captionnumfigure}{alph}}{
		\renewcommand{\thefigure}{\sectionobjectnumfig\alph{figure}}
	}{
	\ifthenelse{\equal{\captionnumfigure}{Alph}}{
		\renewcommand{\thefigure}{\sectionobjectnumfig\Alph{figure}}
	}{
	\ifthenelse{\equal{\captionnumfigure}{roman}}{
		\renewcommand{\thefigure}{\sectionobjectnumfig\roman{figure}}
	}{
	\ifthenelse{\equal{\captionnumfigure}{Roman}}{
		\renewcommand{\thefigure}{\sectionobjectnumfig\Roman{figure}}
	}{
		\throwbadconfig{Tipo numero figura desconocido}{\captionnumfigure}{arabic,alph,Alph,roman,Roman}}}}}
	}
	
	\ifthenelse{\equal{\captionnumsubfigure}{arabic}}{
		\renewcommand{\thesubfigure}{\arabic{subfigure}}
	}{
	\ifthenelse{\equal{\captionnumsubfigure}{alph}}{
		\renewcommand{\thesubfigure}{\alph{subfigure}}
	}{
	\ifthenelse{\equal{\captionnumsubfigure}{Alph}}{
		\renewcommand{\thesubfigure}{\Alph{subfigure}}
	}{
	\ifthenelse{\equal{\captionnumsubfigure}{roman}}{
		\renewcommand{\thesubfigure}{\roman{subfigure}}
	}{
	\ifthenelse{\equal{\captionnumsubfigure}{Roman}}{
		\renewcommand{\thesubfigure}{\Roman{subfigure}}
	}{
		\throwbadconfig{Tipo numero subfigura desconocido}{\captionnumsubfigure}{arabic,alph,Alph,roman,Roman}}}}}
	}
	
	\ifthenelse{\equal{\captionnumtable}{arabic}}{
		\renewcommand{\thetable}{\sectionobjectnumtab\arabic{table}}
	}{
	\ifthenelse{\equal{\captionnumtable}{alph}}{
		\renewcommand{\thetable}{\sectionobjectnumtab\alph{table}}
	}{
	\ifthenelse{\equal{\captionnumtable}{Alph}}{
		\renewcommand{\thetable}{\sectionobjectnumtab\Alph{table}}
	}{
	\ifthenelse{\equal{\captionnumtable}{roman}}{
		\renewcommand{\thetable}{\sectionobjectnumtab\roman{table}}
	}{
	\ifthenelse{\equal{\captionnumtable}{Roman}}{
		\renewcommand{\thetable}{\sectionobjectnumtab\Roman{table}}
	}{
		\throwbadconfig{Tipo numero tabla desconocido}{\captionnumtable}{arabic,alph,Alph,roman,Roman}}}}}
	}
	
	\ifthenelse{\equal{\captionnumsubtable}{arabic}}{
		\renewcommand{\thesubtable}{\arabic{subtable}}
	}{
	\ifthenelse{\equal{\captionnumsubtable}{alph}}{
		\renewcommand{\thesubtable}{\alph{subtable}}
	}{
	\ifthenelse{\equal{\captionnumsubtable}{Alph}}{
		\renewcommand{\thesubtable}{\Alph{subtable}}
	}{
	\ifthenelse{\equal{\captionnumsubtable}{roman}}{
		\renewcommand{\thesubtable}{\roman{subtable}}
	}{
	\ifthenelse{\equal{\captionnumsubtable}{Roman}}{
		\renewcommand{\thesubtable}{\Roman{subtable}}
	}{
		\throwbadconfig{Tipo numero subtabla desconocido}{\captionnumsubtable}{arabic,alph,Alph,roman,Roman}}}}}
	}
	
	\ifthenelse{\equal{\predocpageromannumber}{true}}{
		\renewcommand{\thepage}{\arabic{page}}}{
	}
	
	\ifthenelse{\equal{\predocresetpagenumber}{true}}{
		\setcounter{page}{1}}{
	}
	
	\setcounter{section}{0}
	\setcounter{footnote}{0}
	
	\ifthenelse{\equal{\showlinenumbers}{true}}{
		\linenumbers}{
	}
	
	\titleclass{\subsubsubsection}{straight}[\subsection]
	
}

\def\authorshf {Madariaga, Mondschein \& Torres} 

\addauthor{Benjamín Madariaga}{Department of Mathematical Engineering, University of Chile, Santiago, Chile.}{}
\addauthor{Susana Mondschein$^{(*)}$}{Department of Industrial Engineering, University of Chile, Santiago, Chile.} {}
\addauthor{\hspace{-11pt}}{Instituto Sistemas Complejos de Ingeniería (ISCI), Santiago, Chile.}{}
\addauthor{Soledad Torres}{Centro Integral de la Mama, Clínica Las Condes, Santiago, Chile.}{}

\begin{document}
	
\templatePagecfg

\templateFinalcfg


\inserttitle

$(*)$ Corresponding author: Susana Mondschein, \url{susana.mondschein@uchile.cl}.

\begin{abstract}
\noindent \textbf{Introduction:} Breast cancer (BC) is one of the most common cancers in women worldwide and in Chile. Due to the lack of a Chilean national cancer registry, there is partial information on the status of breast cancer in the country. \\
\noindent \textbf{Aims:} To study inequities in BC health care outcomes for Chilean women, including case fatality rates and survival rates in Chilean women, stratified by type of health care insurance provider and geographical area. A secondary goal is to estimate breast cancer incidence and mortality rates by health care providers and region. \\
\noindent \textbf{Methods:} We used two public anonymized databases provided by the Department of Health Statistics and Information of the Ministry of Health: the national death and hospital discharges datasets. For calculations of incidence and mortality rates, we selected all patients whose primary diagnosis was breast cancer, according to the CDI-10 diagnostic code. For survival analysis, we used the Kaplan Meier product--limit estimator with a 95\% confidence interval and the Cox proportional hazards model for studying the effects of covariates with null--hypothesis significance testing of $p>0.001$.  \\
\noindent \textbf{Results:} We considered a cohort of 58,254 and 16,615 BC hospital discharges and deaths for the period 2007--2018. The new cases of BC increased by 43.6\%, from 3,785 in 2007 to 5,435 in 2018. Total BC deaths increased by 33.6\% from 1,158 to 1,547 during the same time period. Age--adjusted incidence rates were stable over time, with an average rate of 44.0 and a standard deviation of 2.2 during the study period. There were considerable differences in age--adjusted incidence rates among regions,  with no clear geographical trend. Women affiliated to a private provider (ISAPRE) have an average age adjusted incidence rate of 60.6 compared to 38.8 for women affiliated with the public provider (FONASA). Age--adjusted mortality rates did not change significantly during the study period, with an average of 10.5 and a standard deviation of 0.4. The national fatality rate has remained relatively constant over time, with a mean of 26.8 and a standard deviation of 1.1. However, women affiliated with ISAPRE had a considerably lower fatality rate during the period under study, with an average of 15.7 compared to 27.5 for women affiliated with FONASA. The 95\% confidence intervals for the one-year survival rate were [0.931 ± 0.002] for FONASA, [0.972 ± 0.003] for the ISAPRE patients, and [0.940 ± 0.002] when considering all women combined. The 95\% confidence intervals for the five-year survival rates were [0.806 ± 0.004], [0.901 ± 0.007], and [0.827 ± 0.004] for FONASA, ISAPRE and all women, respectively. Women from the Metropolitan region have higher survival rates than women from other regions. Survival rates obtained using the Cox proportional hazards model have a similar behavior to those obtained by the Kaplan--Meier estimations.
\\
\noindent \textbf{Discussion:} Despite the inclusion of breast cancer within the set of pathologies selected for the GES plan in 2005 with the goal of providing  greater equity for all patients, there are still significant differences in case fatality and survival rates for patients affiliated with a private provider compared to those in the public system, a choice that is directly associated with socioeconomic level. The same disparity is observed between patients in the Metropolitan and other regions. Further studies are required to determine the causes of these disparities, such as differences in compliance with screening recommendations and general health status.  

\end{abstract}
\vspace{0.5cm}



\section{Introduction}

Breast cancer (BC) is one of the most common cancers in women worldwide, with 2.26 million new cases and 685 thousand deaths globally in 2020 and 7.8 million women alive with breast cancer diagnosed in the past 5 years \cite{WHO}. In 2020, age-standardized world breast cancer incidence and mortality rates were 47.8/100,000 and 13.6/100,000 women, respectively \cite{globocan_data}, while the figures reported for Chile by the Ministry of Health (Minsal) for 2018 were 44.1/100,000 and 11.3/100,000 women for incidence and mortality rates, respectively \cite{minsal_informe_incidencia,minsal_informe_mortalidad}. 

The Chilean health care system consists of a hybrid of public and private insurance plans as well as providers. The three main insurance systems are the National Health Fund (FONASA) for 78\% of the Chilean population, private health care insurers (ISAPREs) for 14\% of the Chilean population, and  the Military and Police Forces’ health system for 3\% of the Chilean population. The remaining 5\% of the population either has no health care insurer or their insurance status is unknown. Privately insured patients can solely access private providers, with a variety of coverage, depending on their health care plan and monthly payments. On the other hand, FONASA patients – paying, on average, a lower monthly price compared to those in the private system, might access public and private hospitals, depending on the health care plan within FONASA, with different amounts of copayments. Thus, the FONASA has four different segments of patients (A, B, C, and D), with A and D being those plans with the lowest- and highest- income patients, respectively. FONASA also provides coverage for attending private health centers to patients from the B, C and D segments \cite{informe_mle}.

The selection of the health care provider is determined mainly based on the individual's economic income \cite{Gallardo_Varas_Gallardo_2017,Castillo-Laborde2017}; as a consequence, profound inequalities in terms of quality of life and education, among others, have existed between users of the private and public affiliates.
Health care indicators have been significantly better for those with private health insurance compared to those for publicly insured patients. For example, in 2017, 27.8\% of FONASA affiliates declared having problems obtaining health care, while this percentage decreased to 12.6\% for ISAPRE affiliates  \cite{Casen_salud}.

Consequently, as an effort to provide better health care services for all citizens, in 2004, Chile implemented a profound health reform, the Explicit Health Guarantees (GES), which aimed to achieve a more equitable and fairer system \cite{lenz2007proceso}. The objective of this reform was to ensure access, quality, opportunity and financial protection all Chilean citizens requiring care for a set of most common pathologies to all Chilean citizens.
Due to budget constraints, the program started in 2005 with 25 pathologies, including breast cancer (for people aged 15 years and over with suspected, diagnosed or recurrent breast cancer), and has gradually grown to currently cover 85 pathologies. As stated by the Ministry of Health, Pedro García, in 2005, the GES plan will mean a qualitative leap toward greater equity in the health care sector, and the country will enter "a path of no return toward greater dignity for the citizens of our country and for that we are proud" \cite{cooperativa2005}.
Seventeen years after its implementation, there is a great deal of  important evidence that the GES plan has partly 
reached the goals of opportunity, with 92.8\% of services requested in 2017 fulfilled within the expected time \cite{monitoreoges2018}, and is constantly monitoring the quality of health care centers \cite{cuentapublicacancer}. In terms of access, among people from the FONASA who received health care in 2017, 27.8\% reported having received health care with some access difficulty, while the remaining 72.2\% received health care without access difficulties. The same figures for ISAPRE affiliates are 12.6\% and 87.4\%, respectively \cite{Casen_salud}.

However, to the best of our knowledge, there are no studies showing the effectiveness of the GES plan in terms of health care outcomes such as survival and case fatality (mortality/incidence)
rates. In particular, the focus of this paper is to study whether the inclusion of breast cancer in the GES program, including these four guarantees for all women, has resulted in a reduction (or elimination) of the gap in health outcomes, including case fatality (CFR) and survival rates among Chilean women.
To achieve this objective, reliable data for new cases and deaths due to breast cancer are needed to estimate both case fatality and survival rates. Unfortunately, in Chile, there is no national cancer registry, and thus, incidence rates have been estimated based on projections on the number of breast cancer cases diagnosed from 1998 to 2012, with follow-ups until 2015, from four regional population-based registries (Antofagasta and Los Ríos regions and Concepción and Biobío provinces). The  estimates for the rest of the country and from 2013 onward are based on statistical models, with assumptions such as a constant fatality rate across regions and over time, and in some instances, using neighboring country data \cite{globocan_methods, minsal_informe_incidencia}. However, these assumptions might lead to unreliable estimates when ignoring significant differences in ethnicity, urbanization, socioeconomic composition and health coverage among regions \cite{Casen_salud,Casen_ingresos,Casen_pueblos_indigenas,Casen_rural} and advances in breast cancer treatment over time. Therefore, fatality rates might largely vary among regions due to differences in incidence as well as in mortality rates and over time due to, for example, decreases in mortality rates.

On the other hand,  mortality rate estimates are much more reliable because of the existence of a high-quality national death registry that includes a series of demographic variables along with the cause of death of every decedent in Chile. The death registry has been used in studies such as those by Icaza, Nuñez and Bugueño \cite{ICAZA2017}, where the authors exploit the potential of this registry by studying breast cancer mortality, including ecological analysis by sociodemographic variables.

Thus, the primary goal of this paper is to study inequities in breast cancer outcomes such as lethality and survival rates in Chilean women, stratified by type of health care insurance provider and geographical area. The lack of reliable incidence rate estimates has led to a secondary objective for this study, which is the estimation of breast cancer incidence and mortality rates using
an alternative methodology, based on a
complete anonymized public database of national registry of hospital discharges, which includes information and diagnosis at the patient level, and the national death registry, with primary and secondary causes of death through a unique ID classification, compiled by the Department of Health Statistics and Information (DEIS) of the Ministry of Health.

\section{Methods and Materials}

\textbf{Ethics statement}

This work used publicly available data from the Chilean Ministry of Health through the Department of Health Statistics and Information. All data were protected, and personal information was anonymized; therefore, no consent from participants was needed.\\

\subsection{Data}
\label{data}

Two public anonymized databases were provided by DEIS. The first is the National Death Registry, which includes 2,549,800 deaths from January 1990 to December 2018. For each death entry in the registry, the patient’s ID (identifying code), date of birth, date of death, gender, town and region of residence, marital status, occupation and cause of death code according to the International Statistical Classification of Diseases and Related Health Problems (CDI-10) were available. There were 2,984 death registries without an ID, which corresponds to 0.1\% of the complete database.
The second database includes all discharges from  public and private health care facilities in the country, which consists of 32,443,591 registries from January 2001 to December 2020. Each registry has 39 fields, such as the patient’s ID (same as the national death database), date of birth, gender, town and region of residence, ethnicity, health insurance, length of stay, condition at discharge, and primary and secondary diagnoses according to CDI-10 classification, among others.\\

\noindent \textbf{\large{Inclusion and exclusion criteria}}

\vspace*{2mm}

The cohort for our study corresponds to all women who were diagnosed with breast cancer from 2007 to 2018. First, we chose 2007 as the starting year  to guarantee that the first discharge in the database is, in fact, the time of diagnosis. 
Thus, all women with breast cancer discharges from 2007 onward, but none from 2001 to 2007, were included as newly diagnosed patients. Second, we did not include data from 2019 onward because it is not reliable due to the political and social uprising in Chile (October--December 2019) \cite{Chile2019} and the COVID-19 pandemic, which largely influenced hospital discharges and deaths. Accordingly, we considered the same period of time for the national death registry.

A discharge or death registry is considered to correspond to breast cancer if its CDI-10 diagnostic code belongs to the categories C50 and D05 (\textit{malignant neoplasm of breast} and \textit{carcinoma in situ of breast}, respectively).

For calculations of mortality rates, we selected all females whose cause of death was directly associated with breast cancer from the 2,549,800 registries in the mortality database, resulting in 16,615 BC deaths after removing 18 death registries with missing IDs, during the study period.

\begin{figure}[htb]
    \centering
    \includegraphics[scale=0.7]{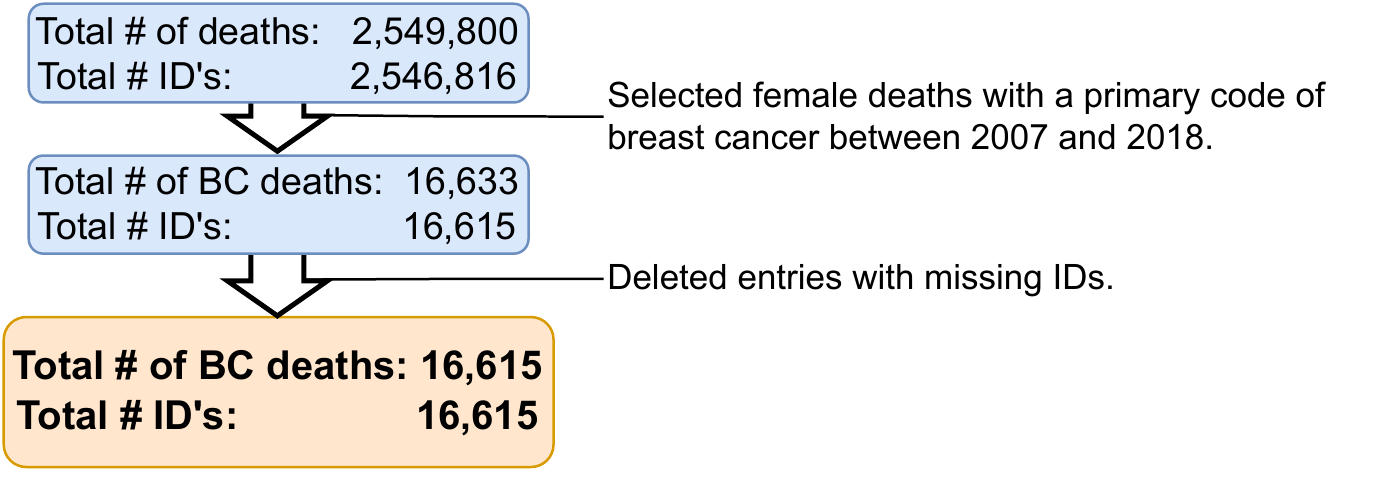}
    \caption{\textbf{ Inclusion and exclusion criteria for the  breast cancer death database (2007--2018).}}
    \label{fig:death_inclusion_exclusion}
\end{figure}

\begin{figure}[htb]
    \centering
    \includegraphics[scale=0.7]{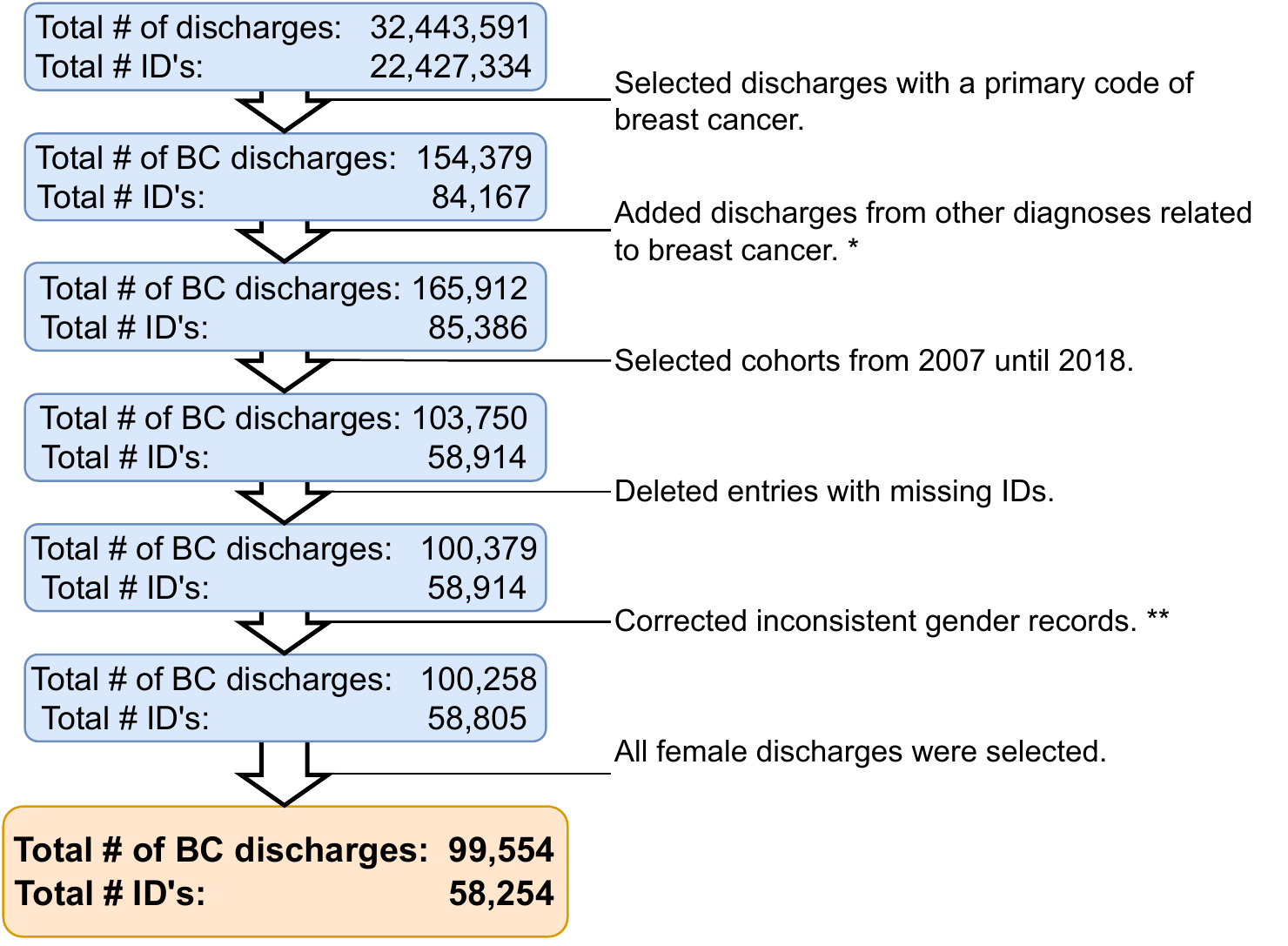}
    \caption{\textbf{Inclusion and exclusion criteria for the  breast cancer hospital discharge database (2007--2018).}\\
    * Several diagnoses were included considering health problems that could arise due to breast cancer progression or its treatment for patients who died due to breast cancer. See appendix \ref{app:other_diagnostics} for further explanation.\\
    ** There were patients with records who identified them both as male and female. It was considered that this was an error from the database; therefore, it was corrected.}
    \label{fig:discharges_inclusion_exclusion}
\end{figure}

For calculations of incidence rates, we selected all patients whose primary diagnosis was breast cancer from the 32,443,591 hospital discharges, resulting in 154,379 discharges. There were 5,230 deaths that did not have a  C50 and D05 discharge in the resulting database including 154,379 discharges. Thus, a set of diagnoses related to breast cancer was considered for such patients (see Appendix \ref{app:other_diagnostics} for a detailed explanation), leading to a total of 165,912 breast cancer-related discharges. Then, the cohorts from 2007 until 2018 were selected, resulting in 103,750 records corresponding to 58,914 patients. Out of the 103,750 discharge registries, there were 3,371 with a missing ID. Hence, we do not know if these discharges are associated with new patients or with those already in the cohort of 58,914 patients. Thus, discharge records with missing IDs were removed from the database. However, a sensitivity analysis regarding these registries is performed in the Results section. From the remaining 58,914 patients identified by their IDs, there were 987 patients who had inconsistencies in their gender, i.e.,these patients had hospital discharges where they were identified both as male and female patients. These gender inconsistencies were corrected by considering the gender information from the mortality database and considering the mode of all records associated with such patients. Thus, the gender of 815 people was corrected in the discharges database, while 109 cases for which the mode was inconclusive were deleted. 
Finally, all female patients were selected, resulting in a total of 99,554 discharge records, corresponding to 58,254 patients, with an average of 1.7 discharges per patient.
Figures \ref{fig:death_inclusion_exclusion} and \ref{fig:discharges_inclusion_exclusion} summarize the inclusion and exclusion criteria for the death and discharge databases, respectively. 

\subsection{Methods}
\label{meth}


We calculated both crude and age adjusted incidence and mortality rates for breast cancer at the national and regional levels. Population estimates and projections provided by the Chilean National Institute of Statistics (INE) \cite{ine_poblacion,ine_poblacion_comunas} were used. For the age-standardized incidence and mortality rates, we used the International Agency for Research on Cancer standard population \cite{IARC_local_data}.
We also calculated crude and age-standardized incidence and mortality rates by year, health insurer (public and private systems), and geographical region. 

The female population affiliated with the public health system was obtained using the data provided by the FONASA in its yearly statistical bulletin, which is available online from 2009 onward \cite{fonasa_data}. Prior to this, FONASA bulletins had incomplete information, without segmentation by age intervals or gender. Therefore, we used linear regression to estimate the 2007 and 2008 populations from the available data. 
Information from the private health system beneficiary population was obtained using the data provided by the Health Superintendence in its yearly statistics publication for ISAPRE's beneficiaries \cite{isapre_data}.

Case fatality rates, i.e., the percentage of people who died as a fraction of the newly diagnosed per year, were calculated for each year and strata (region and health care system) as \textit{Crude Mortality$\backslash$Crude Incidence.}


For the survival analysis, we used the standard estimator of the survival function proposed by Kaplan and Meier \cite{kaplan58}, namely, the product-limit estimator, considering right-censored data. In our study, censored data consisted of breast cancer patients who either died of other causes during the timeline of the study or survived from then on. To estimate survival, we considered time-on-study survival as our variable of interest, or equivalently, the time from the first breast cancer diagnosis to death.
Without loss of generality, we measured time in months, and therefore, the survival time was recorded as months difference; for example, a patient diagnosed by 1/1/2007 who died on 01/31/2007 had a 0 month survival, while a patient diagnosed by 1/31/2007 that died on 2/01/2007 had a one month survival. The Kaplan-Meier survival curves shown with their 95\% confidence intervals in the Results section.

Finally, we used the Cox proportional hazards model to study the effects of covariates \cite{cox72}. We used health insurance provider, age, year of discharge, region of residence and FONASA benefit segments as covariates. These covariates were processed, transforming some of them into dummy variables, resulting in a total of 26 covariates from which to select those relevant for the model. Selection was based on Akaike’s information criterion \cite{akaike73} and on the p value for each variable’s significance (p<0.001). See Appendix \ref{app:cox_regression} for a description of the model selection procedure. 

\vspace{3mm}

\noindent \textbf{Management of death data without hospital discharges}

There were 4,011 IDs from the death database during the study period without an associated breast cancer discharge, and therefore, the time of diagnosis was unknown. In Table \ref{tab:database_characteristics_def}, column \textit{Death without discharge} shows the distribution of these cases by year, with an average of 334 deaths by year with a standard deviation of 20.6. The fact that the number of deaths is stable over time implies that any assumption regarding the time of diagnosis would lead to similar estimates for yearly incidence rates. Hence, for computational purposes, we assumed that this set of 4,011 women had a random survival between one and twelve months. When generating such a random survival period, 3,839 out of the 4,011 cases had an incidence within the 2007--2018 period. This adds 3,839 registries to the discharge database, obtaining a total of 62,093 diagnosed patients. 

We also note that for these 4,011 deaths, women have a mean age of death of 75.4, which is  9 years older than the average age of all deaths (66.1). Additionally, out of the 4,011 deaths, there are 2,418 with a discharge for causes other than breast cancer in the discharge database.
For those patients, we retrieved their health care insurance (which is not included in the death certificate), resulting in 81.6\% affiliated with FONASA, 7.0\% affiliated with ISAPRE, 2.0\% affiliated with Armed Forces, and the remaining 9.4\% having no health care insurance or it was unknown.
The three most repeated diagnosis categories were "pneumonia, unspecified organism," "cholelithiasis" and "heart failure." Thus, these women are older and sicker, which might explain why they  died without being hospitalized due to  breast cancer treatments.

Table \ref{tab:database_characteristics_def} shows number of deaths, average age of death, and standard deviation by year. The last two columns show the percentage of deaths that do not have a corresponding breast cancer discharge registry, this is, these patients died without hospitalization due to breast cancer, and their average age at death with standard deviation by year.
Table \ref{tab:database_characteristics_inc} shows the number of newly diagnosed breast cancer cases, with the  average age of diagnosis and standard deviation.  

\enabletablerowcolor[2] 
\begin{table}
    \centering
    \begin{tabular}{lrrrr}
    \toprule
     &  \multicolumn{2}{c}{Total Deaths} &  \multicolumn{2}{c}{Deaths without discharge* } \\

    Year &  Cases & Mean Age (std) &  Cases(\%) & Mean Age (std) \\
    \midrule
    2007    &   1,158 &    65.4 (15.6) &  28.2 &    73.6 (14.8) \\
    2008    &   1,228 &    65.3 (15.6) &  29.2 &    72.8 (15.1) \\
    2009    &   1,337 &    65.8 (15.6) &  24.8 &    75.5 (15.0) \\
    2010    &   1,298 &    64.9 (15.6) &  24.3 &    73.8 (15.3) \\
    2011    &   1,349 &    66.2 (15.5) &  22.4 &    75.2 (15.0) \\
    2012    &   1,371 &    66.4 (15.6) &  23.2 &    74.9 (15.2) \\
    2013    &   1,391 &    65.8 (15.6) &  24.5 &    75.9 (14.7) \\
    2014    &   1,424 &    66.3 (16.0) &  22.9 &    75.9 (15.2) \\
    2015    &   1,512 &    66.5 (15.6) &  23.1 &    76.5 (14.4) \\
    2016    &   1,492 &    66.4 (15.3) &  21.2 &    76.4 (14.7) \\
    2017    &   1,508 &    66.8 (15.3) &  24.3 &    76.7 (14.9) \\
    2018    &   1,547 &    67.1 (16.0) &  23.3 &    77.0 (14.5) \\
    \midrule
    Total   &  16,615 &    66.1 (15.6) &  24.1 &    75.4 (14.9) \\
    \bottomrule
    \end{tabular}
    \caption{Total breast cancer deaths and percentage of breast cancer deaths without breast cancer   discharge by year for the period 2007 -- 2018.  \\
    * Women died due to breast cancer, but there were no hospital discharges associated with a breast cancer diagnosis. For the calculation of incidence, it is assumed that these women had a random uniform survival period between 1 and 12 months.}
    \label{tab:database_characteristics_def}
\end{table}
\disabletablerowcolor

\enabletablerowcolor[2] 
\begin{table}
    \centering
    \begin{tabular}{lrr}
    \toprule
    Year & New cases & Mean Age (std) \\
    \midrule
    2007       &   3,785 &    57.4 (14.1) \\
    2008       &   3,938 &    57.5 (14.1) \\
    2009       &   4,576 &    57.3 (13.9) \\
    2010       &   4,415 &    57.9 (14.1) \\
    2011       &   4,685 &    57.7 (14.0) \\
    2012       &   4,995 &    57.3 (13.8) \\
    2013       &   5,172 &    57.7 (13.7) \\
    2014       &   4,932 &    58.1 (13.7) \\
    2015       &   5,427 &    57.8 (13.6) \\
    2016       &   5,569 &    57.5 (13.8) \\
    2017       &   5,325 &    58.1 (14.0) \\
    2018       &   5,435 &    58.1 (13.9) \\
    \midrule
    Total      &  58,254 &    57.7 (13.9) \\
    \bottomrule
    \end{tabular}
    \caption{New breast cancer cases by year for 2007 -- 2018.}
    \label{tab:database_characteristics_inc}
\end{table}
\disabletablerowcolor


\section{Results}

The number of new breast cancer diagnoses increased by 43.6\%, from 3,785 in 2007 to 5,435 in 2018. Total breast cancer deaths increased by 33.6\% from 1,158 to 1,547 during the same period of time. We noticed a slight increasing trend in the mean age at death, with a 2.6\% growth from 65.4 to 67.1, while there was no significant change in the mean age at diagnosis. We noted that the female population in Chile grew 13.3\% from 8,390,194 to 9,506,921 from 2007 to 2018. According to these records, each woman had an average of 1.67 hospital discharges per patient. In what follows, the results of incidence, mortality, fatality and survival rates are shown.

\subsection{Incidence rates}

\begin{figure}
    \centering
    \includegraphics{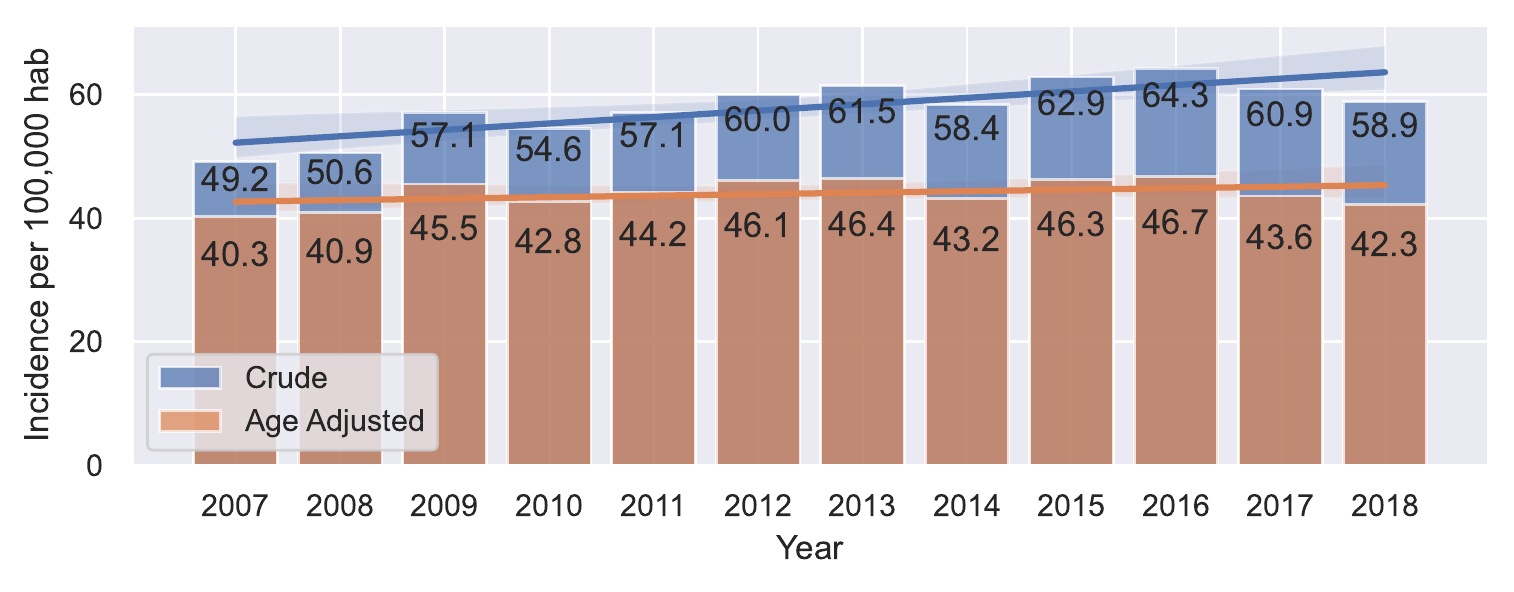}
    \caption{Crude and age--adjusted incidence rates by year (cases/100,000 women).}
    \label{fig:incidence_both_chile}
\end{figure}

As explained in Subsection \ref{meth}, incidence rates are calculated based on a set of 62,093 patients. In total, 58,254 patients were identified through the discharge registries plus the 3,839 breast cancer death registries without a discharge registry. Figure \ref{fig:incidence_both_chile} shows both crude and age--adjusted incidence rates (cases/100,000 women). We observe that the crude incidence rates are higher than the age--adjusted rates and have a growing trend over time, with an increase of 19.7\%, from 49.2 in 2007 to 58.9 in 2018. The age--adjusted rates are stable over time, with an average rate of 44.0 and a standard deviation of 2.2 during the study period.

It is important to note that there were 3,371 discharge registries that were deleted for having no ID and, therefore, it cannot be determined if these registries correspond to patients who have been already identified in the cohort or if they correspond to new diagnoses. To evaluate the impact of these cases on the incidence estimates, we considered two possible scenarios. First, we considered the worst--case scenario, where all 3,371 registries are new breast cancer diagnoses. Table \ref{tab:missing_ids_incidence_worst} shows the crude incidence rates calculated, accounting for each of these registries as a new diagnosis. Table \ref{tab:missing_ids_incidence_likely} shows a more likely scenario, where some of these discharges do not belong to a new patient. Women from the resulting database had an average of 1.67 hospital discharges; then, as a more likely scenario, we considered that there is a newly diagnosed patient for every 1.67 discharges with missing IDs.

\begin{table}
    \centering
    \begin{tabular}{lrrrrrrrrrrrr}
    \toprule
    Year &    2007 &   2008 &   2009 &   2010 &   2011 &   2012 &   2013 &   2014 &  2015 &  2016 &  2017 &  2018 \\
    \midrule
    Registries with missing IDs         &  1286.0 &  528.0 &  276.0 &  241.0 &  225.0 &  245.0 &  206.0 &  164.0 &  28.0 &  28.0 &  47.0 &  97.0 \\
    Crude Incidence        &    64.5 &   56.8 &   60.4 &   57.4 &   59.7 &   62.7 &   63.8 &   60.2 &  63.2 &  64.6 &  61.4 &  59.9 \\
    Variation (\%) &    23.8 &   11.0 &    5.3 &    4.8 &    4.3 &    4.4 &    3.6 &    3.0 &   0.5 &   0.5 &   0.8 &   1.7 \\
    \bottomrule
    \end{tabular}
    \caption{Registries with missing IDs and crude incidence rates (cases/100,000 women) calculated considering these as new cases (worst case scenario). The last row shows the variation between these crude rates and those shown in Figure \ref{fig:incidence_both_chile}. }
    \label{tab:missing_ids_incidence_worst}
\end{table}

\begin{table}
    \centering
    \begin{tabular}{lrrrrrrrrrrrr}
    \toprule
    Year &    2007 &   2008 &   2009 &   2010 &   2011 &   2012 &   2013 &   2014 &  2015 &  2016 &  2017 &  2018 \\
    \midrule
    Registries with missing IDs         &  772.0 &  317.0 &  166.0 &  145.0 &  135.0 &  147.0 &  124.0 &  98.0 &  17.0 &  17.0 &  28.0 &  58.0 \\
    Crude Incidence        &   58.4 &   54.3 &   59.1 &   56.3 &   58.6 &   61.6 &   62.9 &  59.5 &  63.1 &  64.4 &  61.2 &  59.5 \\
    Variation (\%) &   15.8 &    6.9 &    3.3 &    3.0 &    2.6 &    2.7 &    2.2 &   1.8 &   0.3 &   0.3 &   0.5 &   1.0 \\
    \bottomrule
    \end{tabular}
    \caption{Registries with missing IDs and crude incidence rates (cases/100,000 women) calculated considering only a portion of these as new cases (likely scenario). This equals one new diagnosis every 1.67 discharges. The last row shows the variation among these crude rates and those shown in Figure \ref{fig:incidence_both_chile}.}
    \label{tab:missing_ids_incidence_likely}
\end{table}

There was a large number of registries with missing IDs in 2007, accounting for more than a third of all registries with missing IDs. The years prior to 2007 also show a large number of discharges with missing IDs, with a yearly average of 1,820 for the 2001-2006 period. This may be due to registration errors during the first years that were corrected, resulting in a much lower number of missing IDs in the later years of the period.

As expected, the crude incidence rates obtained in both cases are higher than those shown in Figure \ref{fig:incidence_both_chile}, with the highest differences in 2007 and 2008. In the worst--case scenario, apart from these two years, all variations are lower than 5.3\%. The mean crude rate from 2007--2018 was 61.2/100,000 women, with a standard deviation of 2.6. The mean crude incidence rate previously found is 57.9/100,000 women, with a variation between the two crude rates of 5.3\%, which represents an upper bound considering the worst--case scenario for the error on the estimates shown in Figure
\ref{fig:incidence_both_chile}.

In the more likely scenario shown in Table \ref{tab:missing_ids_incidence_likely}, in addition to 2007 and 2008, all crude rates have a variation lower than 3.3\%. The mean crude rate from 2007--2018 was 59.9/100,000 women, with a standard deviation of 2.9. This represents a variation with respect to the mean crude rate found in Figure \ref{fig:incidence_both_chile} of 3.3\%, which represents an upper bound considering a more likely scenario for the error.

\begin{figure}
    \centering
    \includegraphics{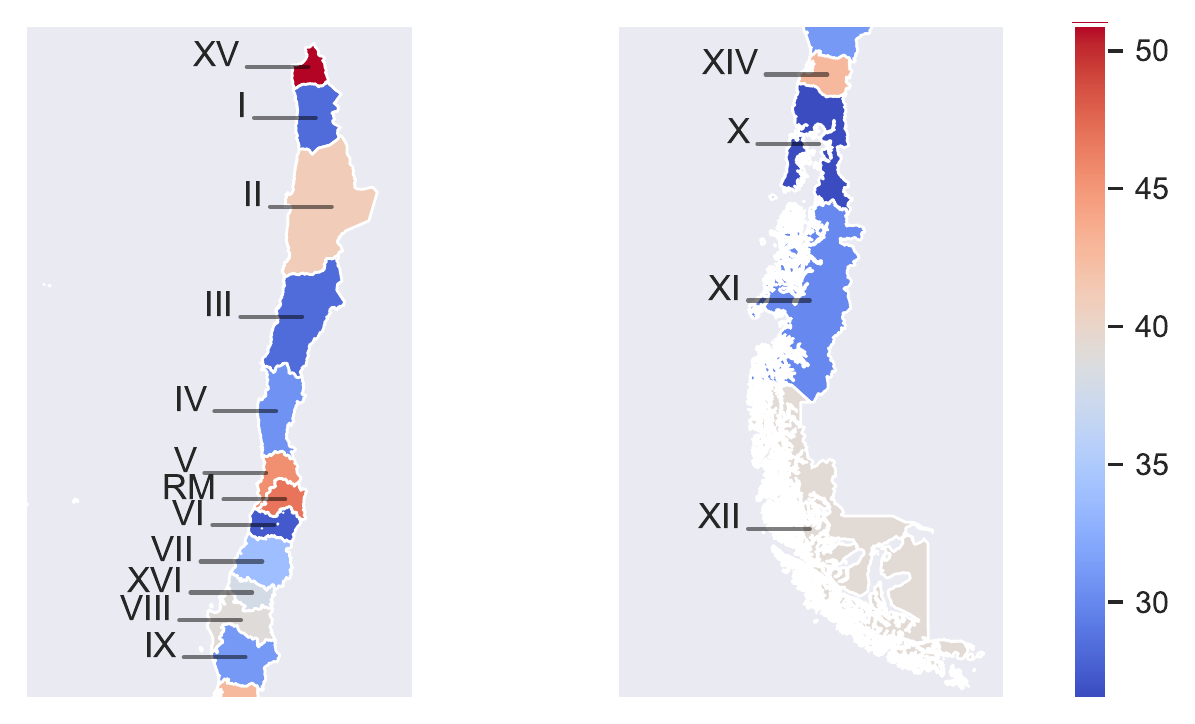}
    \caption{Mean age--adjusted incidence rates (cases/100,000 women) during 2007-2018 by geographical region. Detailed incidence values can be found in Table \ref{tab:incidence_and_mortality_by_region}.}
    \label{fig:incidence_adjusted_map}
\end{figure}

There are considerable differences in age--adjusted incidence rates among regions, as shown in Figure \ref{fig:incidence_adjusted_map}, with no clear geographical trend. Higher incidence rates are found in the northern part of the country, the Arica and Parinacota region (XV), and in the central part of the country, corresponding to the Metropolitan region (RM) and Valparaiso region (V), with age--standardized incidence rates of 50.9, 46.8 and 45.5, respectively. Interestingly, two of the regions with the lowest rates were also found in the northern area, the Atacama (III) and Tarapacá (I) regions, with incidences of 28.2 and 28.3, respectively, surpassed only by a southern and a central region, Los Lagos (X)  and O'Higgins (VI), with age--adjusted incidence rates of 26.5 and 27.3, respectively. The rest of the regions have an age--adjusted rate between 30.0 and 42.7.

Table \ref{tab:incidence_by_health_insurance} shows crude and age adjusted incidence rates by year for women affiliated with the private and public health systems. For this analysis, 56,250 out of the 62,093 registries were considered, i.e., a total of 90.6\% of all diagnosed women in the 2007--2018 period had an affiliation with the private and public health systems. Of the remaining 5,843 women, 1,944 women had armed forces health insurance, 885 women had no health insurance, and 3,014 women had missing information.


We observed that women affiliated with a private provider (ISAPRE) had higher incidence rates during the period under study, having an average age--adjusted incidence rate of 60.6/100,000 compared to 38.8/100,000 for women affiliated with the public provider (FONASA). The age--standardized incidence rate for the private health system shows greater fluctuations over time (sd of 11.0) than that for the public sector (sd of 2.0).


\enabletablerowcolor[2] 
\begin{table}
	\centering
    \begin{tabular}{lrrrr}
    \toprule
     & \multicolumn{2}{c}{ISAPRE} & \multicolumn{2}{c}{FONASA} \\
    Year & Crude & Age adjusted & Crude & Age adjusted \\
    \midrule
    2007       &         36.3 &         37.2 &        46.7 &        36.7 \\
    2008       &         41.5 &         41.9 &        49.1 &        37.9 \\
    2009       &         72.3 &         71.1 &        54.9 &        42.9 \\
    2010       &         66.0 &         65.3 &        50.5 &        37.4 \\
    2011       &         61.1 &         58.4 &        54.2 &        39.2 \\
    2012       &         64.2 &         61.2 &        56.4 &        40.6 \\
    2013       &         63.1 &         59.2 &        57.8 &        40.6 \\
    2014       &         71.3 &         66.5 &        53.3 &        36.5 \\
    2015       &         72.9 &         67.0 &        59.1 &        40.7 \\
    2016       &         84.0 &         75.1 &        57.6 &        39.3 \\
    2017       &         73.0 &         62.8 &        55.2 &        37.5 \\
    2018       &         72.3 &         61.2 &        53.6 &        36.7 \\
    \midrule
    Mean (Std) &  64.8 (13.6) &  60.6 (11.0) &  54.0 (3.7) &  38.8 (2.0) \\
    \bottomrule
    \end{tabular}
	\caption{Crude and age--adjusted incidence rates (cases/100,000 women) for women with private and public health insurance.}
	\label{tab:incidence_by_health_insurance}
\end{table}

\subsection{Mortality rates}

\begin{figure}
    \centering
    \includegraphics{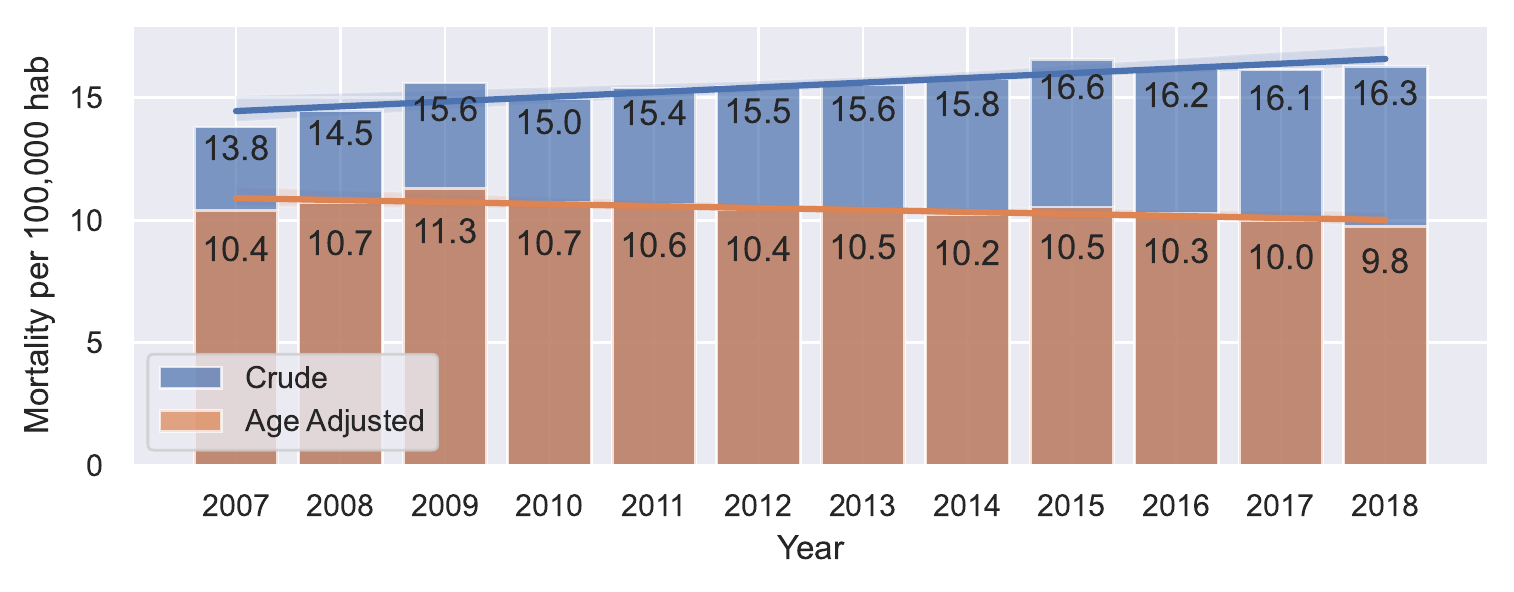}
    \caption{Crude and age--adjusted mortality rates (deaths/100,000 women) by year.}
    \label{fig:mortality_both_chile}
\end{figure}

Figure \ref{fig:mortality_both_chile} shows both crude and age--adjusted mortality rates (deaths/100,000 women) for the period under study. Crude mortality rates increased 18.1\%, from 13.8 per 100,000 women in 2007 to 16.3 per 100,000 women in 2018. On the other hand, age--adjusted mortality rates did not have significant changes during the study period, with an average standardized mortality rate of 10.5 per 100,000 women and a standard deviation of 0.4. 

\begin{figure}
    \centering
    \includegraphics{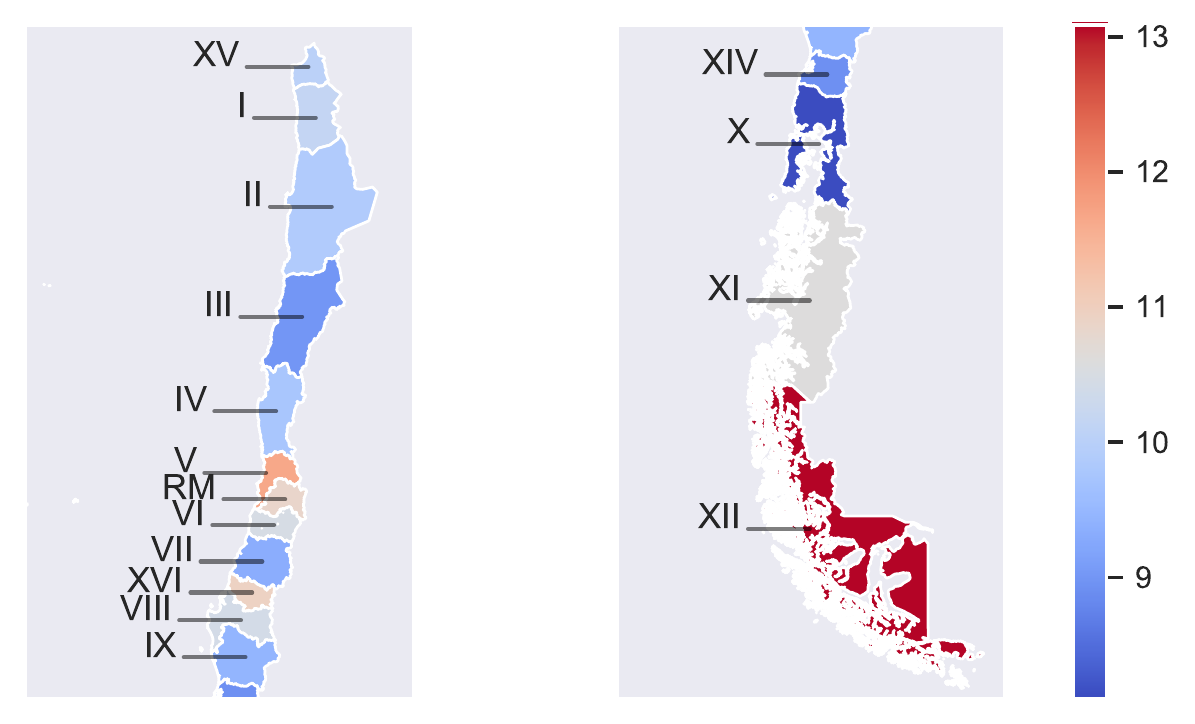}
    \caption{Mean age--adjusted mortality rates (deaths/100,000 women) during 2002-2018 by geographical region. Detailed mortality values can be found in Table \ref{tab:incidence_and_mortality_by_region}.}
    \label{fig:mortality_adjusted_map}
\end{figure}

Figure \ref{fig:mortality_adjusted_map} shows that the southern cities of Magallanes and Antártica Chilena (XII) have the highest age--adjusted mortality rate of 13.1, followed by the Valparaiso (V), Ñuble (XVI) and Metropolitan (RM) regions, with average age--adjusted rates of 11.7, 10.9, and 10.8, respectively. On the other hand, the regions with the lowest mortality rates are the southern regions of Los Lagos (X) and Los Ríos (XIV), with age--adjusted mortality rates of 8.1 and 8.9, respectively. The rest of the regions have similar mortality rates, averaging between 9.0 and 10.6 deaths per 100,000 women.

Table \ref{tab:mortality_by_health_insurance} shows mortality rates by year for women affiliated with the private and public health systems. As the national death registry does not include the health insurance of the deceased patients, we obtained them by searching the discharge database with the corresponding patient ID. Information was found for 89.3\% of all women from the death database, corresponding to 14,837 women, though the rest did not have any hospital discharge data. From this set of women, 14,057 women belong to either the FONASA or ISAPRE. The last column of Table \ref{tab:mortality_by_health_insurance}
shows  the percentage of women from the death database from which it was not possible to recover their health care provider.

We observe, in terms of crude rates, that FONASA patients have a higher mortality rate than ISAPRE patients, with rates of 14.8 and 9.8 per 100,000 women, respectively. When adjusted by age groups, FONASA rates decreased, while ISAPRE rates increased, resulting in FONASA patients having a slightly lower age--adjusted mortality rate of 9.5 compared to 10.1 for ISAPRE patients. Both systems had a rather constant trend in terms of age--standardized rates throughout the study period. 

\enabletablerowcolor[2] 
\begin{table}
	\centering
    \begin{tabular}{lrrrrr}
    \toprule
     & \multicolumn{2}{c}{ISAPRE} & \multicolumn{2}{c}{FONASA} &\\
    Year & Crude & Age adjusted & Crude & Age adjusted & Missing (\%)\\
    \midrule
    2007       &        8.4 &         9.8 &        12.4 &        9.0 &    18.0 \\
    2008       &        8.0 &         8.6 &        13.0 &        9.2 &    16.9 \\
    2009       &        9.2 &        10.3 &        15.4 &       11.0 &    13.1 \\
    2010       &        9.4 &        10.2 &        13.9 &        9.5 &    13.3 \\
    2011       &        9.8 &        10.5 &        14.7 &        9.5 &    10.8 \\
    2012       &        8.7 &         8.8 &        15.0 &        9.3 &     9.5 \\
    2013       &       10.1 &        10.6 &        15.0 &        9.5 &     8.4 \\
    2014       &        9.7 &        10.1 &        15.5 &        9.4 &     8.7 \\
    2015       &       11.7 &        11.9 &        16.1 &        9.6 &     9.3 \\
    2016       &       10.3 &         9.8 &        16.1 &        9.6 &     7.7 \\
    2017       &       11.3 &        10.5 &        15.5 &        9.2 &     8.0 \\
    2018       &       11.1 &         9.8 &        15.5 &        8.9 &     8.3 \\
    \midrule
    Mean (Std) &  9.8 (1.2) &  10.1 (0.9) &  14.8 (1.2) &  9.5 (0.5) &    10.7 \\
    \bottomrule
    \end{tabular}
	\caption{Crude and age--adjusted mortality rates (deaths/100,000 women) for women with private and public health insurance. The last columns show the percentage of registries with no health insurance data.}
	\label{tab:mortality_by_health_insurance}
\end{table}

\subsection{Case fatality rates}

\begin{figure}
    \centering
    \includegraphics{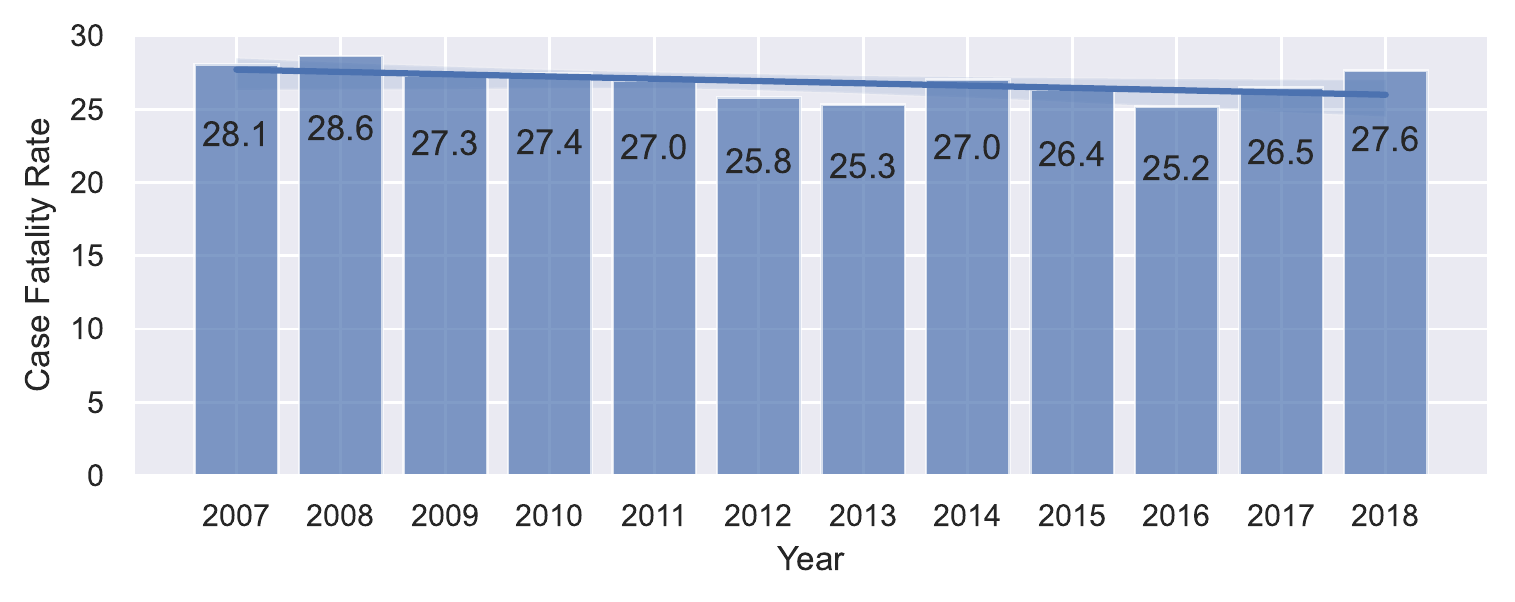}
    \caption{Case fatality rate (\%) by year for the 2007--2018 period.}
    \label{fig:fatality_chile}
\end{figure}

Case fatality rates (\%) by year for the 2007--2018 period are shown in Figure \ref{fig:fatality_chile}. The overall fatality rate remained relatively constant over time, with a mean of 26.8 and a standard deviation of 1.1.

\begin{figure}
    \centering
    \includegraphics{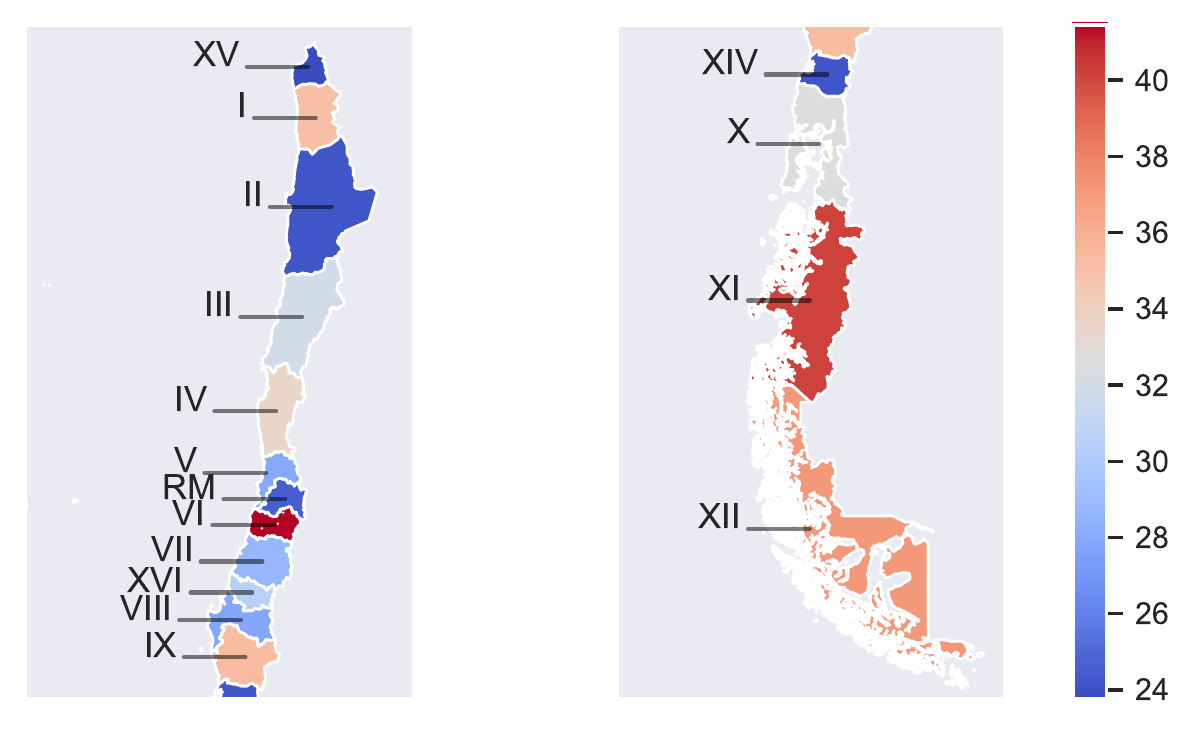}
    \caption{Mean case fatality rate (\%) over 2007-2018 by geographical region. Detailed fatality values can be found in Table \ref{tab:incidence_and_mortality_by_region}. }
    \label{fig:fatality_region_mapped}
\end{figure}

Figure \ref{fig:fatality_region_mapped} shows considerable differences in case fatality rates across regions. The regions with the highest rates are O'Higgins (VI) and Aysén (XI), with fatality rates of 41.4\% and 40.1\%, respectively. On the other hand, the regions with the lowest rates are Metropolitan (RM), Los Rios (XIV), Antofagasta (II) and Arica and Parinacota (XV), with average fatality rates of 24.5\%, 24.2\%, 24.1\% and 23.8\%, respectively. The rest of the countries' fatality rates are between 27.8\% and 37.1\%.

Table \ref{tab:fatality_by_health_insurance} shows case fatality rates for women affiliated with the private and public health systems. There is a set of women from the death database for whom there is no information about their health care provider, as shown in  Table \ref{tab:mortality_by_health_insurance}. This problem does not occur for the incidence rate estimations because the discharge database provides information on the patients' health care insurance.
 

As fatality rates are calculated based on crude mortality and incidence rates, CFRs by health insurer are also affected by this missing information. Assuming that the death registries for which this information was not available are unbiased in terms of health care insurance, the case fatality rates shown in Table \ref{tab:fatality_by_health_insurance} are good estimates. Women affiliated with ISAPREs have a considerably lower fatality rate during the period under study, with an average  of 15.7 compared to 27.5 for women affiliated with the FONASA.

Neither of the systems had a statistically significant trend (increasing or decreasing), though the CFRs from ISAPRE affiliates had greater fluctuations, with a standard deviation of 3.1, while FONASA had a standard deviation of 1.0.

\begin{table}
    \centering
    \begin{tabular}{lcc}
    \toprule
    Year &  ISAPRE &  FONASA \\
         &  (private) & (public)\\
    \midrule
    2007       &        23.3 &        26.5 \\
    2008       &        19.3 &        26.4 \\
    2009       &        12.7 &        28.1 \\
    2010       &        14.3 &        27.4 \\
    2011       &        16.1 &        27.1 \\
    2012       &        13.5 &        26.6 \\
    2013       &        16.0 &        26.0 \\
    2014       &        13.6 &        29.2 \\
    2015       &        16.1 &        27.2 \\
    2016       &        12.2 &        27.9 \\
    2017       &        15.5 &        28.1 \\
    2018       &        15.3 &        29.0 \\
    \midrule
    Mean (std) &  15.7 (3.1) &  27.5 (1.0) \\
    \bottomrule
    \end{tabular}
    \caption{Case fatality rate (\%) by health insurance}
    \label{tab:fatality_by_health_insurance}
\end{table}

\subsection{Survival rates}

As discussed in Section \ref{data}, we considered the 58,254 patients who had a first discharge during the period under study to calculate survival rates. The 3,839 patients who died during the study period without hospital discharge were not considered because the onset of the disease was unknown. Although these patients' data might reduce the survival rates (assuming that they did not live long enough to receive treatment), this set of women has a distribution of health care insurance similar to that of the overall set of patients, and therefore, its inclusion would influence the survival rates of patients from ISAPRE and FONASA evenly, causing the survival rate difference between both groups to be kept. 
Finally, 19 patients were found to have inconsistencies when matching both databases, with discharge dates later than their death date, and therefore, were eliminated from the survival analysis, resulting in a  survival database of 58,235 records.  

\subsubsection{Kaplan--Meier estimations}

\begin{figure}
    \centering
    \includegraphics{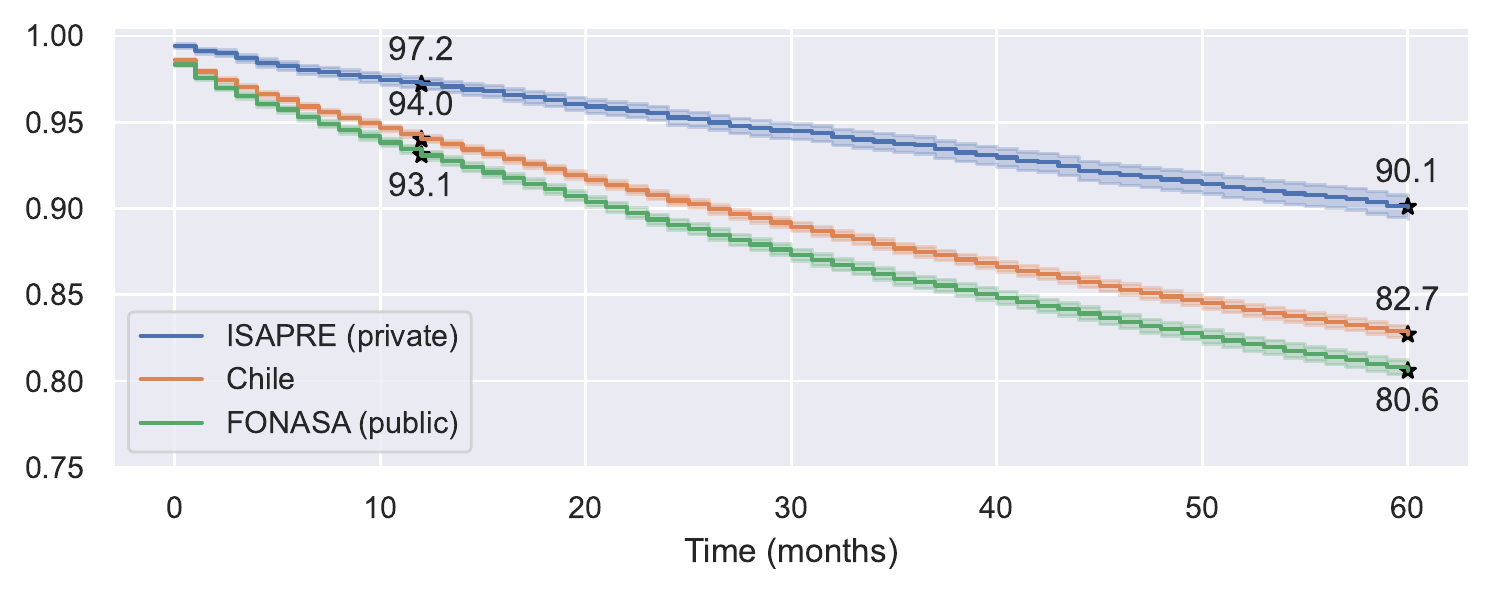}
    \caption{Kaplan--Meier survival curves for Chilean women separated by health insurance system (public vs. private).}
    \label{fig:KM_prevision}
\end{figure}

The Kaplan--Meier estimator was computed considering 58,235 observations, of which 50,126 were right-censored observations. Figure \ref{fig:KM_prevision} shows the survival rates when considering all patients under the study period and separated by private and public health care insurance systems.
The log rank test confirmed that the survival curves for the private and the public insured patients  and all women were significantly different (p<0.001). 

The estimated one--year survival rate 95\% confidence intervals are [0.931 $\pm$ 0.002] for FONASA, [0.972 $\pm$ 0.003] for ISAPRE patients and [0.940 $\pm$ 0.002] when considering all women. The 95\% confidence intervals for the five--year survival rate were [0.806 $\pm$ 0.004], [0.901 $\pm$ 0.007], and [0.827 $\pm$ 0.004] for FONASA, ISAPRE and all women, respectively. Thus, women in the private system have a five--year survival rate that is 12\% higher than that for women in the public system.

\begin{figure}
    \centering
    \includegraphics{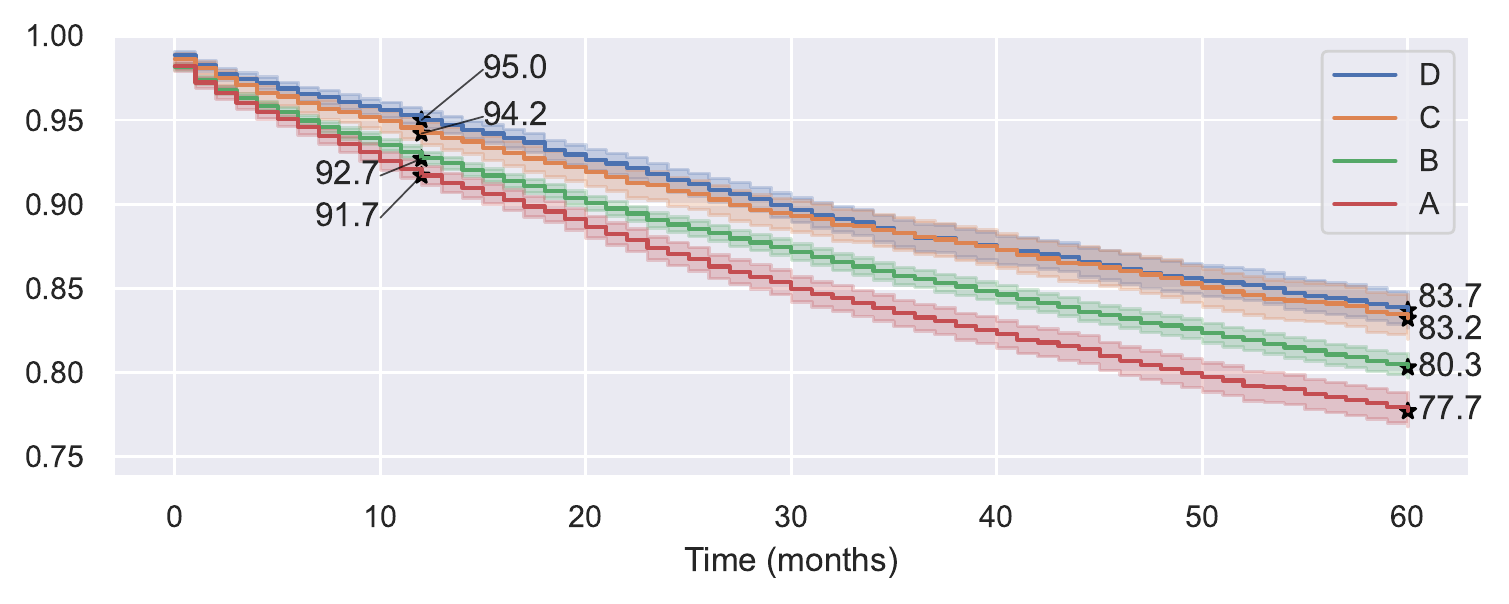}
    \caption{Kaplan--Meier survival curves for FONASA patients by benefit segment.}
    \label{fig:KM_tramo_fonasa}
\end{figure}

A similar result was obtained when analyzing the survival curves for patients in each segment within the public health system as shown in Figure \ref{fig:KM_tramo_fonasa}.
We observe that women with segments C and D health insurance benefits have similar survival rates, which are higher than those for women in FONASA segments A and B.
Women in benefit segment A have the worst survival rates out  of the four groups. The 5-year survival rates are 0.837, 0.832, 0.803 and 0.777 for benefit segments D, C, B and A, respectively. We notice that all these survival rates are lower than those for women in the private system. 

\begin{figure}
    \centering
    \includegraphics{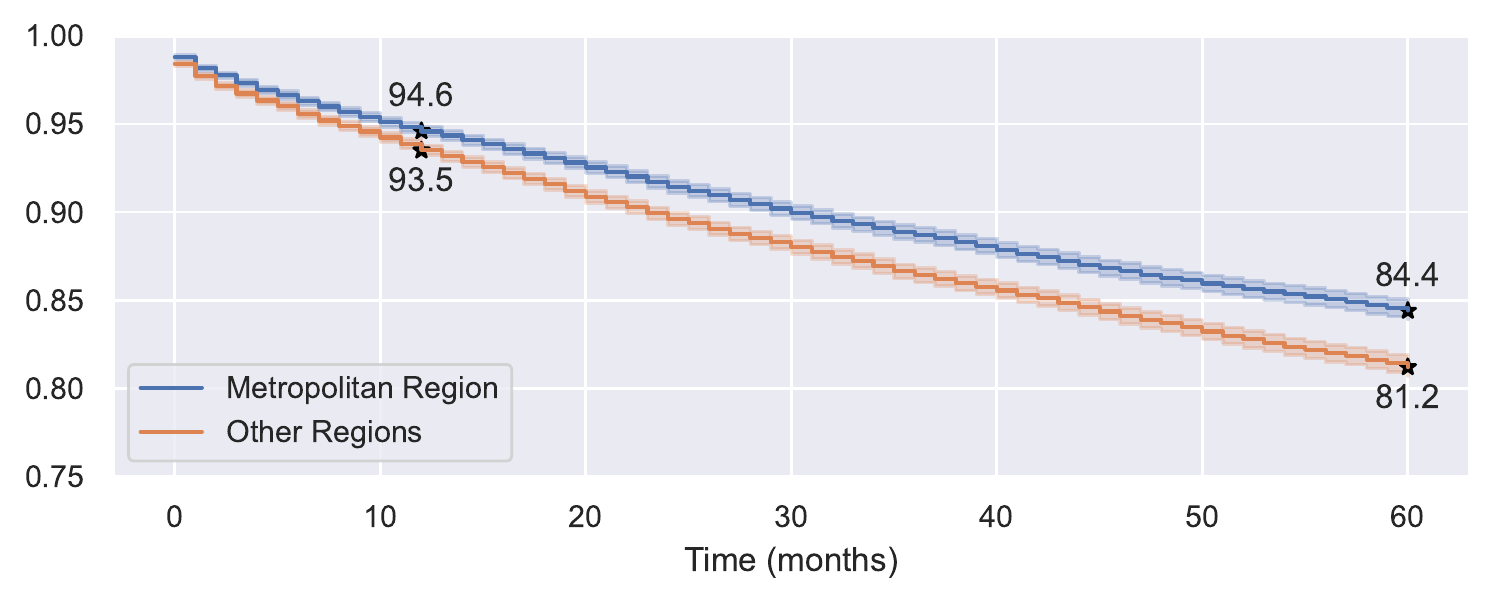}
    \caption{Kaplan--Meier survival curves for Chilean women, separated by their region of residence.}
    \label{fig:KM_region}
\end{figure}

Women from the Metropolitan region have a higher survival rate than women from other regions. We observe this in Figure \ref{fig:KM_region}, with five--year survival rates of [0.844 $\pm$ 0.005] and [0.812 $\pm$ 0.005] for women from the Metropolitan region and other regions, respectively. The two curves shown in this figure are significantly different (p<0.001).

\subsubsection{The Cox proportional hazards model}

Using the procedure described in Appendix \ref{app:cox_regression}, we estimated a Cox proportional hazards model with 11 selected variables out of 26 considered. The results are summarized in Table \ref{tab:cox_coefs}, where the second column contains the coefficient with its associated confidence interval and the third column shows the p value for the null hypothesis corresponding to equality of the base and the affected covariate. 

The Cox model allows us to evaluate the survival rate of a patient identified through these 11 selected covariates, such as dummy variables for private and public health insurance, benefit segment A, age, squared age, year of diagnosis and residence in some regions. The Cox model also allowed us to compare the survival curves by modifying selected variables while keeping the rest constant. For the results below, the variables that remain unmodified are set as the median of the data.

\begin{table}
    \centering
    \begin{tabular}{lll}
    \toprule
    Variable                    &      Coefficient & p value \\
    \midrule
    Year of diagnosis    &  -0.051 $\pm$ 0.007 &  <0.005 \\
    Age                  &  -6.438 $\pm$ 0.903 &  <0.005 \\
    Squared age          &   7.119 $\pm$ 0.745 &  <0.005 \\
    FONASA               &   0.259 $\pm$ 0.093 &  <0.005 \\
    ISAPRE               &  -0.285 $\pm$ 0.112 &  <0.005 \\
    FONASA beneficiary A &   0.251 $\pm$ 0.054 &  <0.005 \\
    Region XV            &  -0.477 $\pm$ 0.206 &  <0.005 \\
    Region II            &  -0.244 $\pm$ 0.145 &  <0.005 \\
    Region V             &  -0.147 $\pm$ 0.071 &  <0.005 \\
    Region XIII          &   -0.203 $\pm$ 0.05 &  <0.005 \\
    Region VI            &   0.165 $\pm$ 0.107 &  <0.005 \\
    \bottomrule
    \end{tabular}
    \caption{Results for the Cox proportional hazard regression model.}
    \label{tab:cox_coefs}
\end{table}

Hazard ratios (or odds ratios) between two sets of characteristics can be evaluated through Cox regression. Table \ref{tab:cox_ratios} shows the hazard ratio for some specific variables. For example, we observe that women 40 and 50 years old have almost the same hazard and that women 60 years old have a slightly higher hazard of dying due to breast cancer than women 40 years old. The hazard ratios for an increase of 10 and 20 years of age increase; thus, we observe that the hazard of 60--year--old women is 1.15 times that of 50--year--old women and that the hazard of 70--year--old women is 1.52 times that of 50--year--old women. The hazard of FONASA beneficiaries from the benefit segment B-C-D is 1.72 times that of ISAPRE beneficiaries. FONASA patients how benefit from segment A have an even higher hazard, which is 1.29 times higher than that of FONASA patients who benefit from segments B-C-D. We observe that women diagnosed in 2018 have almost half the hazard (0.57) of women diagnosed in 2007.

\begin{table}
    \centering
    \begin{tabular}{lr}
    \toprule
    Coefficient &  Hazard ratio \\
    \midrule
    Age 40 $\rightarrow$ Age 50                           &          1.00 \\
    Age 40 $\rightarrow$ Age 60                           &          1.15 \\
    Age 50 $\rightarrow$ Age 60                           &          1.15 \\
    Age 50 $\rightarrow$ Age 70                           &          1.52 \\
    ISAPRE $\rightarrow$ FONASA Beneficiary BCD           &          1.72 \\
    FONASA Beneficiary BCD $\rightarrow$ FONASA Beneficiary A &          1.29 \\
    ISAPRE $\rightarrow$ FONASA Beneficiary A             &          2.22 \\
    Year 2007 $\rightarrow$ Year 2010                     &          0.86 \\
    Year 2007 $\rightarrow$ Year 2013                     &          0.74 \\
    Year 2007 $\rightarrow$ Year 2018                     &          0.57 \\
    Metropolitan Region $\rightarrow$ Region XV                     &          0.76 \\
    Metropolitan Region $\rightarrow$ Region II                     &          0.96 \\
    Metropolitan Region $\rightarrow$ Region V                      &          1.06 \\
    Metropolitan Region $\rightarrow$ Region VI                     &          1.44 \\
    Metropolitan Region $\rightarrow$ Any other region              &          1.22 \\
    \bottomrule
    \end{tabular}
    \caption{Hazard ratios obtained through the Cox model.}
    \label{tab:cox_ratios}
\end{table}

\begin{figure}
    \centering
    \includegraphics{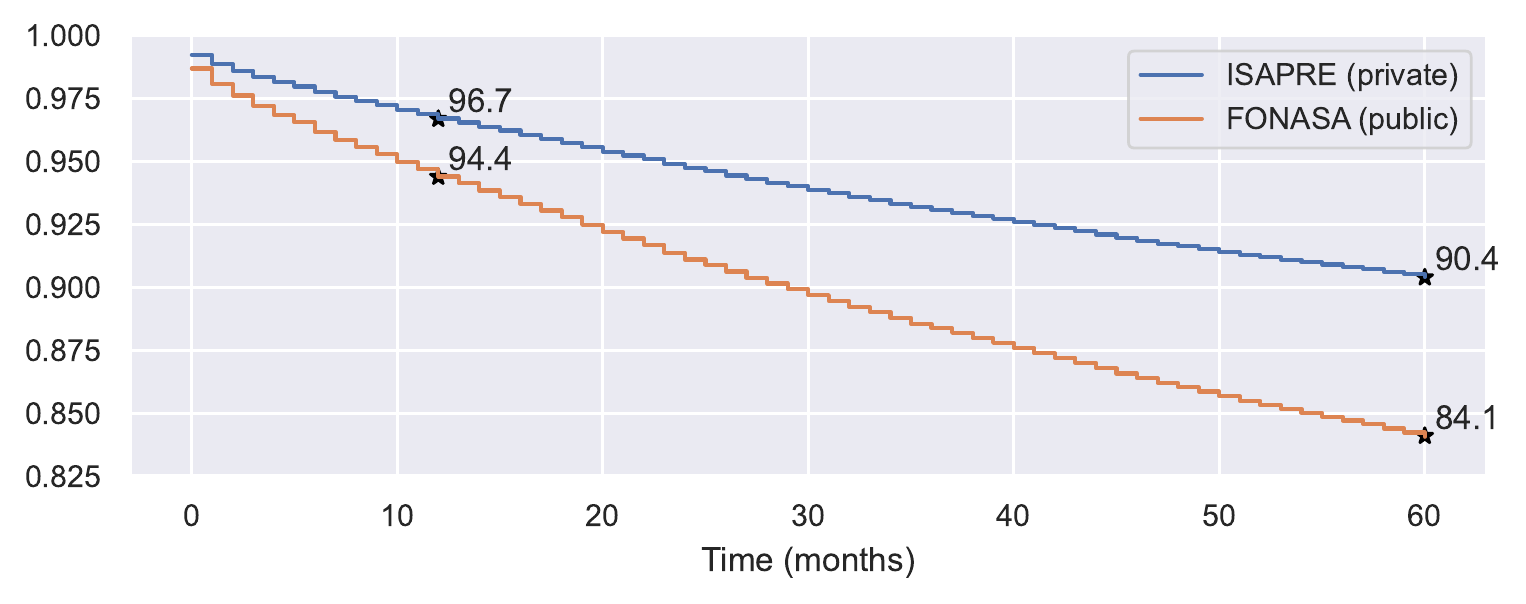}
    \caption{Cox survival curves adjusted by health insurance: FONASA (public) vs. ISAPRE (private) health plans.}
    \label{fig:cox_prevision}
\end{figure}

Figure \ref{fig:cox_prevision} presents the Cox survival curves for women affiliated with public and private health insurance systems. We observe that the survival rate trends have similar behavior as those obtained by the Kaplan--Meier estimations, with a survival curve for the patients in the private system that is significantly higher than that for patients affiliated with the public health insurance. The Cox regression estimates the one--year survival rates to be 0.967 and 0.944 for  patients in the private and public sectors, respectively. On the other hand, the Cox regression estimates the five--year survival rates to be 0.904 and 0.841, respectively. 

\begin{figure}
    \centering
    \includegraphics{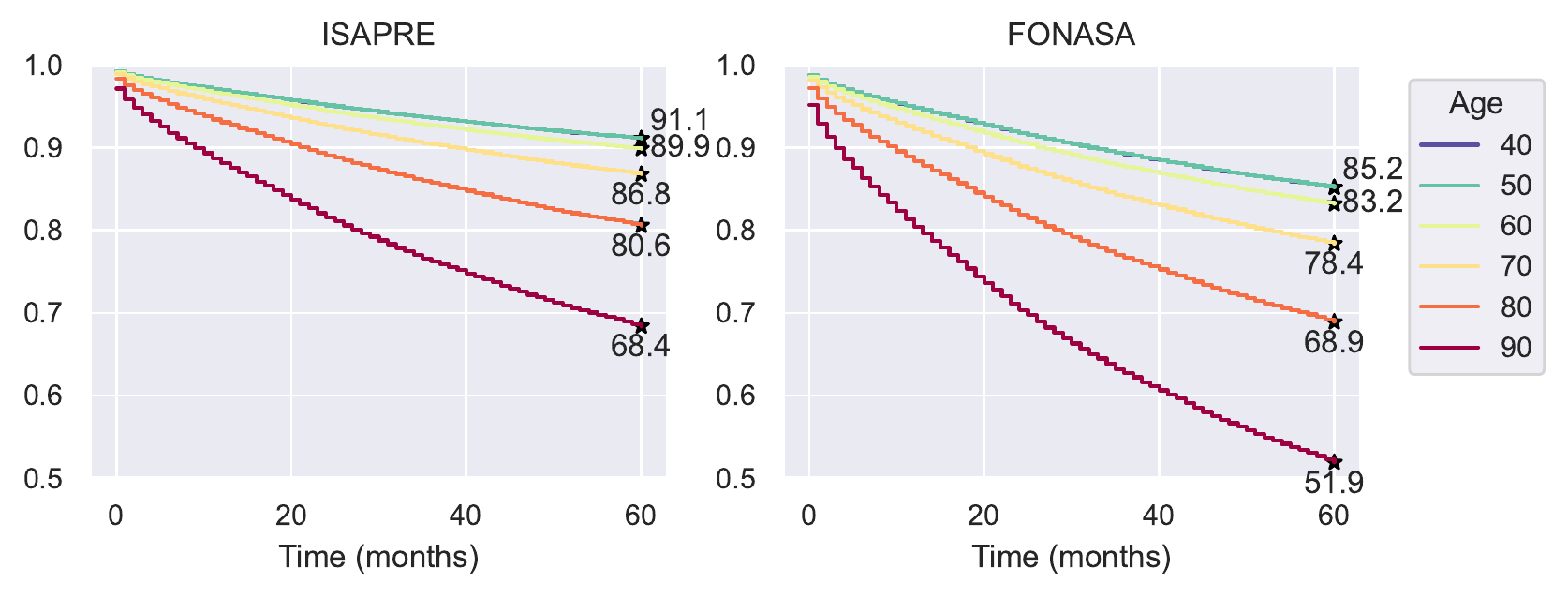}
    \caption{Survival curves obtained by Cox regression, adjusted by age, for each type of health insurance. The five--year survival rates for 50--year--old women were omitted from the graph because they were very similar to those of 40--year--old women.}
    \label{fig:cox_age}
\end{figure}

Figure \ref{fig:cox_age} shows the survival curves for different ages of women for both health insurance systems. We noticed that the difference in survival rates between ISAPRE and FONASA patients increased for older people. The five--year survival rates for women aged 40 were 0.91 and 0.85 for the ISAPRE and FONASA systems, respectively, with a survival difference of 0.06. On the other hand, the five--year survival rates for women aged 80 are 0.81 for ISAPRE and 0.69 for FONASA, with a survival difference of 0.12. The five--year survival rates are shown in Table \ref{tab:cox_age}.

\begin{table}
    \centering
    \begin{tabular}{lrrrrrr}
    \toprule
    Age &     40 &     50 &     60 &     70 &     80 &     90 \\
    \midrule
    ISAPRE &  0.911 &  0.911 &  0.899 &  0.868 &  0.806 &  0.684 \\
    FONASA &  0.852 &  0.852 &  0.832 &  0.784 &  0.689 &  0.519 \\
    \bottomrule
    \end{tabular}
    \caption{Five--year survival rate predicted by the Cox model by age and health care system.}
    \label{tab:cox_age}
\end{table}

\section{Discussion}

Breast cancer is one of the two cancers, together with gallbladder cancer \cite{minsal2016}, that most frequently affects Chilean women. It is also the leading cause of death from cancer for Chilean women  \cite{minsal19plancancer}. In 1995, the Cancer Unit of the Ministry of Health formed the National Breast Cancer Commission to focus on the design of a National Breast Cancer Program to better organize the care for breast diseases \cite{minsal98}. The commission defined referral, diagnostic, and treatment guidelines and carried out several activities across the country to improve breast cancer patients' care. 
As of 2004, a major reform in the Chilean health system was introduced: the Universal Access Plan for Explicit Guarantees (AUGE), later renamed GES, which aimed to decree timely and quality health care without discrimination and with financial protection to all Chilean residents. Thus, the Ministry of Health, through this plan, guarantees the coverage of a  number of diseases (It started in 2005 with 25 pathologies and grew to 85 in 2022). Breast cancer was part of the initial pathologies included in 2005; in addition to the quality, financial, and opportunity guarantees, it also defined the complete protocol of  medical treatments, including procedures and drugs to diagnose and treat breast cancer in all its stages. This protocol was updated in 2010 and 2015 according to the new treatment options that had been developed over time. In particular, in 2015, the high--cost drug trastuzumab was included for all Chilean women with HER2--positive breast cancer. The GES plan activates the above guarantees at the moment of a suspicion of breast cancer for people aged 15 years and over with suspected, diagnosed or recurrent breast cancer \cite{minsal19auge}).

\vspace{1mm}





Regarding early breast cancer detection, the Ministry of Health currently \textit{recommends} an annual mammogram for women 50 to 69 years old (www.diprece.minsal.cl). However, despite this recommendation, it only guarantees triannual screening mammograms free of charge for all women  50 to 59 years old, independent of their health insurer \cite{minsal19auge}. The National Socioeconomic Characterization Survey, CASEN 2017 \cite{Casen_2017}, shows that in 2017, only 53.4\% and 33.1\% of women aged 50 or more in the private and public health systems had a screening mammogram during the last year, respectively. These figures increase to 76.3\% and 57.1\% for a screening mammogram in the last three years.

These results might have changed with the implementation, as of July 2019, of a MINSAL strategy of increasing the coverage of mammograms in public primary health care, with important investments in new mammographers and the hiring of human resources in Family Health Centers (CESFAM) and community hospitals. One specific goal of this plan was to improve the coverage of breast cancer screening mammograms. For this purpose, the exams are taken by medical technologists in the aforementioned health care facilities and are sent electronically to the \textit{mammography cell} (digital hospital), where the report is made by radiologists specialized in breast imaging, totaling 22\% of the mammography reports in the country from January to September 2021 \cite{hospital_digital}.

However, we note that in Chile there is no screening program as defined by the World Health Organization \cite{world2020screening}. In particular, there is no information system in place linked to population registries that would allow the screenings to be offered to the eligible population using a call and recall system and have a registry of all screenings performed in the country.



\vspace{1mm}

With regard to epidemiological indicators, Chile does not have a national cancer registry. Hence, key estimates, such as incidence and case fatality rates, have been extrapolated from population records created in 1998 and updated until 2012 in the regions of Antofagasta, Biobío, Valdivia and Concepción \cite{minsal_informe_rpc}. These regional data are insufficient to have a comprehensive view of the evolution of cancer at the national level due to differences in health care coverage, ethnicity, and regional particularities, among others. Furthermore, there is no reliable information from 2012 onward. Thus, we observe that most of the published research focuses on mortality indicators, of which there is a reliable national death database. However, regarding incidence rates, there are few studies with assumptions that might lead to inaccurate estimates. For example, the MINSAL study in 2019 \cite{minsal_informe_incidencia} reported a constant drop in incidence rates, which is inconsistent with the data of the current state of breast cancer in Chile. This drop is obtained by the actual decrease in mortality rates and the assumption of a constant fatality rate over the study period, without considering the improvements in breast cancer treatments over time. These estimates feed the Chilean data in GLOBOCAN  \cite{globocan_methods}.

In what follows, we discuss the main results of our research. For this, we divide the discussion into three areas: estimates of epidemiological indicators, the impact of the GES plan on key health care indicators, and limitations and future research of this project.
 
\vspace{5mm}

\subsection{Incidence, mortality, case fatality, and survival rates}
 
The incidence age-standardized rates found for the 2007--2018 period are lower than crude rates for the same period. This is largely explained by the demographic composition of the Chilean population. Breast cancer worldwide is significantly more common in older women and is less likely in women younger than 30 years old. This distribution is also observed in Chile. As Chile has a higher proportion of older women than the standardized population \cite{ine_poblacion,IARC_local_data}, it is expected to have higher crude incidence rates. A similar result is found for mortality rates in the 2007--2018 period. The Global Cancer Observatory (GLOBOCAN) reported the Chilean incidence and mortality age--adjusted rates as 37.4 and 10.2 per 100,000 women in 2020, respectively \cite{globocan_data}. Our study estimates incidence and mortality age adjusted rates as 44.0 and 10.5 per 100,000 women, respectively, with a constant trend over the 2007--2008 period. We observe that our estimates of the mortality rate are consistent with those reported by GLOBOCAN, while our estimation of the incidence rate is higher than that of GLOBOCAN. The trends in age--standardized incidence reported by GLOBOCAN show that almost all countries have either had a constant trend (in countries such as Canada, Italy and Poland) or a growing trend (in countries such as Germany, Thailand or Korea) in the last 20 years. We remark that the constant trend found in our study for age--standardized incidence rates at the national level is consistent with this global trend observed by GLOBOCAN and differs from the decreasing trend found by the Ministry of Health \cite{minsal_informe_incidencia}. It is important to note that the GLOBOCAN platform reports data on incidence and mortality based on information provided by countries and has no responsibility over their quality \cite{ferlay2019estimating}.


\vspace{1mm}

We also observe from Table \ref{tab:incidence_by_health_insurance} a much higher age--adjusted incidence rate in privately insured patients (ISAPRE) compared to those with public health care insurance. As discussed in the Introduction, these two groups of patients have important differences, such as income, lifestyle, diet composition, comorbidities and use of hormonal therapies, among others, which might explain these incidence differences \cite{McTiernan2005}. Another plausible explanatory hypothesis for this difference might rely on the  higher use of spontaneous screening among privately insured patients \cite{Casen_2017}, increasing the detection at early stages and at early ages, or even in older women who otherwise would not be tested \cite{Puliti2012}.

The estimated incidence and mortality rates shown by region in Figures \ref{fig:incidence_adjusted_map} and \ref{fig:mortality_adjusted_map} show the great differences in breast cancer epidemiology along the Chilean territory. These differences indicate that the reality of each region must be taken into account when designing public health measures. Moreover, public measures and other related policies may be misdirected if they are designed and evaluated only through a few local registries, such as those from Antofagasta, Biobío and Los Rios regions. This highlights the need for a national--level cancer registry from which adequate data can be obtained.

\vspace{1mm}

Our study shows an estimate of case fatality rate of 26.8\% at a national level, with a constant trend over the study period. Similar to incidence and mortality, there is a high variation among regions, without any clear distribution. More importantly, there is a clear difference in the case fatality rates of the FONASA and ISAPRE, with averages for the 2007--2018 period of 27.5\% and 15.7\%, respectively. This is because patients with public insurance have a fatality rate that is 1.75 times higher than those with private insurance. 

\vspace{1mm}

We estimated the five--year survival to be 0.827 ($\pm$0.004) at the national level. For the public and private health care sectors, we estimated five--year survival rates of 0.806 ($\pm$0.004) and 0.901 ($\pm$0.007), respectively. We also found the five year survival rate for women in the Metropolitan region to be 0.844 ($\pm$0.005). On the other hand, the five--year survival for women from other regions is 0.812 ($\pm$0.005).


Survival differences were also found comparing women from the Metropolitan region to women from other regions. Patients from the Metropolitan region had better survival rates. This difference may be associated with the availability of specialized health care centers and specialists, which are known to be concentrated in metropolitan areas \cite{GUILLOU2011}.

The hazard ratios obtained in our study show, as expected, that there is an important impact of age on survival, which grows for older ages. Thus, we observe that the hazard ratio for a 20--year difference between 40- and 60-year-old women is 1.15 (that is, women of 60 years old have a slightly higher risk of death due to breast cancer), while the same 20-year difference between 50- and 70-year-old women leads to a hazard ratio of 1.52. It is also relevant to highlight the hazard ratio between private and public insured patients of 1.72, i.e., women from the public health care system have a risk 1.7 times higher than that of women from the private health care system.

Hazard ratios associated with year of diagnosis show an improvement in survival from 2007 onward. This may be attributable to advances in breast cancer treatment due to the systems organization and the effect of the health care reforms implemented in previous years. One such policy is the Ricarte Soto law, which is related to access to high-cost drugs and treatments.

By looking at hazard ratios between the Metropolitan region and other regions, we observe that the women in the Metropolitan region have a similar risk to the regions of Antofagasta (II) and Valparaiso (V), with hazard ratios close to 1. The northern region of Arica y Parinacota shows the smallest breast cancer risk. In addition to these regions, the Metropolitan region has a lower risk than any of the other regions.

\subsection{GES plan and health care inequities }
\label{dis_ges}

Most of the evaluations of the GES plan have focused on compliance of the explicit guarantees offered by the plan: access, opportunity, financial protection, and quality \cite{monitoreoges2018}. In terms of opportunity, in 2017, 99.6\% of the GES services fulfilled this guarantee, with 92.8\% of them before the established  deadline. On the other hand, inspections by the Comptroller General of Chile (CGR) ensure that a series of quality measures are fulfilled by the health care facilities. However, ultimately, the effectiveness of the GES must be reflected in the improvement of health indicators, such as patients survival rates and case fatality rates. As discussed in the Results section, we observed a marked difference in survival rates between insured patients in the public and private health systems. Thus, according to the Cox regression, at the age of 60, the five--year survival rates for FONASA and ISAPRE patients were 0.83 and 0.90, respectively, with a difference of 0.07. Meanwhile, at the age of 80, the five--year survival rates for FONASA and ISAPRE patients were 0.69 and 0.81, respectively, increasing the difference to 0.12. All these differences were statistically significant (p<0.001).

Multiple reasons contribute to this inequity between private and publicly insured patients. In what follows, we discuss some of them; however, they should be investigated in greater depth to develop and implement appropriate public policies to reduce this gap.

\begin{itemize}
    \item ISAPRE and FONASA patients not only differ in the health insurer of their choice, but also conform to two different sociodemographic group norms in terms of economic income and education and eventually comorbidities and age. For example, more than 85\% of the people from the five lower income deciles belong to FONASA, while only 25\% of people in the highest income decile belong to FONASA \cite{Casen_salud}. Likewise, income is also related to education, where the three lower income deciles have an average of less than 10 years of schooling, while people from the highest decile have an average of more than 15 years of schooling \cite{Casen_educacion}. These differences might translate into differences in the opportunity to consult a doctor, in adherence to treatments, and in treatment outcomes due to comorbidities, among others.
 
    \item The GES plan guarantees an adequate diagnosis and treatment, defined in clinical guidelines, and is periodically updated to incorporate new validated therapeutic options. However, there might be a lag in the opportunity of when these updates are incorporated. For example, the high-cost drug trastuzumab was included for all Chilean women with HER2--positive breast cancer in 2015, nine years after it was approved for its use by the FDA. Consequently, although quality treatment is guaranteed for all breast cancer patients, there is a time gap between the appearance of new drugs adequately supported by good evidence and their incorporation in the GES plan. Therefore, patients having a higher purchasing power most likely complement their treatments with these new drugs outside of the GES plan, leading to higher survival rates.
    


    \item Early detection is a key factor for survival rates: cancer detected in stages 1 and 2 have a five--year survival rate of 99\%, while survival rates when cancer is detected in stages 3 and 4 are 86\% and 28\%, respectively \cite{acs21}. According to data from 2017, 55.1\% of women in the public health system had a screening mammogram in the last three years compared to 71.4\% for women in the private system \cite{Casen_salud}, which might translate into significant inequities in the cancer stage at the moment of diagnosis between women in the two insurance systems. As mentioned before, Chile does not have a national cancer registry, and therefore, staging of breast cancer at diagnosis is not available to confirm this hypothesis. As part of our ongoing research, we are estimating cancer staging at diagnosis using databases of private and public hospitals.
        
\end{itemize}

\subsection{Study strength and limitations}

To the best of our knowledge, this is the first study on breast cancer incidence and survival rates at the national level using publicly available data from several national registries on death records and hospital discharges. Incidence has been reported by GLOBOCAN, estimating a 2020 incidence rate of 37.4. GLOBOCAN's estimate is lower than the estimation of this study, and even more so considering the constant trend shown in Figure \ref{fig:incidence_both_chile}, with an average incidence of 42.3. Furthermore, this study highlights the important differences in key health care indicators for women across regions, health care insurance plans, and age.

This study also has some limitations. First is the limitation of the quality of the national database used for hospital discharges and deaths. As discussed previously, there are missing and conflicting data and ambiguity in the cause of death in some death certificates. Furthermore, we used a hospital discharge database, and therefore, there might be some patients diagnosed with breast cancer who are never hospitalized when receiving treatment (for example, those who do not have surgery). Thus, incidence rates might be underestimated.

\section{Conclusions}

The methodology used in this study presents an alternative for obtaining incidence, mortality, case fatality, and survival rates from public databases, which generates results consistent with other measures, specially in the absence of a national cancer registry, such is the case in Chile.


The reasons that explain some of the significant  differences found  in survival rates, especially between private and publicly insured patients, are part of our ongoing research. However, these results have already  highlighted critical inequalities in terms of health outcomes between these two groups of patients. This is  particularly relevant when a new health system is being discussed for Chile.

\subsectionanum{Acknowledgement}

The authors gratefully acknowledge financial support from ANID PIA/APOYO AFB180003 and CMM-Conicyt PIA AFB170001.

\bibliography{main}

\newpage
\begin{appendixd}

\section{Other breast cancer diagnoses}\label{app:other_diagnostics}

After selecting discharges with a primary diagnosis code of breast cancer, there were still a considerable number of breast cancer death registries that did not have any associated discharge registry (5,230 deaths). However, there are discharge registries for causes other than breast cancer that are not typically included in a breast cancer incidence analysis. These are health problems that can arise due to the progression of the disease or its treatment (surgery, chemotherapy, radiotherapy) and therefore will be included under certain specific conditions. Thus, these diagnostics were taken into consideration only when they belonged to a patient who died because of breast cancer. We group these diagnostics into three groups:

\begin{enumerate}
    \item Discharge registries directly attributable to breast cancer and its treatment. This item includes examination, treatment by chemotherapy and radiotherapy, and different breast--related diseases and issues, which may be due to breast cancer misdiagnosis. These diagnostics are D486, D24X, Z123, Z803, Z853, Z031, Z080, Z081, Z082, Z087, Z088, Z089, Z129, Z400, Z510, Z511, Z512, Z515, Z809, Z859, and Z860.
    \item Diagnostics of other cancers and malignancies associated with breast cancer. This item includes secondary tumors and tumors of unspecified places. These diagnostics are C798, C782, C795, C793, C787, C786, C412, C800, D382, C792, C709, C383, C799, D059, C770, D420, C796, C414, C500, C413, C771, C728, C967, C779, D383, C399, C773, C781, C783, and C700.
    \item Other diagnostics might be attributable to symptoms of breast cancer and its treatment. They are considered only if such discharge is close enough to the death registry. Each diagnosis has a different period of time to be associated with death and is shown in Table \ref{tab:other_diagnostics}. The absence of a period means that the diagnosis is always included independent of its gap from the death registry.
\end{enumerate}

\enabletablerowcolor[2]
\begin{table}
    \centering
    \begin{tabular}{lr}
    \toprule
    Diagnostic Code &  Relation period (years) \\
    \midrule
    D649   &                  2 \\
    G039   &                  1 \\
    G540   &                  2 \\
    G939   &                  1 \\
    G952   &                  2 \\
    I495   &                 - \\
    I891   &                  1 \\
    I972   &                 - \\
    J80X   &                  2 \\
    J90X   &                  2 \\
    J948   &                  2 \\
    J960   &                  2 \\
    J969   &                  1 \\
    J984   &                  2 \\
    J989   &                  2 \\
    M532   &                  4 \\
    M544   &                  4 \\
    M546   &                  4 \\
    M549   &                  4 \\
    M808   &                  4 \\
    M844   &                  4 \\
    N63X   &                 - \\
    N645   &                 - \\
    N648   &                 - \\
    N649   &                 - \\
    \bottomrule
    \end{tabular}
     \hspace{2cm} \begin{tabular}{lr}
    \toprule
    Diagnostic Code &  Relation period (years) \\
    \midrule
    N850   &                 - \\
    R060   &                  1 \\
    R17X   &                  1 \\
    R18X   &                  1 \\
    R51X   &                  1 \\
    R53X   &                  1 \\
    S220   &                  4 \\
    S320   &                  2 \\
    S323   &                  4 \\
    S325   &                  4 \\
    S327   &                  4 \\
    S328   &                  4 \\
    S423   &                  4 \\
    S720   &                  4 \\
    S721   &                  4 \\
    S722   &                  4 \\
    S723   &                  4 \\
    S724   &                  4 \\
    S728   &                  4 \\
    S729   &                  4 \\
    T08X   &                  4 \\
    T12X   &                  4 \\
    T142   &                  4 \\
    T932   &                  4 \\
    \bottomrule
    \end{tabular}
    \caption{CDI-10 codes and relation periods of the health problems to be considered as the breast cancer debut. }
    \label{tab:other_diagnostics}
\end{table}
\disabletablerowcolor

Through the addition of these diagnoses, 11,533 discharge registries were added to the discharge database, which corresponds to 1,219 patients.

\begin{landscape}

\section{Incidence, Mortality and Fatality Results}\label{app:result_tables}

\subsection{Breast cancer incidence and mortality by geographical region}

In table \ref{tab:incidence_and_mortality_by_region} displays the mean incidence over the period 2007--2018 and mortality over the period 2007--2018 for each of the 16 Chilean regions. Both are presented as age--adjusted and crude rates. They are sorted from north to south. 

\enabletablerowcolor[2] 
\begin{table}
	\centering
	\caption{Average case fatality rate (\%), incidence and mortality (cases/100,000 women) over the period 2007--2018 by region.}
	\begin{tabular}{llrrrrr}
    \toprule
     &  &  & \multicolumn{2}{c}{Incidence} & \multicolumn{2}{c}{Mortality} \\
    Region           & Region Name   & Case Fatality Rate  &   Crude  &   Age adjusted  &   Crude   &   Age adjusted  \\
    \midrule
    XV           &                         Arica y Parinacota &  23.8 &   68.8 &   50.9 &   14.1 &   10.0 \\
    I            &                                   Tarapacá &  35.1 &   33.9 &   28.3 &   11.8 &   10.1 \\
    II           &                                Antofagasta &  24.1 &   50.0 &   41.0 &   11.8 &    9.9 \\
    III          &                                    Atacama &  31.9 &   37.0 &   28.2 &   11.6 &    9.0 \\
    IV           &                                   Coquimbo &  33.5 &   43.3 &   30.7 &   14.2 &    9.7 \\
    V            &                                 Valparaíso &  27.9 &   70.5 &   45.5 &   19.6 &   11.7 \\
    RM           &                  Metropolitana de Santiago &  24.5 &   66.6 &   46.8 &   16.2 &   10.8 \\
    VI           &      Libertador General Bernardo O'Higgins &  41.4 &   39.7 &   27.3 &   15.3 &   10.5 \\
    VII          &                                      Maule &  28.6 &   48.6 &   33.8 &   13.7 &    9.3 \\
    XVI          &                                      Ñuble &  30.5 &   56.6 &   37.8 &   17.2 &   10.9 \\
    VIII         &                                     Biobío &  27.8 &   55.6 &   39.0 &   15.3 &   10.4 \\
    IX           &                               La Araucanía &  35.3 &   44.4 &   31.0 &   14.0 &    9.4 \\
    XIV          &                                   Los Ríos &  24.2 &   62.5 &   42.7 &   13.4 &    8.9 \\
    X            &                                  Los Lagos &  32.5 &   36.4 &   26.5 &   11.6 &    8.1 \\
    XI           &  Aysén del General Carlos Ibáñez del Campo &  40.1 &   38.4 &   30.0 &   13.2 &   10.6 \\
    XII          &       Magallanes y de la Antártica Chilena &  37.1 &   58.0 &   39.3 &   20.3 &   13.1 \\
    \bottomrule
    \end{tabular}
	\label{tab:incidence_and_mortality_by_region}
\end{table}
\disabletablerowcolor 

\subsection{Breast cancer incidence by age group}

Table \ref{tab:incidence_crude_age_year} presents age--specific incidence rates for each age interval for each year from 2007 to 2018.

\enabletablerowcolor[2] 
\begin{table}
	\centering
	\caption{Crude incidence rate per year and age group (cases/100,000 women).}
    \begin{tabular}{lrrrrrrrrrrrrl}
    \toprule
    Age &   2007 &   2008 &   2009 &   2010 &   2011 &   2012 &   2013 &   2014 &   2015 &   2016 &   2017 &   2018 &    Mean (Std) \\
    \midrule
    0-19  &    0.6 &    0.6 &    0.7 &    0.8 &    0.6 &    0.7 &    0.8 &    0.6 &    0.8 &    0.5 &    0.7 &    0.5 &     0.7 (0.1) \\
    20-24 &    2.0 &    3.2 &    3.8 &    2.3 &    3.3 &    3.0 &    3.8 &    2.0 &    2.2 &    3.5 &    4.5 &    3.1 &     3.1 (0.8) \\
    25-29 &    7.1 &    8.0 &    5.1 &    7.7 &    6.1 &    5.6 &    6.5 &    5.8 &    6.4 &    8.5 &    6.8 &    8.2 &     6.8 (1.0) \\
    30-34 &   11.6 &   12.1 &   14.0 &   14.0 &   13.0 &   18.9 &   16.0 &   15.2 &   17.0 &   20.5 &   17.8 &   17.9 &    15.7 (2.7) \\
    35-39 &   29.1 &   28.6 &   36.5 &   30.9 &   35.2 &   36.7 &   36.9 &   34.6 &   38.8 &   36.6 &   36.0 &   35.7 &    34.6 (3.1) \\
    40-44 &   60.5 &   61.2 &   66.9 &   59.7 &   67.1 &   74.4 &   72.9 &   68.1 &   70.5 &   80.3 &   72.1 &   65.2 &    68.2 (5.9) \\
    45-49 &   96.7 &   92.8 &  111.3 &  102.4 &  111.5 &  117.7 &  108.3 &  103.0 &  113.7 &  121.7 &   98.7 &  105.9 &   107.0 (8.3) \\
    50-54 &  104.2 &  108.1 &  125.1 &  113.5 &  113.7 &  125.0 &  122.2 &  110.0 &  133.0 &  130.7 &  108.5 &  118.0 &   117.7 (9.1) \\
    55-59 &  121.6 &  113.3 &  140.7 &  120.7 &  122.1 &  122.4 &  140.9 &  131.4 &  142.1 &  130.9 &  131.0 &  124.9 &   128.5 (8.9) \\
    60-64 &  150.6 &  156.5 &  164.7 &  160.0 &  161.1 &  164.7 &  158.3 &  146.3 &  142.9 &  143.7 &  146.6 &  138.9 &   152.9 (8.7) \\
    65-69 &  160.1 &  179.5 &  172.1 &  172.4 &  182.6 &  186.0 &  197.3 &  175.4 &  181.4 &  170.5 &  174.6 &  150.6 &  175.2 (11.5) \\
    70-74 &  147.3 &  145.8 &  175.1 &  181.8 &  177.8 &  179.8 &  187.7 &  195.8 &  197.0 &  188.1 &  187.8 &  186.8 &  179.2 (15.9) \\
    75-79 &  154.1 &  164.4 &  176.3 &  160.7 &  172.3 &  178.3 &  172.8 &  165.8 &  190.1 &  198.6 &  182.6 &  171.4 &  174.0 (11.9) \\
    80-84 &  178.6 &  163.7 &  164.9 &  189.2 &  168.8 &  163.1 &  179.7 &  172.2 &  165.8 &  168.2 &  148.5 &  143.6 &  167.2 (12.0) \\
    85+   &  235.9 &  236.0 &  228.9 &  206.2 &  209.6 &  172.6 &  178.4 &  171.4 &  147.7 &  165.6 &  171.2 &  107.0 &  185.9 (37.2) \\
    \bottomrule
    \end{tabular}
	\label{tab:incidence_crude_age_year}
\end{table}
\disabletablerowcolor 

\subsection{Breast cancer mortality by age group}

Table \ref{tab:mortality_crude_age_year} presents age--specific mortality rates for each age interval for each year from 2007 to 2018.

\enabletablerowcolor[2] 
\begin{table}
	\centering
	\caption{Crude mortality rate by year and age group (cases/100,000 women).}
    \begin{tabular}{lrrrrrrrrrrrrl}
    \toprule
    Age &   2007 &   2008 &   2009 &   2010 &   2011 &   2012 &   2013 &   2014 &   2015 &   2016 &   2017 &   2018 &    Mean (Std) \\
    \midrule
    0-19  &    0.0 &    0.0 &    0.0 &    0.0 &    0.0 &    0.0 &    0.0 &    0.0 &    0.0 &    0.0 &    0.0 &    0.0 &     0.0 (0.0) \\
    20-24 &    0.0 &    0.0 &    0.1 &    0.1 &    0.0 &    0.1 &    0.0 &    0.0 &    0.1 &    0.0 &    0.1 &    0.1 &     0.1 (0.1) \\
    25-29 &    0.5 &    1.2 &    0.9 &    0.9 &    1.1 &    0.3 &    1.2 &    0.9 &    0.4 &    0.5 &    0.3 &    1.0 &     0.8 (0.3) \\
    30-34 &    1.1 &    3.0 &    1.4 &    2.0 &    2.0 &    1.9 &    2.0 &    2.1 &    2.6 &    2.1 &    2.9 &    3.4 &     2.2 (0.6) \\
    35-39 &    4.6 &    5.3 &    5.5 &    6.0 &    4.0 &    4.9 &    5.0 &    7.0 &    6.2 &    4.8 &    4.1 &    5.2 &     5.2 (0.8) \\
    40-44 &   11.1 &    9.3 &    9.7 &   11.4 &    9.6 &   12.3 &   11.1 &   10.2 &   10.6 &   10.7 &    8.7 &   12.8 &    10.6 (1.2) \\
    45-49 &   18.8 &   16.6 &   19.6 &   18.9 &   17.2 &   17.2 &   19.7 &   17.7 &   17.5 &   17.3 &   14.8 &   14.1 &    17.4 (1.6) \\
    50-54 &   23.3 &   24.7 &   25.7 &   25.4 &   26.7 &   22.0 &   21.2 &   24.5 &   22.5 &   24.7 &   25.3 &   18.8 &    23.7 (2.2) \\
    55-59 &   33.4 &   29.0 &   32.2 &   31.0 &   29.0 &   29.6 &   28.9 &   27.8 &   32.8 &   30.0 &   29.7 &   29.5 &    30.3 (1.7) \\
    60-64 &   34.4 &   40.8 &   42.4 &   39.1 &   38.0 &   36.4 &   38.1 &   29.1 &   33.4 &   31.7 &   38.0 &   30.8 &    36.0 (4.0) \\
    65-69 &   45.4 &   48.0 &   53.2 &   47.8 &   48.0 &   44.9 &   43.3 &   46.4 &   47.6 &   47.2 &   38.7 &   39.7 &    45.8 (3.7) \\
    70-74 &   52.2 &   55.1 &   52.5 &   47.5 &   60.3 &   56.9 &   63.4 &   56.9 &   54.4 &   62.1 &   60.1 &   49.2 &    55.9 (4.8) \\
    75-79 &   60.8 &   72.6 &   70.7 &   69.0 &   55.6 &   68.9 &   61.3 &   66.3 &   74.3 &   69.1 &   71.6 &   83.2 &    68.6 (6.9) \\
    80-84 &   83.2 &   81.8 &   91.7 &   85.8 &  102.4 &   88.8 &   95.6 &   84.7 &   99.6 &   77.5 &   73.6 &   90.8 &    88.0 (8.3) \\
    85+   &  185.3 &  175.1 &  192.0 &  153.2 &  170.4 &  173.4 &  139.1 &  164.8 &  150.5 &  139.6 &  142.1 &  144.3 &  160.8 (17.7) \\
    \bottomrule
    \end{tabular}
	\label{tab:mortality_crude_age_year}
\end{table}
\disabletablerowcolor 

\section{Survival Records}\label{app:survival}

The Kaplan--Meier curves for women in FONASA are made with the events observed and censored shown in table \ref{tab:KM_public}. On the other hand, table \ref{tab:KM_private} shows the events observed and censored used to make the Kaplan--Meier curves for women in ISAPRE. The event table for both FONASA and ISAPRE women can be obtained by adding both tables.

Event tables for other Kaplan--Meier curves shown in the results section are not included due to their extension.

\enabletablerowcolor[2] 
\begin{table}
    \centering
    \begin{tabular}{lrrrr}
    \toprule
    Time (months) &  Removed &  Observed &  Censored &   At risk \\
    \midrule
    0        &      796 &       712 &        84 &    43,025 \\
    1        &      692 &       333 &       359 &    42,229 \\
    2        &      614 &       247 &       367 &    41,537 \\
    3        &      545 &       195 &       350 &    40,923 \\
    4        &      466 &       189 &       277 &    40,378 \\
    5        &      508 &       142 &       366 &    39,912 \\
    6        &      516 &       187 &       329 &    39,404 \\
    7        &      537 &       157 &       380 &    38,888 \\
    8        &      499 &       142 &       357 &    38,351 \\
    9        &      483 &       134 &       349 &    37,852 \\
    10       &      550 &       154 &       396 &    37,369 \\
    11       &      456 &       153 &       303 &    36,819 \\
    12       &      475 &       140 &       335 &    36,363 \\
    13       &      437 &       128 &       309 &    35,888 \\
    14       &      440 &       129 &       311 &    35,451 \\
    15       &      433 &       121 &       312 &    35,011 \\
    16       &      427 &       126 &       301 &    34,578 \\
    17       &      469 &       114 &       355 &    34,151 \\
    18       &      434 &       121 &       313 &    33,682 \\
    19       &      479 &       134 &       345 &    33,248 \\
    20       &      482 &       128 &       354 &    32,769 \\
    21       &      389 &       114 &       275 &    32,287 \\
    22       &      477 &       112 &       365 &    31,898 \\
    23       &      444 &       137 &       307 &    31,421 \\
    24       &      427 &       104 &       323 &    30,977 \\
    25       &      444 &        87 &       357 &    30,550 \\
    26       &      418 &       109 &       309 &    30,106 \\
    27       &      423 &       105 &       318 &    29,688 \\
    28       &      376 &        88 &       288 &    29,265 \\
    29       &      418 &        93 &       325 &    28,889 \\
    30       &      425 &       104 &       321 &    28,471 \\
    \bottomrule
    \end{tabular}
     \hspace{2cm} \begin{tabular}{lrrrr}
    \toprule
    Time (months) &  Removed &  Observed &  Censored &   At risk \\
    \midrule
    31       &      349 &        91 &       258 &    28,046 \\
    32       &      415 &        89 &       326 &    27,697 \\
    33       &      408 &        73 &       335 &    27,282 \\
    34       &      400 &        91 &       309 &    26,874 \\
    35       &      376 &        89 &       287 &    26,474 \\
    36       &      393 &        63 &       330 &    26,098 \\
    37       &      333 &        56 &       277 &    25,705 \\
    38       &      410 &        74 &       336 &    25,372 \\
    39       &      392 &        63 &       329 &    24,962 \\
    40       &      352 &        72 &       280 &    24,570 \\
    41       &      402 &        69 &       333 &    24,218 \\
    42       &      378 &        59 &       319 &    23,816 \\
    43       &      385 &        59 &       326 &    23,438 \\
    44       &      339 &        66 &       273 &    23,053 \\
    45       &      355 &        69 &       286 &    22,714 \\
    46       &      379 &        59 &       320 &    22,359 \\
    47       &      342 &        65 &       277 &    21,980 \\
    48       &      356 &        49 &       307 &    21,638 \\
    49       &      295 &        52 &       243 &    21,282 \\
    50       &      322 &        61 &       261 &    20,987 \\
    51       &      340 &        52 &       288 &    20,665 \\
    52       &      314 &        54 &       260 &    20,325 \\
    53       &      333 &        42 &       291 &    20,011 \\
    54       &      282 &        48 &       234 &    19,678 \\
    55       &      288 &        45 &       243 &    19,396 \\
    56       &      327 &        46 &       281 &    19,108 \\
    57       &      290 &        36 &       254 &    18,781 \\
    58       &      312 &        51 &       261 &    18,491 \\
    59       &      258 &        42 &       216 &    18,179 \\
    60       &    17,921 &        48 &     17,873 &    17,921 \\
    \bottomrule
    \end{tabular}
    \caption{Event table for the survival curve for patients in the public health system.}
    \label{tab:KM_public}
\end{table}
\disabletablerowcolor 

\newpage
\enabletablerowcolor[2] 
\begin{table}
    \centering
    \begin{tabular}{lrrrr}
    \toprule
    Time (months) &  Removed &  Observed &  Censored &   At risk \\
    \midrule
    0        &       70 &        64 &         6 &    11,147 \\
    1        &      143 &        33 &       110 &    11,077 \\
    2        &      140 &        14 &       126 &    10,934 \\
    3        &      143 &        31 &       112 &    10,794 \\
    4        &      109 &        29 &        80 &    10,651 \\
    5        &      141 &        20 &       121 &    10,542 \\
    6        &      105 &        23 &        82 &    10,401 \\
    7        &       94 &        13 &        81 &    10,296 \\
    8        &      107 &        17 &        90 &    10,202 \\
    9        &      110 &        15 &        95 &    10,095 \\
    10       &      101 &        17 &        84 &     9,985 \\
    11       &       83 &        12 &        71 &     9,884 \\
    12       &      104 &        11 &        93 &     9,801 \\
    13       &      119 &        15 &       104 &     9,697 \\
    14       &      112 &        17 &        95 &     9,578 \\
    15       &      119 &         9 &       110 &     9,466 \\
    16       &      132 &        20 &       112 &     9,347 \\
    17       &      112 &        15 &        97 &     9,215 \\
    18       &      116 &        15 &       101 &     9,103 \\
    19       &      129 &        19 &       110 &     8,987 \\
    20       &       99 &        15 &        84 &     8,858 \\
    21       &       78 &        11 &        67 &     8,759 \\
    22       &       96 &        11 &        85 &     8,681 \\
    23       &       68 &        13 &        55 &     8,585 \\
    24       &      117 &        23 &        94 &     8,517 \\
    25       &      121 &         9 &       112 &     8,400 \\
    26       &      126 &        16 &       110 &     8,279 \\
    27       &      113 &        16 &        97 &     8,153 \\
    28       &      116 &        11 &       105 &     8,040 \\
    29       &      133 &        12 &       121 &     7,924 \\
    30       &      110 &         4 &       106 &     7,791 \\
    \bottomrule
    \end{tabular}
     \hspace{2cm} \begin{tabular}{lrrrr}
    \toprule
    Time (months) &  Removed &  Observed &  Censored &   At risk \\
    \midrule
    31       &      112 &        10 &       102 &     7,681 \\
    32       &      119 &        18 &       101 &     7,569 \\
    33       &      115 &        10 &       105 &     7,450 \\
    34       &      127 &         9 &       118 &     7,335 \\
    35       &       76 &        10 &        66 &     7,208 \\
    36       &      126 &         7 &       119 &     7,132 \\
    37       &      109 &        18 &        91 &     7,006 \\
    38       &      120 &        13 &       107 &     6,897 \\
    39       &       88 &        11 &        77 &     6,777 \\
    40       &       85 &        10 &        75 &     6,689 \\
    41       &       95 &        14 &        81 &     6,604 \\
    42       &       96 &         6 &        90 &     6,509 \\
    43       &      102 &        14 &        88 &     6,413 \\
    44       &      105 &        19 &        86 &     6,311 \\
    45       &       90 &        10 &        80 &     6,206 \\
    46       &      119 &         8 &       111 &     6,116 \\
    47       &       76 &         8 &        68 &     5,997 \\
    48       &      109 &         8 &       101 &     5,921 \\
    49       &      105 &         7 &        98 &     5,812 \\
    50       &      101 &         8 &        93 &     5,707 \\
    51       &      122 &        12 &       110 &     5,606 \\
    52       &       67 &         6 &        61 &     5,484 \\
    53       &       86 &         8 &        78 &     5,417 \\
    54       &       81 &         8 &        73 &     5,331 \\
    55       &       97 &         6 &        91 &     5,250 \\
    56       &       73 &         6 &        67 &     5,153 \\
    57       &       94 &         6 &        88 &     5,080 \\
    58       &       91 &        11 &        80 &     4,986 \\
    59       &       68 &        11 &        57 &     4,895 \\
    60       &     4,827 &         5 &      4,822 &     4,827 \\
    \bottomrule
    \end{tabular}
    \caption{Event table for the survival curve for patients in the private health system.}
    \label{tab:KM_private}
\end{table}
\disabletablerowcolor

\end{landscape}

\section{Cox Regression Analysis}\label{app:cox_regression}

The variables were processed prior to the selection. We added the squared age as a new variable to add more flexibility to the hazard modeling as a function of age and normalized both age and squared age, to a maximum age of 100 years. We transformed the categorical variables (health insurance, region of residence and FONASA benefit segments) into dummy variables. This process resulted in a total of 26 covariables: 3 health insurance dummy variables accounting for public, private and armed forces health care insurance; 4 benefit segment dummy variables, namely segments A, B, C and D; 16 region dummy variables; year of diagnosis; normalized age and squared normalized age. The selection of variables was performed following a greedy algorithm based on Akaike’s information criterion and on the p value for each variable’s significance; that is, we sequentially selected variables by adding the ones that mostly improved Akaike’s information criterion while maintaining all variables’ p values as statistically significant.

More formally, given a set of n variables from which to choose which ones generate the best model, we proceeded based on each variable’s p value and the models' Akaike information criterion as follows.

\begin{sourcecode}{pseudocode}{Pseudocode for greedy variable selection for Cox model.}
Variable selection:
    Set the current best variables as an empty list.
    Set the current best Akaike value as infinity.
    Set the n variables as a pending list.
    While there are still variables in the pending list, do:
        For each variable from the pending list, generate a Cox model with it and the selected variables.
        For each such model, check if all their variables have significant p values ($\leq$0.05). If none of the models comply with it, break.
        For each model that meets the above:
            Calculate the Akaike information criterion.
            If the calculated Akaike value is lower than the current best, set such value as the current best Akaike value and add such variable to the current best variables, removing it from the pending list. 
        If none of the above models achieves an Akaike value better than the current best, break.
    Return the model generated with the current best variables.
\end{sourcecode}

For the 26 variables mentioned above, the previous process selected the following 11 variables: normalized age, square normalized age, private health insurance dummy, public health insurance dummy, discharge year, A FONASA beneficiary classification dummy, residence in the Arica y Parinacota region dummy, residence in the Antofagasta region dummy, residence in the Valparaíso region dummy, residence in the Metropolitan region dummy and residence in the Libertador O’Higgins region dummy.

\end{appendixd}
\end{document}